\documentclass[3p]{elsarticle}
\journal{}

\usepackage{graphicx}
\usepackage{amsmath}
\usepackage{amssymb}
\usepackage{amsthm}
\usepackage{latexsym}
\usepackage{bm}
\usepackage{color}
\usepackage{stmaryrd}
\usepackage{mathrsfs}
\usepackage{array}
\usepackage{algorithm}
\usepackage{algpseudocode}
\usepackage{multirow}
\usepackage{url}
\usepackage{mdframed}

\SetSymbolFont{stmry}{bold}{U}{stmry}{m}{n}
\newtheorem{remark}{Remark}
\numberwithin{equation}{section}
\newcolumntype{P}[1]{>{\centering\arraybackslash}p{#1}}
\usepackage{etoolbox}

\newtheorem{assumption}{Assumption}
\AfterEndEnvironment{assumption}{\noindent\ignorespaces}

\makeatletter
\newcommand*{\rom}[1]{\expandafter\@slowromancap\romannumeral #1@}
\makeatother

\newenvironment{myenv}[1]
  {\mdfsetup{
    frametitle={\colorbox{white}{\space#1\space}},
    innertopmargin=10pt,
    frametitleaboveskip=-\ht\strutbox,
    frametitlealignment=\center
    }
  \begin{mdframed}
  }
  {\end{mdframed}}

\begin{document}

\title{A reduced unified continuum formulation for \\ vascular fluid-structure interaction}

\author[1]{Ingrid S. Lan}
\ead{ingridl@stanford.edu}

\author[2,3]{Ju Liu\corref{cor1}}
\ead{liuj36@sustech.edu.cn}

\author[4]{Weiguang Yang}
\ead{wgyang@stanford.edu}

\author[1,4,5]{Alison L. Marsden}
\ead{amarsden@stanford.edu}

\cortext[cor1]{Corresponding author}
\address[1]{Department of Bioengineering, Stanford University, Stanford, CA 94305, USA}
\address[2]{Department of Mechanics and Aerospace Engineering, Southern University of Science and Technology, Shenzhen, Guangdong 518055, P.R China}
\address[3]{Guangdong-Hong Kong-Macao Joint Laboratory for Data-Driven Fluid Mechanics and Engineering Applications, Southern University of Science and Technology, Shenzhen, Guangdong 518055, P.R. China}
\address[4]{Department of Pediatrics (Cardiology), Stanford University, Stanford, CA 94305, USA}
\address[5]{Institute for Computational and Mathematical Engineering, Stanford University, Stanford, CA 94305, USA}

\begin{abstract}
We recently derived the unified continuum and variational multiscale formulation for fluid-structure interaction (FSI) using the Gibbs free energy as the thermodynamic potential. Restricting our attention to vascular FSI, we now reduce this formulation in arbitrary Lagrangian-Eulerian (ALE) coordinates by adopting three common modeling assumptions for the vascular wall. The resulting semi-discrete formulation, referred to as the reduced unified continuum formulation, achieves monolithic coupling of the FSI system in the Eulerian frame through a simple modification of the fluid boundary integral. While ostensibly similar to the semi-discrete formulation of the coupled momentum method introduced by Figueroa et al., its underlying derivation does not rely on an assumption of a fictitious body force in the elastodynamics sub-problem and therefore represents a direct simplification of the ALE method. Furthermore, uniform temporal discretization of the entire FSI system is performed via the generalized-$\alpha$ scheme. In contrast to the predominant approach yielding only first-order accuracy for pressure, we collocate both pressure and velocity at the intermediate time step to achieve uniform second-order temporal accuracy. In conjunction with quadratic tetrahedral elements, our methodology offers higher-order temporal and spatial accuracy for quantities of clinical interest, including pressure and wall shear stress. Furthermore, without loss of consistency, a segregated predictor multi-corrector algorithm is developed to preserve the same block structure as for the incompressible Navier-Stokes equations in the implicit solver's associated linear system. Block preconditioning of a monolithically coupled FSI system is therefore made possible for the first time. Compared to alternative preconditioners, our three-level nested block preconditioner, which achieves improved representation of the Schur complement, demonstrates robust performance over a wide range of physical parameters. We present verification of our methodology against Womersley's deformable wall theory and additionally develop practical modeling techniques for clinical applications, including tissue prestressing. We conclude with an assessment of our combined FSI technology in two patient-specific cases.
\end{abstract}


\maketitle

\section{Introduction}
\label{sec:introduction}

Fluid-structure interaction (FSI) problems present the challenge of coupling a deformable structural problem to a fluid problem posed on a domain moving in accordance with the deforming structure. In the last four decades, both interface-tracking and interface-capturing methods have been developed to account for the deforming fluid domain. In interface-tracking methods, the coupling interface is resolved by the mesh, and the arbitrary Lagrangian-Eulerian (ALE) formulation is adopted to describe mechanics problems posed on a moving domain \cite{Hirt1974,Hughes1981,Donea1982}; in interface-capturing methods, including the immersed boundary \cite{Peskin1972,Mittal2005} and fictitious domain methods \cite{Baaijens2001}, the interface is described implicitly on a background mesh. Whereas applications in cardiac mechanics involving valve leaflet motion largely employ the interface capturing method \cite{Borazjani2013,Griffith2009,Kamensky2015,Hart2003,Loon2006}, ventricular and vascular wall deformation are typically modeled with the ALE method \cite{Wu2014,Hsu2014,Liu2018,Liu2020b}, allowing for hemodynamic attributes near the wall to be accurately resolved for clinical implications.

In addition to this classification of FSI formulations, FSI coupling strategies can also be categorized into monolithic and partitioned approaches. In monolithic approaches, the coupling conditions, namely the continuity of velocity and stress at the fluid-solid interface, are exactly satisfied \cite{Fernandez2011}. Despite their superior robustness, the resulting system is highly nonlinear \cite{Nobile2013,Wu2014}, requires novel algorithms for the coupled system, and necessitates additional implementation efforts. On the other hand, partitioned methods are generally favored for their modularity, as existing fluid and structure codes can be independently used and loosely coupled via transmission conditions at the fluid-solid interface. Partitioned methods, however, were initially developed for aeroelastic problems \cite{Piperno2000}, in which the structural density is much larger than the fluid density. Numerical instabilities arise in problems involving fluid and structural densities of comparable magnitudes. This so-called added mass effect \cite{Badia2008,Causin2005,Guidoboni2009} does not vanish with time step refinement and is particularly pronounced in hydroelastic problems such as cardiovascular FSI problems, where the fluid and structural densities are almost identical. Many approaches, such as generalized Robin-to-Robin transmission conditions \cite{Badia2008a}, have been proposed to improve the stability of partitioned algorithms under the added mass effect. Yet, recent results also suggest that this improved stability may actually be at the expense of critical dynamic characteristics of the structural sub-problem \cite{Kadapa2021}, signifying an alarming issue regarding partitioned approaches for hydroelastic problems.

In this work on vascular FSI, we adopt our recently developed unified continuum and variational multiscale (VMS) formulation \cite{Liu2018}, a monolithically coupled ALE method. Derived using the Gibbs free energy rather than the Helmholtz free energy as the thermodynamic potential, the formulation bridges the conventionally diverging approaches for computational fluid and solid mechanics. Its ability to naturally recover important continuum models, including viscous fluids and hyperelastic solids, through appropriate constitutive modeling drastically simplifies monolithic FSI coupling. Furthermore, the formulation is well-behaved in both compressible and incompressible regimes, enabling simulation of structural dynamics with a Poisson's ratio up to $0.5$. Given the nontrivial computational expense associated with an ALE formulation, we apply three common modeling assumptions concerning the strain magnitude, geometry, and constitutive model of the vascular wall--the infinitesimal strain, thin-walled, and linear elastic membrane assumptions, respectively--to arrive at our so-called \textit{reduced unified continuum formulation}. The resulting semi-discrete formulation presents a monolithically coupled FSI system posed in an Eulerian frame of reference, in which the structural velocity degrees of freedom are reduced to the fluid velocity degrees of freedom at the fluid-solid interface. 

Despite its ostensible similarity to the semi-discrete formulation of the coupled momentum method (CMM), first introduced by Figueroa et al. \cite{Figueroa2006} and recently extended to a nonlinear rotation-free shell formulation \cite{Nama2020}, the FSI coupling in CMM relies on an assumption of a fictitious body force in the elastodynamics sub-problem, defined in relation to the fluid traction on the wall. While this coupling approach was inspired by Womersley's derivation of an analytical solution for axisymmetric flow in an elastic pipe \cite{Womersley1955,Zamir2000}, we believe this assumption of a fictitious body force is unnecessary. Since its introduction, CMM has been implemented in the open-source blood flow simulation software packages SimVascular \cite{Updegrove2017,Lan2018} and CRIMSON \cite{Arthurs2020} and extensively used in clinical applications ranging from interventions for coronary artery disease \cite{Williams2010, Gundert2011, Taylor2013} and aortic coarctation \cite{Coogan2010} to single-ventricle physiology \cite{Yang2010}, and Alagille syndrome \cite{Yang2016,Yang2017}. It has also been validated against experimental measurements from compliant in vitro phantom models \cite{Kung2011, Kung2011a} and Womersley's analytical solution for axisymmetric flow in a thin, linear elastic pipe subject to an oscillating pressure gradient \cite{Figueroa2006a,Filonova2019}. While the studies found good agreement for pressure, flow, pulse wave propagation, and wall displacement, Filonova et al. \cite{Filonova2019} documented large errors in radial velocity.

In this work, stabilized spatial discretization is performed with the residual-based variational multiscale formulation \cite{Bazilevs2007}, which retains numerical consistency across all scales and exhibits superior performance as a large eddy simulation turbulence model when compared to approaches employing traditional stabilized formulations. We further note that integration-by-parts is not adopted for the divergence operator in the continuity equation for two reasons. First, from an energy perspective, the additional boundary integral term produced from integration-by-parts could pollute the energy dissipation structure in the discrete scheme. In addition, integration-by-parts yields a contradiction in the regularity of the pressure function space in the Galerkin formulation, though this contradiction seems not to produce apparent numerical issues when the stabilized formulation is invoked.

The generalized-$\alpha$ method was initially proposed for structural dynamics as an unconditionally stable and second-order accurate implicit scheme for temporal discretization with user-specified levels of high-frequency dissipation \cite{Chung1993}. When Jansen et al. first applied the generalized-$\alpha$ method to the compressible Navier-Stokes equations \cite{Jansen2000}, the pressure primitive variables were uniformly evaluated at the intermediate time step $t_{n+\alpha_f}$. The predominant approach in the computational fluid dynamics (CFD) and FSI communities today \cite{Figueroa2006, Bazilevs2007, Bazilevs2008, Bazilevs2012, Joshi2018, Kang2012}, however, is to evaluate velocity at the intermediate step but pressure at time step $t_{n+1}$. While the community continues to reference the second-order temporal accuracy of the generalized-$\alpha$ method, we recently demonstrated that this particular dichotomous approach yields only first-order accuracy in pressure. Concurrent evaluation of velocity and pressure at the intermediate step, as in our approach, recovers second-order accuracy for pressure \cite{Liu2020a}.

In contrast to the use of the Newmark-$\beta$ method \cite{Newmark1959} in CMM \cite{Figueroa2006} for temporal integration of membrane displacements, we adopt the fully implicit generalized-$\alpha$ method for uniform temporal discretization of both the fluid and solid sub-problems, enabling second-order accuracy and high-frequency dissipation simultaneously in the full FSI system. With a segregated predictor-multicorrector algorithm we previously used for the unified continuum and VMS formulation \cite{Liu2018,Liu2020}, the three-by-three block structure for the matrix problem in the consistent Newton-Raphson procedure can be reduced to a two-by-two block structure identical to that of the incompressible Navier-Stokes equations. Not only does this segregated algorithm preserve the consistency of the Newton-Raphson method, but it also enables the use of existing CFD solvers with only minimal modifications. We exploit this preserved two-by-two block structure for preconditioning of the linear system with our three-level nested block preconditioner \cite{Liu2020}, which attains improved representation of the Schur complement with a ``matrix-free" technique to algorithmically define the action of the Schur complement on a vector. Our nested block preconditioning technique is thus robust for cardiovascular simulations involving several contributing terms in the Schur complement of widely varying orders of magnitude, associated with convection, diffusion, vascular wall stiffness, and reduced models at the outlets representing downstream vasculature. We further note that our study represents the first in which block preconditioning is performed for a monolithically coupled FSI system.

The body of this work is organized as follows. In Section \ref{sec:formulation}, the unified continuum and VMS formulation is simplified to our reduced unified continuum formulation for vascular FSI via three modeling assumptions. The spatiotemporal discretization methods and the associated predictor multi-corrector algorithm are also presented. In Section \ref{sec:iterative_solution_method}, the preconditioning technique for the associated linear system is developed. In Section \ref{sec:verification}, verification of our reduced unified continuum formulation is performed against the rigid and deformable Womersley benchmark cases using both linear and quadratic tetrahedral elements. Verification of CMM with linear elements is also presented for comparison. In Section \ref{sec:practical_modeling_techniques}, we discuss practical modeling techniques for capturing physiological behavior in patient-specific clinical applications. Among these practical modeling techniques is tissue prestressing to account for the nonzero internal stress state of the vascular wall at imaging, which we iteratively update via fixed-point iterations while solving a modified problem over the vascular wall under a fluid traction corresponding to the cardiac phase at imaging. We additionally present a centerline-based approach for variable wall thickness assignment to avoid unphysiological thicknesses produced by previous Laplacian approaches at regions of sharp local changes in geometry. Finally, we conclude with an assessment of our combined FSI technology with two patient-specific cases in Section \ref{sec:clinical_applications}.

\section{Governing equations and their spatiotemporal discretization}
\label{sec:formulation}
In this section, we introduce the strong and weak forms of the elastodynamic and incompressible Newtonian fluid problems following the unified continuum formulation \cite{Liu2018} and outline the assumptions yielding our reduced unified continuum formulation in the Eulerian description. This monolithically coupled FSI system is then integrated in time using the generalized-$\alpha$ method, which is solved by a segregated predictor multi-corrector algorithm.

\subsection{Strong-form problem}
We consider a domain $\Omega \subset \mathbb R^3$ admitting a non-overlapping subdivision $\overline{\Omega} = \overline{\Omega^f \cup \Omega^s}$, $\emptyset = \Omega^f \cap \Omega^s$, in which $\Omega^f$ and $\Omega^s$ represent the sub-domains occupied by the fluid and solid materials, respectively. The fluid-solid interface is a two-dimensional manifold denoted by $\Gamma_I$, and the boundary $\Gamma := \partial \Omega$ can be partitioned into four non-overlapping subdivisions:
\begin{align*}
\Gamma = \overline{\Gamma^f_g \cup \Gamma^f_h \cup \Gamma^s_g \cup \Gamma^s_h}, \mbox{ and } \emptyset = \Gamma_g^f \cap \Gamma_h^f  = \Gamma_g^f \cap \Gamma_g^s = \Gamma_g^f \cap \Gamma_h^s = \Gamma_h^f \cap \Gamma_h^s = \Gamma_h^f \cap \Gamma_g^s = \Gamma_g^s \cap \Gamma_h^s.
\end{align*}
In the above, the four subdivisions represent the Dirichlet part of the fluid boundary, the Neumann part of the fluid boundary, the Dirichlet part of the solid boundary, and the Neumann part of the solid boundary, respectively. We note that since the present theory involves multiple unknowns in $\mathbb R^3$, the boundary $\Gamma$ should in fact be generalized to admit a different decomposition for each component of each unknown \cite[p.77]{Hughes1987}. To simplify our presentation, however, we consider the same partition of $\Gamma$ for all unknowns here and note that practical problems would require generalization. We demand $\Gamma$ to be at least Lipschitz such that the outward normal vector $\bm n$ is well-defined almost everywhere. We also assume that the interior fluid-solid interface $\Gamma_I$ is sufficiently smooth such that its outward normal vector is well-defined. In particular, we use $\bm n^f$ and $\bm n^s$ to represent the unit outward normal vector on $\Gamma_{I}$ relative to $\Omega^f$ and $\Omega^s$ respectively, such that $\bm n^f = - \bm n^s$. Let the time interval of interest be denoted by $(0,T) \subset \mathbb R$, with $T>0$. With this geometric configuration in mind, we state the strong-form sub-problems separately for the two sub-domains.

Under the Stokes' hypothesis and isothermal condition, the initial-boundary value problem in the solid sub-domain $\Omega^s$ can be stated as follows in the Lagrangian description \cite{Liu2018}. Given the body force per unit mass $\bm b^s$, Dirichlet data $\bm g^s$, boundary traction $\bm h^s$, and initial displacement and velocity fields $\bm u_0^s$ and $\bm v_0^s$, find the solid displacement $\bm u^s$, pressure $p^s$, and velocity $\bm v^s$, such that
\begin{align}
\label{eq:ela_kinematics}
& \bm 0 = \frac{d \bm u^s}{d t} - \bm v^s, && \mbox{ in } \Omega^s \times (0,T), \displaybreak[2] \\
\label{eq:ela_mass}
& 0 = \beta^s_{\theta}(p^s) \frac{dp^s}{dt} + \nabla \cdot \bm v^s, && \mbox{ in } \Omega^s \times (0,T), \displaybreak[2] \\
\label{eq:ela_mom}
& \bm 0 = \rho^s(p^s) \frac{d \bm v^s}{d t}  - \nabla \cdot \bm \sigma^s_{\mathrm{dev}} + \nabla p^s - \rho^s(p^s) \bm b^s, && \mbox{ in } \Omega^s \times (0,T), \displaybreak[2] \\
\label{eq:dirichlet_bc_s}
& \bm u^s = \bm g^s, && \mbox{ on } \Gamma_g^s \times (0,T), \displaybreak[2] \\
\label{eq:neumann_bc_s}
&\bm \sigma^s \bm n = \bm h^s,  && \mbox{ on } \Gamma_h^s \times (0,T), \displaybreak[2] \\
\label{eq:initial_condition_s}
&\bm u^s(\cdot, 0) = \bm u_0^s(\cdot), && \mbox{ in } \bar{\Omega}^s, \displaybreak[2] \\
&\bm v^s(\cdot, 0) = \bm v_0^s(\cdot), && \mbox{ in } \bar{\Omega}^s.
\end{align}
Here, $\beta^s_{\theta}$ is the isothermal compressibility coefficient, $\rho^s$ is the solid density, and $\bm \sigma^s_{\mathrm{dev}}$ is the deviatoric component of the Cauchy stress. To characterize the material behavior, constitutive relations for $\beta^s_{\theta}$, $\rho^s$, and $\bm \sigma^s_{\mathrm{dev}}$ must be provided. Interested readers may refer to \cite[Sec.~2.4]{Liu2018} for an overview of various constitutive relations for $\beta_{\theta}^s$ and $\rho^s$ and their relations with different forms of volumetric free energies.
\begin{assumption}
\label{assumption:small_strain}
The solid deformation is small enough such that the infinitesimal strain theory is valid. 
\end{assumption}
Under the infinitesimal strain assumption, the reference and current frames coincide, as do the total ($d/dt$) and partial ($\partial / \partial t$) time derivatives in \eqref{eq:ela_kinematics}-\eqref{eq:ela_mom}. The density $\rho^s(p^s)$ takes the value in the reference configuration, denoted $\rho^s$.  Furthermore, one may show that $\beta_{\theta}(p^s) = 1/\kappa^s$, where $\kappa^s$ is the solid bulk modulus \cite[p.941]{Liu2019}. Integrating \eqref{eq:ela_mass} in time then yields
\begin{align}
\label{eq:small_strain_pressure_constitutive}
p^s = - \kappa^s \nabla \cdot \bm u^s.
\end{align}
The infinitesimal strain tensor is given by
\begin{align*}
\bm \epsilon(\bm u^s) := \frac12 \left( \nabla \bm u^s + \left( \nabla \bm u^s \right) ^T \right) = \bm \epsilon_{\mathrm{dev}}(\bm u^s) + \frac{1}{3} \nabla \cdot \bm u^s \bm I,
\end{align*}
where $\bm \epsilon_{\mathrm{dev}}$ is the strain deviator and $\bm I$ is the second-order identity tensor. Given a strain energy function $W(\bm \epsilon_{\mathrm{dev}})$, the stress deviator is then
\begin{align*}
\bm \sigma^s_{\mathrm{dev}} = \frac{\partial W(\bm \epsilon_{\mathrm{dev}})}{\partial \bm \epsilon} = \frac{\partial W(\bm \epsilon_{\mathrm{dev}})}{\partial \bm \epsilon_{\mathrm{dev}}} : \frac{\partial \bm \epsilon_{\mathrm{dev}}}{\partial \bm \epsilon} = \mathbb P^T \frac{\partial W(\bm \epsilon_{\mathrm{dev}})}{\partial \bm \epsilon_{\mathrm{dev}}}, \qquad \mathbb P := \mathbb I - \frac{1}{3} \bm I \otimes \bm I,
\end{align*}
where $\mathbb P$ is the deviatoric projector, and $\mathbb I$ is the fourth-order symmetric identity tensor. The Cauchy stress for the solid body thus takes the following form,
\begin{align*}
\bm \sigma^s := \bm \sigma^s_{\mathrm{dev}} - p^s \bm I = \mathbb P^T \frac{\partial W(\bm \epsilon_{\mathrm{dev}})}{\partial \bm \epsilon_{\mathrm{dev}}} + \kappa^s \nabla \cdot \bm u^s \bm I.
\end{align*}

\begin{remark}
As will be revealed, Assumption \ref{assumption:small_strain} renders the eventual FSI formulation implementationally and computationally appealing. While several promising results have been reported in the literature \cite{Colciago2014,Filonova2019}, its validity must be judiciously assessed under various physiological settings in both health and disease.
\end{remark}

While the fluid sub-problem in an ALE formulation is indeed posed on a moving domain that tracks the solid deformation, Assumption \ref{assumption:small_strain} guarantees this geometry adherence and renders mesh motion unnecessary. The initial-boundary value problem for the incompressible Newtonian fluid in the fluid sub-domain $\Omega^f$ can thus be stated as follows. Given the body force per unit mass $\bm b^f$, Dirichlet data $\bm g^f$, boundary traction $\bm h^f$, and divergence-free initial velocity field $\bm v^f_0$, find the fluid velocity $\bm v^f$ and pressure $p^f$, such that
\begin{align}
\label{eq:ns_mom}
& \bm 0 = \rho^f \frac{\partial \bm v^f}{\partial t}  + \rho^f  \bm v^f \cdot \nabla \bm v^f - \nabla \cdot \bm \sigma^f_{\mathrm{dev}} + \nabla p^f - \rho^f \bm b^f, && \mbox{ in } \Omega^f \times (0,T), \\
\label{eq:ns_mass}
& 0 = \nabla \cdot \bm v^f, && \mbox{ in } \Omega^f \times (0, T), \\
\label{eq:dirichlet_bc_f}
& \bm v^f = \bm g^f && \mbox{ on } \Gamma_g^f \times (0,T), \\
\label{eq:neumann_bc_f}
&\bm \sigma^f \bm n = \bm h^f  && \mbox{ on } \Gamma_h^f \times (0,T), \\
\label{eq:initial_condition_f}
&\bm v^f(\cdot, 0) = \bm v_0^f(\cdot), && \mbox{ in } \bar{\Omega}^f,
\end{align}
wherein
\begin{align}
\label{eq:ns_sigma}
\bm \sigma^f_{\mathrm{dev}} := 2\mu^f \bm \varepsilon_{\mathrm{dev}}(\bm v^f), \qquad \bm \varepsilon_{\mathrm{dev}}(\bm v^f) := \frac12 \left( \nabla \bm v^f + \left( \nabla \bm v^{f} \right)^T \right) - \frac13 \nabla \cdot \bm v^f \bm I.
\end{align}
Here, $\rho^f$ is the fluid density, $\bm \sigma^f_{\mathrm{dev}}$ is the deviatoric component of the Cauchy stress for a Newtonian fluid, $\mu^f$ is the dynamic viscosity, and $\bm \varepsilon_{\mathrm{dev}}$ is the deviatoric component of the rate-of-strain tensor.

The strong-form FSI problem can be completed with the following kinematic condition enforcing the continuity of velocity on $\Gamma_I$, 
\begin{align}
\label{eq:coupling-kinematic}
\bm v^f = \bm v^s, \qquad \hspace{1.3cm} \mbox{on } \Gamma_I,
\end{align}
and the following dynamic condition enforcing the continuity of stress,
\begin{align}
\label{eq:coupling-traction}
\bm \sigma^f \bm n^f = -\bm \sigma^s \bm n^s, \qquad \mbox{ on } \Gamma_I.
\end{align}
Together, Equations \eqref{eq:ela_kinematics}-\eqref{eq:coupling-traction} constitute the coupled strong-form FSI problem, in which the solid problem is restricted to small-strain elastodynamics.

\subsection{Semi-discrete formulation}
\label{subsec:semi-discrete-formulation}
In this section, we present the semi-discrete formulations for the two coupled sub-problems separately. By invoking two more assumptions for the vascular wall, we then reduce the elastodynamics formulation to a thin-walled, linear elastic membrane formulation, yielding a convenient FSI formulation that does not explicitly require solid degrees of freedom. We further note that the reduction to a membrane formulation conveniently bypasses the troublesome procedure of modeling the vascular wall, which current medical imaging techniques largely remain unable to accurately resolve \cite{Liu2020b}.

\subsubsection{Semi-discrete formulation for elastodynamics}
Let $\mathcal S_{\bm v}^s$ be the trial solution space for the solid velocity; let $\mathcal S_{\bm u}^s$ and $\mathcal V_{\bm u}^s$ denote the trial solution and test function spaces for the solid displacement. We can then state the semi-discrete elastodynamics formulation in $\Omega^s$ as follows. Find
\begin{align*}
\bm y^s_h(t):= \Big\lbrace \bm v^s_h(t), \bm u^s_h(t) \Big\rbrace^T \in \mathcal S^s_{\bm v} \times \mathcal S^s_{\bm u}
\end{align*}
such that
\begin{align*}
& \mathbf B^{s}_{\mathrm{k}}\Big( \dot{\bm y}^s_h, \bm y^s_h \Big) = \bm 0, && \displaybreak[2] \\
& \mathbf B^{s}_{\mathrm{m}}\Big( \bm w^s_h ;  \dot{\bm y}^s_h, \bm y^s_h \Big) = 0, && \forall \bm w^s_h \in \mathcal V^s_{\bm u},
\end{align*}
where
\begin{align}
\label{eq:semi-discrete-kinematic}
& \mathbf B^{s}_{\mathrm{k}} \Big( \dot{\bm y}_h^s, \bm y_h^s \Big) := \frac{d\bm u^s_h}{dt} - \bm v^s_h, \displaybreak[2] \\
\label{eq:semi-discrete-u}
& \mathbf B^{s}_{\mathrm{m}} \Big( \bm w_h^s ;  \dot{\bm y}_h^s, \bm y_h^s \Big) := \int_{\Omega^s} \bm w_h^s \cdot \rho^s \left( \frac{d \bm v_h^s}{d t} - \bm b^s \right) d\Omega  + \int_{\Omega^s} \bm \epsilon(\bm w_h^s) : \bm \sigma^s(\bm u_h^s)  d\Omega - \int_{\Gamma_h^s} \bm w_h^s \cdot \bm h^s d\Gamma,
\end{align}
with $\bm y^s_h(0) = \left\lbrace \bm v^s_0, \bm u^s_0 \right\rbrace^T$. Here, $\bm v^s_0$ and $\bm u^s_0$ are $\mathcal L^2$ projections of the initial velocity and displacement fields onto the discrete spaces $\mathcal S_{\bm v}^s$ and $\mathcal S_{\bm u}^s$, respectively.

\begin{remark}
In contrast to the conventional ``acceleration form" in which only displacement degrees-of-freedom are utilized, acceleration is represented here as the first time derivative of velocity via the kinematic relation \eqref{eq:semi-discrete-kinematic} \cite{Hulbert2017}. While this ``momentum form"  ostensibly introduces three additional velocity degrees of freedom on each node in $\Omega^s$, we will later show that \eqref{eq:semi-discrete-kinematic} does not enter the implicit solution procedure for the fully discrete formulation. Furthermore, as will be discussed later, this first-order structural dynamics formulation is favorable for temporal discretization via the generalized-$\alpha$ method.
\end{remark}
\noindent
Restricting our discussion to vascular FSI, we now introduce our second assumption pertaining to the vascular wall geometry. 
\begin{assumption}
\label{assumption:thin-wall}
$\Omega^s$ is thin in one direction and can thus be parameterized by the fluid-solid interface $\Gamma_I$ and a through-thickness coordinate in the unit outward normal direction.
\end{assumption}
To simplify our presentation, let $\Gamma_I$ be parameterized by a single chart $\Xi \subset \mathbb R^2$, a bounded open set. Let $\bm \chi(\xi,\eta)$ be a smooth one-to-one mapping of $(\xi,\eta) \in \Xi$ onto the fluid-solid interface $\bm \chi \in \Gamma_I$, where $\bm \chi$ represents the position vector of a generic point on $\Gamma_I$. The unit outward normal vector to $\Omega^f$ can be represented by
\begin{align*}
\bm n^f = \frac{\bm e_{\xi} \times \bm e_{\eta}}{\| \bm e_{\xi} \times \bm e_{\eta} \|}, \quad \mbox{ where } \quad \bm e_{\xi} := \frac{\partial \bm \chi}{\partial \xi} / \left \|\frac{\partial \bm \chi}{\partial \xi} \right\|, \quad \bm e_{\eta} := \frac{\partial \bm \chi}{\partial \eta} / \left \|\frac{\partial \bm \chi}{\partial \eta} \right\|.
\end{align*}
Given this thin-walled assumption, we can introduce the following diffeomorphism from $\bm \xi := \left\lbrace \xi, \eta, \zeta \right\rbrace \in \Xi \times (0,1)$ to $\bm x \in \Omega^s$,
\begin{align}
\label{eq:3Dshell-parameterization}
\bm x(\bm \xi) = \bm x(\xi,\eta,\zeta) := \bm \chi(\xi,\eta) + \zeta h^s(\xi,\eta) \bm n^f,
\end{align}
where $\xi$ and $\eta$ are the in-plane parametric coordinates, $h^s$ is the wall thickness as a function of $\xi$ and $\eta$, and $\zeta \in (0,1)$ is the through-thickness parametric coordinate. 
For any fixed $\zeta$, the surface defined by this parameterization of $\Omega^s$ is a lamina, and a corresponding lamina coordinate system $\{\bm e^l_1, \bm e^l_2, \bm e^l_3\}$, denoted with a superscript $l$, may be constructed as follows \cite[Sec.~6.2]{Hughes1987},
\begin{align*}
\bm e_1^l := \frac{\sqrt{2}}{2} \left( \bm e_\alpha - \bm e_\beta \right), \qquad
\bm e_2^l := \frac{\sqrt{2}}{2} \left( \bm e_\alpha + \bm e_\beta \right), \qquad
\bm e_3^l := \bm n^f,
\end{align*}
in which,
\begin{align*}
\bm e_\alpha := \frac12 \left( \bm e_\xi + \bm e_\eta \right) / \left \| \frac12 \left( \bm e_\xi + \bm e_\eta \right) \right \| , \quad
\bm e_\beta := \bm e_3^l \times \bm e_\alpha / \| \bm e_3^l \times \bm e_\alpha \|.
\end{align*}
With these lamina basis vectors $\left\lbrace \bm e_i^l \right\rbrace_{i=1}^{3}$, the coordinate transformation from the global coordinates $\bm x$ to the local lamina coordinates $\bm x^l$ is then given by $\bm x^l = \bm Q \bm x$ with the rotation matrix
\begin{align*}
\bm Q := 
\begin{bmatrix}
\bm e_1^l & \bm e_2^l & \bm e_3^l
\end{bmatrix}^T .
\end{align*}
From the parameterization \eqref{eq:3Dshell-parameterization}, we have
\begin{align*}
j :=& \mathrm{det}\left(\frac{\partial \bm x}{\partial \bm \xi}\right) = h^s \bm n ^f \cdot \left(\frac{\partial \bm x}{\partial \xi} \times \frac{\partial \bm x}{\partial \eta} \right) = h^s \bm n ^f \cdot \left( \left(\frac{\partial \bm \chi}{\partial \xi} + \zeta \frac{\partial h^s}{\partial \xi} \bm n^f \right) \times \left( \frac{\partial \bm \chi}{\partial \eta} + \zeta \frac{\partial h^s}{\partial \eta} \bm n^f  \right) \right) \nonumber \\
=& h^s \bm n ^f \cdot \left(\frac{\partial \bm \chi}{\partial \xi} \times \frac{\partial \bm \chi}{\partial \eta} \right),
\end{align*}
indicating the following transformation of the volume element from $\Xi \times (0,1)$ to $\Omega^s$,
\begin{align}
\label{eq:shell-volume-element}
d\Omega^s := d\bm x= j d\bm \xi = j d\xi d\eta d\zeta = h^s \bm n ^f \cdot \left(\frac{\partial \bm \chi}{\partial \xi} \times \frac{\partial \bm \chi}{\partial \eta} \right) d\xi d\eta d\zeta = h^s \bm n^f \cdot \bm n^f d\Gamma_I d\zeta = h^s d\Gamma_I d\zeta,
\end{align}
where we have utilized the transformation of the area element from $\Xi$ to $\Gamma_I$,
\begin{align*}
\bm n^f d\Gamma_I = \left(\frac{\partial \bm \chi}{\partial \xi} \times \frac{\partial \bm \chi}{\partial \eta} \right) d\xi d\eta.
\end{align*}
The volume integral over $\Omega^s$ can thus be simplified in the following manner,
\begin{align}
\label{eq:shell-volume-integral}
\int_{\Omega^s} \left( \cdot \right) d\Omega 
= \int_{\Gamma_I} h^s \int_0^1 \left( \cdot \right) d\zeta d\Gamma.
\end{align}
We finally introduce the following membrane assumption for the vascular wall.
\begin{assumption}
\label{assumption:membrane}
The displacement $\bm u^s$ is a function of the in-plane parametric coordinates $(\xi, \eta)$ only, and the transverse normal stress $\bm \sigma^s_{33}$ is zero in the $\bm e^l_3$ direction of the lamina system.
\end{assumption}
Cardiac pulse wavelengths are at least three orders of magnitude larger than arterial diameters \cite{Alastruey2012}, causing vessels to respond to transverse loading primarily with in-plane stresses rather than bending stresses. Out-of-plane rotations and their corresponding bending effects are thus neglected under this membrane assumption, minimizing the number of degrees of freedom and facilitating convenient fluid-solid coupling. In addition, to avoid thickness locking, also known as Poisson thickness locking in classical shell theories, the transverse normal stress is assumed to vanish, which has been well-substantiated over time \cite{Bazilevs2009a, Bischoff2004}. Furthermore, it is commonly known that when the linear, constant strain triangle is used to model membrane components experiencing transverse loads in three-dimensional structures, it suffers from severe transverse and in-plane shear locking, thereby demonstrating overly stiff behavior \cite{Bergan1985, Jun2018}. Transverse shear modes are therefore added to stabilize the linear membrane.

We now define the solid constitutive relation in the lamina coordinate system to enforce the zero transverse normal stress condition. Considering the strain energy for isotropic linear elasticity,
\begin{align*}
W(\bm \epsilon_{\mathrm{dev}}) = \mu \bm \epsilon_{\mathrm{dev}} : \bm \epsilon_{\mathrm{dev}},
\end{align*}
the constitutive relation is given by
\begin{align*}
\bm \sigma^{s, l}_{\mathrm{dev}} = 2\mu^s \bm \epsilon_{\mathrm{dev}}(\bm u^{s, l}).
\end{align*}
Recalling from \eqref{eq:small_strain_pressure_constitutive} that the hydrostatic component of the Cauchy stress is already given by $p^s = -\kappa^s \nabla \cdot \bm u^{s, l}$, the Cauchy stress can be written as
\begin{align*}
& \bm \sigma^{s, l} = \bm \sigma^{s, l}_{\mathrm{dev}} - p^s \bm I = \mathbb C^{s, l}  \bm \epsilon^{l}(\bm u^{s,l}), \quad \mbox{ with } \quad \mathbb C^{s, l} := 2 \mu^s(\bm x^l) \mathbb I + \lambda^s(\bm x^l) \bm I \otimes \bm I,
\end{align*}
wherein
\begin{align*}
&\bm \sigma^{s, l} = \left\lbrace \sigma^{s, l}_{I} \right\rbrace = \left[ \sigma_{11}^{s,l}, \sigma_{22}^{s,l}, \sigma_{12}^{s,l}, \sigma_{23}^{s,l} , \sigma_{31}^{s,l} \right]^T, \displaybreak[2] \\
&\bm \epsilon^{l}(\bm u^{s, l}) = \left\lbrace \epsilon^{l}_{I} \right\rbrace = \left[ \epsilon_{11}^{l}, 
\epsilon_{22}^{l}, 2 \epsilon_{12}^{l}, 2 \epsilon_{23}^{l}, 2 \varepsilon_{31}^{l} \right]^T = 
\left[ u_{1,1}^{s,l} , u_{2,2}^{s,l}, u_{1,2}^{s,l} + u_{2,1}^{s,l}, u_{3,2}^{s,l}, u_{3,1}^{s,l} \right]^T, \displaybreak[2] \\
& \mathbb C^{s, l} = \left[ \mathbb C^{s, l}_{IJ} \right] = \frac{E}{(1 - \nu^2)}
\begin{bmatrix}
1 & \nu &  & &  \\[1mm]
\nu & 1 & &  &  \\[1mm]
 &  & \displaystyle \frac{1 - \nu}{2} & &  \\[1mm]
 &  &  & \kappa \displaystyle \frac{(1 - \nu)}{2} &  \\[1mm]
 &  &  &  & \kappa \displaystyle \frac{ (1 - \nu)}{2} \\[1mm]
\end{bmatrix}
\end{align*}
in Voigt notation. Here, $\mu^s$ and $\lambda^s$ are the Lam\'e parameters related to the bulk modulus $\kappa^s$ through the relation $\kappa^s := 2\mu^s/3 + \lambda^s$, $E$ is the Young's modulus, $\nu$ is the Poisson's ratio, and $\kappa = 5/6$ is the shear correction factor \cite[p.391]{Hughes1987}. Now adopting the full tensor notation rather than Voigt notation, the Cauchy stress in the lamina coordinate system can be rotated to the global coordinate system by 
\begin{align*}
\bm \sigma^s = \bm Q^T \bm \sigma^{s,l} \bm Q.
\end{align*}
Assumption \ref{assumption:membrane} further enables evaluation of $\left( \cdot \right)$ in \eqref{eq:shell-volume-integral} at $\zeta = 0$, thereby reducing the volume integral over $\Omega^s$ to a surface integral over $\Gamma_I$,
\begin{align}
\label{eq:shell-volume-integral-to-surface-integral}
\int_{\Omega^s} \left( \cdot \right) d\Omega \approx \int_{\Gamma_I} h^s \left( \cdot \right)|_{\zeta = 0} d\Gamma.
\end{align}

\begin{remark}
The choice of a linear constitutive model can be well justified by experimental canine aortic and pulmonary arterial data exhibiting linearity within the physiological range of pressures \cite{Debes1995, Zhou1997}. Nonetheless, material nonlinearity and anisotropy could instead be considered using an alternative form for the strain energy function $W(\bm \epsilon_{\mathrm{dev}})$ in the above derivation. We note that for problems characterized by large deformation, such as hypertensive clinical cases, Assumptions \ref{assumption:thin-wall} and \ref{assumption:membrane} could still be invoked, yet an ALE description of the fluid sub-problem would be required, necessitating mesh motion and rendering the overall FSI formulation less computationally appealing.
\end{remark}

\subsubsection{Residual-based VMS formulation for an incompressible Newtonian fluid}
Let $\mathcal S_{\bm v}^f$ and $\mathcal S_{p}^f$ denote the trial solution spaces for the fluid velocity and pressure, and let $\mathcal V_{\bm v}^f$ and $\mathcal V_{p}^f$ be their corresponding test function spaces. We can then construct the semi-discrete fluid formulation in $\Omega^f$ using the residual-based VMS formulation \cite{Bazilevs2007} as follows. Find 
\begin{align*}
\bm y_h^f(t):= \left\lbrace \bm v_h^f(t), p_h^f(t) \right\rbrace^T \in \mathcal S_{\bm v}^f \times \mathcal S_{p}^f
\end{align*}
such that
\begin{align}
\label{eq:semi-discrete-v}
& \mathbf B^{f}_{\mathrm{m}} \left( \bm w_h^f ;  \dot{\bm y}_h^f, \bm y_h^f \right) = 0, && \forall \bm w_h^f \in \mathcal V_{\bm v}^f, \\
\label{eq:semi-discrete-p}
& \mathbf B^{f}_{\mathrm{c}}\left( q_h^f; \dot{\bm y}_h^f, \bm y_h^f \right) = 0, && \forall q_h^f \in \mathcal V_{p}^f, 
\end{align}
where
\begin{align}
\label{eq:vms_momentum}
& \mathbf B^{f}_{\mathrm{m}} \left( \bm w_h^f ;  \dot{\bm y}_h^f, \bm y_h^f \right) := \mathbf B_{\mathrm{m}}^{\textup{vol}} \left( \bm w_h^f ;  \dot{\bm y}_h^f, \bm y_h^f \right) + \mathbf B_{\mathrm{m}}^{\mathrm{h}} \left( \bm w_h^f ;  \dot{\bm y}_h^f, \bm y_h^f \right) + \mathbf B_{\mathrm{m}}^{\textup{bf}} \left( \bm w_h^f ;  \dot{\bm y}_h^f, \bm y_h^f \right), \displaybreak[2] \\
& \mathbf B_{\mathrm{m}}^{\textup{vol}} \left( \bm w_h^f ;  \dot{\bm y}_h^f, \bm y_h^f \right) := \int_{\Omega^f} \bm w_h^f \cdot \rho^f \left( \frac{\partial \bm v_h^f}{\partial t} + \bm v_h^f \cdot \nabla \bm v_h^f - \bm b^f \right) d\Omega  - \int_{\Omega^f} \nabla \cdot \bm w_h^f p_h^f d\Omega + \int_{\Omega^f} 2\mu^f  \bm \varepsilon(\bm w_h^f) : \bm \varepsilon(\bm v_h^f)  d\Omega \nonumber \displaybreak[2] \\
& \hspace{28mm}  - \int_{\Omega^{f \prime}} \nabla \bm w_h^f : \left( \rho^f \bm v^{\prime} \otimes \bm v_h^f \right)  d\Omega + \int_{\Omega^{f \prime}} \nabla \bm v_h^f : \left( \rho^f \bm w_h^f \otimes \bm v^{\prime} \right) d\Omega  - \int_{\Omega^{f \prime}} \nabla \bm w_h^f : \left( \rho^f \bm v^{\prime} \otimes \bm v^{\prime} \right) d\Omega \nonumber \displaybreak[2] \\
& \hspace{28mm} - \int_{\Omega^{f \prime}} \nabla \cdot \bm w_h^f p^{\prime} d\Omega,\displaybreak[2] \\
\label{eq:vms_traction}
& \mathbf B_{\mathrm{m}}^{\mathrm{h}} \left( \bm w_h^f ;  \dot{\bm y}_h^f, \bm y_h^f \right) := - \int_{\Gamma_h^f} \bm w_h^f \cdot \bm h^f d\Gamma, \displaybreak[2] \\
\label{eq:back_flow_stabilization}
& \mathbf B_{\mathrm{m}}^{\textup{bf}} \left( \bm w_h^f ;  \dot{\bm y}_h^f, \bm y_h^f \right) := - \int_{\Gamma_h^f}  \rho^f \beta \left(\bm v_h^f \cdot \bm n^f \right)_{-} \bm w_h^f \cdot \bm v_h^f d\Gamma, \displaybreak[2] \\
\label{eq:vms_continuity}
& \mathbf B^{f}_{\mathrm{c}}\left( q_h^f ; \dot{\bm y}_h^f, \bm y_h^f \right) := \int_{\Omega^f} q_h^f \nabla \cdot \bm v_h^f d\Omega  - \int_{\Omega^{f \prime}} \nabla q_h^f \cdot  \bm v^{\prime} d\Omega, \displaybreak[2] \\
& \bm v^{\prime} := -\bm \tau_{M} \left( \rho^f \frac{\partial \bm v_h^f}{\partial t} + \rho^f \bm v_h^f \cdot \nabla \bm v_h^f  + \nabla p_h^f - \mu^f \Delta \bm v_h^f - \rho^f \bm b^f \right), \displaybreak[2] \\
& p^{\prime} := -\tau_C \nabla \cdot \bm v_h^f, \displaybreak[2] \\
\label{eq:vms_def_tau_m}
& \tau_M := \frac{1}{\rho^f}\left( \frac{\mathrm C_{\mathrm T}}{\Delta t^2} + \bm v_h^f \cdot \bm G \bm v_h^f + \mathrm C_{\mathrm I} \left( \frac{\mu^f}{\rho^f} \right)^2 \bm G : \bm G \right)^{-\frac12}, \displaybreak[2] \\
\label{eq:vms_def_tau_c}
& \tau_C := \frac{1}{\tau_M \textup{tr}\bm G}, \displaybreak[2] \\
& G_{ij} := \sum_{k=1}^{3} \frac{\partial y_k}{\partial x_i} M_{kl} \frac{\partial y_l}{\partial x_j}, \displaybreak[2] \\
\label{eq:def_K_for_scale_G}
& \bm M = [ M_{kl} ] = \frac{\sqrt[3]{2}}{2}\begin{bmatrix}
2 & 1 & 1 \\
1 & 2 & 1 \\
1 & 1 & 2
\end{bmatrix}, \displaybreak[2] \\
& \bm G : \bm G := \sum_{i,j=1}^{3} G_{ij} G_{ij}, \displaybreak[2] \\
& \textup{tr}\bm G := \sum_{i=1}^{3} G_{ii}, \displaybreak[2] \\
& \left( \bm v_h^f \cdot \bm n^f \right)_{-} := \frac{\bm v_h^f \cdot \bm n^f - |\bm v_h^f \cdot \bm n^f|}{2} = 
\begin{cases}
\bm v_h^f \cdot \bm n^f & \quad \mbox{ if } \bm v_h^f \cdot \bm n^f < 0, \\
0 & \quad \mbox{ if } \bm v_h^f \cdot \bm n^f \geq 0.
\end{cases}
\end{align}
Here, $\bm y = \left\lbrace y_i \right\rbrace_{i=1}^{3}$ are natural coordinates in the parent domain; $\mathrm C_{\mathrm I}$ depends on the polynomial order of the finite element basis functions, taking the values of $36$ and $60$ for linear and quadratic interpolations, respectively \cite{Figueroa2006, Franca1992}; and $\mathrm C_{\mathrm T}$ is taken to be $4$ \cite{Liu2018, Liu2020}. $\mathbf B_{\mathrm{m}}^{\textup{bf}}$ is an additional convective traction shown to be robust in overcoming backflow divergence \cite{Bazilevs2009b, Moghadam2011}, a well-known issue in cardiovascular simulations. It can be shown that taking $\beta = 1.0$ guarantees energy stability for the numerical scheme adopted here. In this work, $\beta$ is fixed to be $0.2$ to minimize its impact on the flow field and to improve robustness at larger time steps. 

\begin{remark}
In contrast to CMM \cite{Figueroa2006a,Figueroa2006,Taylor1998}, integration-by-parts is not performed for the divergence operator in the continuity equation, which could otherwise lead to a loss of energy stability in the Galerkin formulation. Interested readers may refer to \cite{Gresho1998} for a thorough discussion of the Galerkin formulation for the Navier-Stokes equations. In addition, we adopt the residual-based variational multiscale formulation \cite{Bazilevs2007}, which has been shown to capture the correct energy spectrum and decay of kinetic energy in isotropic and wall-bounded turbulent flows \cite{Bazilevs2007,Bazilevs2007a,Colomes2015}. The conventional streamline upwind Petrov-Galerkin/pressure-stabilizing Petrov-Galerkin (SUPG/PSPG) method \cite{Brooks1982, Franca1992}, on the other hand, cannot correctly describe the energy spectrum and is thus physically inappropriate as a subgrid-scale model \cite{Hughes2000}. Furthermore, the stabilization parameters are defined to be invariant to cyclic permutations of node numbering \cite{Pauli2017,Danwitz2019}. 
\end{remark}

\subsubsection{Reduced unified continuum formulation for vascular FSI}
Discretization of the entire domain $\Omega$ by a single mesh with continuous basis functions across the fluid-solid interface $\Gamma_I$ immediately guarantees satisfaction of the kinematic coupling condition \eqref{eq:coupling-kinematic} in the semi-discrete formulation. The implied relation 
\begin{align}
\label{eq:fsi-semidiscrete-3D-test-fun}
\bm w^f_h = \bm w^s_h, \qquad \mbox{ on } \Gamma_I
\end{align}
also yields weak satisfaction of the traction coupling condition \eqref{eq:coupling-traction}, that is
\begin{align*}
0 = \int_{\Gamma_I} \bm w^f_h \cdot \left( \bm \sigma^f \bm n^f + \bm \sigma^s \bm n^s \right) d\Gamma = \int_{\Gamma_I} \bm w^f_h \cdot \left( \bm \sigma^f \bm n^f - \bm \sigma^s \bm n^f \right) d\Gamma.
\end{align*}
With this mesh choice, the momentum balances \eqref{eq:semi-discrete-u} and \eqref{eq:vms_momentum} over $\Omega^s$ and $\Omega^f$, respectively, can then be combined into a single momentum balance over the entire continuum body $\Omega$,
\begin{align*}
\mathbf B^{s}_{\mathrm{m}}\Big( \bm w^s_h ;  \dot{\bm y}^s_h, \bm y^s_h \Big) + \mathbf B^{f}_{\mathrm{m}} \left( \bm w^f_h ;  \dot{\bm y}^f_h, \bm y^f_h \right) = 0, \qquad \forall \bm w^s_h \in \mathcal V^s_{\bm u} \quad \mbox{ and } \quad \forall \bm w_h^f \in \mathcal V_{\bm v}^f.
\end{align*}

Having applied the outlined assumptions to collapse the three-dimensional elastodynamic problem in $\Omega^s$ to a two-dimensional problem posed on $\Gamma_I$, we now present the reduced semi-discrete FSI formulation. Let $\bm u^{w}_h$ be the membrane displacement on $\Gamma_I$. Using the kinematic coupling condition \eqref{eq:coupling-kinematic}, continuity of test functions on $\Gamma_I$ \eqref{eq:fsi-semidiscrete-3D-test-fun}, and the transformation of volume integrals over $\Omega^s$ \eqref{eq:shell-volume-integral-to-surface-integral}, we can rewrite the kinematic equation \eqref{eq:semi-discrete-kinematic} as 
\begin{align}
\label{eq:semi-discrete-kinematic-reformulated}
& \mathbf B_{\mathrm{k}} \left( \dot{\bm y}_h, \bm y_h \right) := \frac{d\bm u^w_h}{dt} - \bm v^f_h = \bm 0, \qquad \mbox{ on } \Gamma_I,
\end{align}
and the momentum balance \eqref{eq:semi-discrete-u} over $\Omega^s$ as
\begin{align}
\label{eq:semi-discrete-u-reformulated}
& \mathbf B^w_{\mathrm{m}} \left( \bm w_h^f ;  \dot{\bm y}_h, \bm y_h \right) := \int_{\Gamma_I} \bm w_h^f \cdot \rho^s h^s \left( \frac{d \bm v_h^f}{d t} - \bm b^s \right) d\Gamma  + \int_{\Gamma_I} h^s \bm \epsilon(\bm w_h^f) : \bm \sigma^s(\bm u_h^w)  d\Gamma - \int_{\partial \Gamma_I \cap \Gamma^h_s } h^s \bm w_h^f \cdot \bm h^s d\Gamma,
\end{align}
where $\partial \Gamma_I \cap \Gamma^h_s$ constitutes the Neumann partition of the boundary of $\Gamma_I$. Finally, let $\mathcal S^{w}_{\bm u}$ be the trial solution space for $\bm u^{w}_h$. Our reduced unified continuum formulation posed only in the fluid domain $\Omega_f$ is then stated as follows. Find
\begin{align*}
\bm y_h(t) := \left\lbrace \bm u^{w}_h(t), \bm v_h^f(t), p_h^f(t) \right\rbrace^T \in \mathcal S^w_{\bm u} \times \mathcal S_{\bm v}^f \times \mathcal S_{p}^f
\end{align*}
such that
\begin{align}
\label{eq:semidiscrete-fsi-couple}
& \mathbf B_{\mathrm{k}} \left( \dot{\bm y}_h, \bm y_h \right) = \bm 0, &&  \\
\label{eq:semidiscrete-fsi-momentum}
& \mathbf B_{\mathrm{m}} \left( \bm w_h^f ;  \dot{\bm y}_h, \bm y_h \right) := \mathbf B^w_{\mathrm{m}} \left( \bm w_h^f ;  \dot{\bm y}_h, \bm y_h \right) + \mathbf B^f_{\mathrm{m}} \left( \bm w^f_h ;  \dot{\bm y}^f_h, \bm y^f_h \right)  = 0, && \forall \bm w_h^f \in \mathcal V_{\bm v}^f, \\
\label{eq:semidiscrete-fsi-continuity}
& \mathbf B_{\mathrm{c}}\left( q^f_h; \dot{\bm y}_h, \bm y_h \right) := \mathbf B^f_{\mathrm{c}}\left( q^f_h; \dot{\bm y}^f_h, \bm y^f_h \right) = 0, && \forall q_h^f \in \mathcal V_{p}^f.
\end{align}
It is then clear that compared to the fluid sub-problem, the above FSI formulation \eqref{eq:semidiscrete-fsi-couple}-\eqref{eq:semidiscrete-fsi-continuity} consists of an additional coupling relation \eqref{eq:semidiscrete-fsi-couple} and four additional terms corresponding to the vascular wall's mass, body force, stiffness, and boundary traction, all of which are embedded in \eqref{eq:semidiscrete-fsi-momentum} through the form $\mathbf B^w_{\mathrm{m}}$. Importantly, \eqref{eq:semidiscrete-fsi-momentum} represents the semi-discrete formulation for momentum balance over the entire continuum body consisting of both the fluid and vascular wall. This FSI formulation therefore offers a computationally efficient approach for capturing vascular wall deformation on a stationary fluid mesh.

\begin{remark}
Despite the ostensible similarity between our reduced unified continuum formulation and the semi-discrete formulation of CMM, the fluid-solid coupling in CMM was achieved via a fictitious body force assumed to be uniformly distributed through the vessel thickness \cite{Figueroa2006,Figueroa2006a} (see also \cite[p.10]{Figueroa2017} and \cite[p.119]{Taylor2009}). Our recent development of the unified continuum and VMS formulation renders this assumption unnecessary for achieving the desired coupling. Starting from the unified formulation in ALE coordinates, we have instead arrived at a similar reduced FSI formulation simply by invoking the small-strain, thin-walled, and membrane assumptions for the solid sub-problem. We further note that the wall thickness has not been assumed to be uniform over each element and thus appears within the integrals over $\Gamma_I$ in our formulation.
\end{remark}

\subsection{Fully discrete formulation}
To arrive at the fully discrete FSI formulation, we apply the generalized-$\alpha$ method for temporal discretization of the first-order dynamic system. Let the time interval of interest $(0, T)$ be divided into $N_{\mathrm{ts}}$ subintervals of equal size $\Delta t_n := t_{n+1} - t_n$ and delimited by a discrete time vector $\left\lbrace t_n \right\rbrace_{n=0}^{N_{\mathrm{ts}}}$. The approximations of the solution vector and its time-derivative at time step $t_n$ are denoted as
\begin{align*}
\bm y_n := \left\lbrace \bm u^w_n, \bm v^f_n, p^f_n \right\rbrace, \quad \mbox{ and } \quad \dot{\bm y}_n := \left\lbrace \dot{\bm u}^w_n, \dot{\bm v}^f_n, \dot{p}^f_n \right\rbrace.
\end{align*}
Let $N_A$ represent basis functions for all variational spaces, and let $\{\bm e_i\}$ be the Cartesian basis vectors with $i=1,2,3$. We may then define the residual vectors as follows,
\begin{alignat*}{2}
& \boldsymbol{\mathrm R}_{\mathrm{k}}\left( \dot{\bm y}_{n}, \bm y_{n} \right) && := \Big\lbrace \mathbf B_{\mathrm{k}} \Big( \dot{\bm y}_{n}, \bm y_{n} \Big) \Big\rbrace, \nonumber \displaybreak[2] \\
& \boldsymbol{\mathrm R}_{\mathrm{m}}\left( \dot{\bm y}_{n}, \bm y_{n} \right) && := \Big\lbrace \mathbf B_{\mathrm{m}} \left( N_A \bm e_i ;  \dot{\bm y}_{n}, \bm y_{n} \Big) \right\rbrace, \nonumber \displaybreak[2] \\
& \boldsymbol{\mathrm R}_{\mathrm{c}}\left( \dot{\bm y}_{n}, \bm y_{n} \right) && := \Big\lbrace \mathbf B_{\mathrm{c}} \left( N_A ;  \dot{\bm y}_{n}, \bm y_{n} \Big) \right\rbrace.
\end{alignat*}
The fully discrete scheme can be stated as follows. At time step $t_n$, given $\dot{\bm y}_n$, $\bm y_n$, and the time step size $\Delta t_n$, find $\dot{\bm y}_{n+1}$ and $\bm y_{n+1}$ such that
\begin{alignat}{2}
\label{eq:coupling_residual}
& \boldsymbol{\mathrm R}_{\mathrm{k}} \left(  \dot{\bm y}_{n+\alpha_m}, \bm y_{n+\alpha_f} \right) &&= \bm 0, \displaybreak[2] \\
& \boldsymbol{\mathrm R}_{\mathrm{m}} \left(  \dot{\bm y}_{n+\alpha_m}, \bm y_{n+\alpha_f}\right) &&= \bm 0, \displaybreak[2] \\
& \boldsymbol{\mathrm R}_{\mathrm{c}}\left(\dot{\bm y}_{n+\alpha_m}, \bm y_{n+\alpha_f} \right) &&= \bm 0,
\end{alignat}
\begin{alignat}{2}
\label{eq:gen_alpha_def_y_n_alpha_m}
& \dot{\bm y}_{n+\alpha_m} &&= \dot{\bm y}_n + \alpha_m \left(\dot{\bm y}_{n+1} - \dot{\bm y}_n \right), \displaybreak[2] \\
\label{eq:gen_alpha_def_y_n_alpha_f}
& \bm y_{n+\alpha_f} &&= \bm y_n + \alpha_f \left( \bm y_{n+1} - \bm y_n \right), \displaybreak[2] \\
\label{eq:gen_alpha_def_y_n_plus_1}
& \bm y_{n+1} &&= \bm y_n + \Delta t_n \dot{\bm y}_n + \gamma \Delta t_n \left( \dot{\bm y}_{n+1} - \dot{\bm y}_n \right).
\end{alignat}
In the above system, the three parameters $\alpha_m$, $\alpha_f$, and $\gamma$ determine critical numerical properties of the discrete dynamic system. For linear problems, the following parameterization has been shown to achieve second-order accuracy, unconditional stability,  and optimal high frequency dissipation,
\begin{align*}
\alpha_m = \frac12 \left( \frac{3-\varrho_{\infty}}{1+\varrho_{\infty}} \right), \quad \alpha_f = \frac{1}{1+\varrho_{\infty}}, \quad \gamma = \frac12 + \alpha_m - \alpha_f,
\end{align*}
wherein $\varrho_{\infty} \in [0,1]$ is the spectral radius of the amplification matrix at the highest mode \cite{Jansen2000,Chung1993}. In this work, we choose $\varrho_{\infty} = 0.5$.

\begin{remark}
While the Newmark-$\beta$ method \cite{Newmark1959} used to integrate membrane dynamics in CMM \cite{Figueroa2006} has classically been used in structural dynamics and persists in today's solvers, it faces several well-documented issues. First, it cannot simultaneously achieve second-order accuracy and high-frequency algorithmic damping; second, all first-order implementations of the Newmark-$\beta$ method are overly dissipative in the mid-frequency modes \cite[p.501]{Hughes1987}; third, implicit schemes of the Newmark family are ``not designed to conserve energy and also fail to conserve momentum" for nonlinear structural dynamics \cite{Simo1992}. As a result, despite its pervasiveness, the Newmark-$\beta$ method is not recommended for structural dynamics \cite{Hilber1978,Hulbert2017}.
\end{remark}

\begin{remark}
The generalized-$\alpha$ method was initially proposed as an integration scheme for structural dynamics \cite{Chung1993} and has since been applied to fluid dynamics \cite{Jansen2000} as well as FSI problems \cite{Bazilevs2008}. It exhibits all of the desirable attributes of a competitive integration scheme for structural dynamics, as noted by Hilber and Hughes \cite{Hilber1978}. Moreover, when applied to a first-order structural dynamic system, it was recently found not to suffer from the `overshoot' phenomenon, a long-standing issue in computational structural dynamics, and to further possess smaller dissipation and dispersion errors than when applied to a second-order system \cite{Kadapa2017}. The generalized-$\alpha$ method is thus highly recommended for integrating inertial type problems. 
\end{remark}

\begin{remark}
\label{remark:gen-alpha-pressure}
In both CFD and FSI literature, the fluid velocity and pressure are typically treated dichotomously in the generalized-$\alpha$ method for the incompressible Navier-Stokes equations, such that pressure is collocated at time step $t_{n+1}$ rather than the intermediate time step $t_{n+\alpha_f}$ \cite{Figueroa2006,Bazilevs2007a,Taylor1998,Moghadam2013}. Despite the commonly cited second-order accuracy of the generalized-$\alpha$ method, we recently demonstrated that this particular approach yields only first-order temporal accuracy, at least, for pressure. Evaluating pressure at $t_{n+\alpha_f}$ recovers second-order accuracy for the overall algorithm, simplifies the implementation, and resolves a troubling issue in geometric multiscale modeling. Interested readers are referred to \cite{Liu2020a} for details.
\end{remark}

\subsection{A segregated predictor multi-corrector algorithm}
\label{subsec:segregated-predictor-multicorrector}
The fully discrete scheme can be solved iteratively with a predictor multi-corrector algorithm, in which the Newton-Raphson method is used in the multi-corrector iterations to improve the initial prediction. Let $\bm y_{{n+1}, (l)}$ and $\dot{\bm y}_{{n+1}, (l)}$ denote the solution vector and its time derivative at time step $t_{n+1}$ at the $l$-th Newton-Raphson iteration, where $n=0,1, ...,N_{ts} - 1$ and $l=0, 1, ...,l_{\mathrm{max}}$,
\begin{align*}
\bm y_{n+1, (l)} := \lbrace \bm u_{n+1, (l)}^w, \bm v_{n+1, (l)}^f, p_{n+1, (l)}^f \rbrace, \quad \mbox{ and } \quad 
\dot{\bm y}_{n+1, (l)} := \lbrace \dot{\bm u}_{n+1, (l)}^w, \dot{\bm v}_{n+1, (l)}^f, \dot p_{n+1, (l)}^f \rbrace.
\end{align*}
We can then denote the residual vectors at iteration number $l$ as
\begin{alignat*}{2}
& \boldsymbol{\mathrm R}_{(l)} && := \lbrace \boldsymbol{\mathrm R}_{\mathrm{k},(l)}, \boldsymbol{\mathrm R}_{\mathrm{m}, (l)}, \boldsymbol{\mathrm R}_{\mathrm{c}, (l)}\rbrace^T, \\
& \boldsymbol{\mathrm R}_{\mathrm{k},(l)} && := \boldsymbol{\mathrm R}_{\mathrm{k}} \left(  \dot{\bm y}_{n+\alpha_m, (l)}, \bm y_{n+\alpha_f, (l)} \right), \\
& \boldsymbol{\mathrm R}_{\mathrm{m}, (l)} && := \boldsymbol{\mathrm R}_{\mathrm{m}} \left(  \dot{\bm y}_{n+\alpha_m,(l)}, \bm y_{n+\alpha_f,(l)}\right), \\
& \boldsymbol{\mathrm R}_{\mathrm{c}, (l)} && := \boldsymbol{\mathrm R}_{\mathrm{c}} \left(\dot{\bm y}_{n+\alpha_m,(l)}, \bm y_{n+\alpha_f,(l)} \right),
\end{alignat*}
and the consistent tangent matrix as
\begin{align*}
\boldsymbol{\mathrm K}_{(l)} =
\begin{bmatrix}
\boldsymbol{\mathrm K}_{\mathrm{k},(l),\dot{\bm u}^w} & \boldsymbol{\mathrm K}_{\mathrm{k},(l),\dot{\bm v}^f} & \boldsymbol{\mathrm K}_{\mathrm{k},(l),\dot{p}^f} \\
\boldsymbol{\mathrm K}_{\mathrm{m},(l),\dot{\bm u}^w} & \boldsymbol{\mathrm K}_{\mathrm{m},(l),\dot{\bm v}^f} & \boldsymbol{\mathrm K}_{\mathrm{m},(l),\dot{p}^f}  \\
\boldsymbol{\mathrm K}_{\mathrm{c},(l),\dot{\bm u}^w} & \boldsymbol{\mathrm K}_{\mathrm{c},(l),\dot{\bm v}^f} & \boldsymbol{\mathrm K}_{\mathrm{c},(l),\dot{p}^f} 
\end{bmatrix},
\end{align*}
wherein
\begin{align*}
& \boldsymbol{\mathrm K}_{\mathrm{k},(l),\dot{\bm u}^w} := \alpha_m \frac{\partial \boldsymbol{\mathrm R}_{\mathrm{k}} \left(  \dot{\bm y}_{n+\alpha_m, (l)}, \bm y_{n+\alpha_f, (l)} \right)}{\partial \dot{\bm u}_{n+\alpha_m}^w} = \alpha_m \bm I, \displaybreak[2] \\
& \boldsymbol{\mathrm K}_{\mathrm{k},(l),\dot{\bm v}^f} := \alpha_f \gamma \Delta t_n \frac{\partial \boldsymbol{\mathrm R}_{\mathrm{k}} \left(  \dot{\bm y}_{n+\alpha_m, (l)}, \bm y_{n+\alpha_f, (l)} \right)}{\partial \bm v_{n+\alpha_f}^f} = -\alpha_f \gamma \Delta t_n \bm I, \displaybreak[2] \\
& \boldsymbol{\mathrm K}_{\mathrm{k},(l),\dot{p}^f} := \bm 0, \displaybreak[2] \\
& \boldsymbol{\mathrm K}_{\mathrm{m},(l),\dot{\bm u}^w} := \alpha_f \gamma \Delta t_n \frac{\partial \boldsymbol{\mathrm R}_{\mathrm{m}} \left(  \dot{\bm y}_{n+\alpha_m, (l)}, \bm y_{n+\alpha_f, (l)} \right)}{\partial \bm u_{n+\alpha_f}^w}, \displaybreak[2] \\
& \boldsymbol{\mathrm K}_{\mathrm{m},(l),\dot{\bm v}^f} := \alpha_m \frac{\partial \boldsymbol{\mathrm R}_{\mathrm{m}} \left(  \dot{\bm y}_{n+\alpha_m, (l)}, \bm y_{n+\alpha_f, (l)} \right)}{\partial \dot{\bm v}_{n+\alpha_m}^f} + \alpha_f \gamma \Delta t_n \frac{\partial \boldsymbol{\mathrm R}_{\mathrm{m}} \left(  \dot{\bm y}_{n+\alpha_m, (l)}, \bm y_{n+\alpha_f, (l)} \right)}{\partial \bm v_{n+\alpha_f}^f}, \displaybreak[2] \\
& \boldsymbol{\mathrm K}_{\mathrm{m},(l),\dot{p}^f} := \alpha_f \gamma \Delta t_n \frac{\partial \boldsymbol{\mathrm R}_{\mathrm{m}} \left(  \dot{\bm y}_{n+\alpha_m, (l)}, \bm y_{n+\alpha_f, (l)} \right)}{\partial p_{n+\alpha_f}^f}, \displaybreak[2] \\
& \boldsymbol{\mathrm K}_{\mathrm{c},(l),\dot{\bm u}^w} := \bm 0, \displaybreak[2] \\
& \boldsymbol{\mathrm K}_{\mathrm{c},(l),\dot{\bm v}^f} := \alpha_m \frac{\partial \boldsymbol{\mathrm R}_{\mathrm{c}} \left(  \dot{\bm y}_{n+\alpha_m, (l)}, \bm y_{n+\alpha_f, (l)} \right)}{\partial \dot{\bm v}_{n+\alpha_m}^f} + \alpha_f \gamma \Delta t_n \frac{\partial \boldsymbol{\mathrm R}_{\mathrm{c}} \left(  \dot{\bm y}_{n+\alpha_m, (l)}, \bm y_{n+\alpha_f, (l)} \right)}{\partial \bm v_{n+\alpha_f}^f}, \displaybreak[2] \\
& \boldsymbol{\mathrm K}_{\mathrm{c},(l),\dot{p}^f} := \alpha_f \gamma \Delta t_n \frac{\partial \boldsymbol{\mathrm R}_{\mathrm{c}} \left(  \dot{\bm y}_{n+\alpha_m, (l)}, \bm y_{n+\alpha_f, (l)} \right)}{\partial p_{n+\alpha_f}^f}.
\end{align*}
The special block structure in the first row of $\boldsymbol{\mathrm K}_{(l)}$ can be exploited for the following block decomposition \cite{Scovazzi2016, Liu2019a},
\begin{align*}
\boldsymbol{\mathrm K}_{(l)} = 
\begin{bmatrix}
\bm I & \bm 0 & \bm 0 \\[4mm]
\displaystyle\frac{1}{\alpha_m}\boldsymbol{\mathrm K}_{\mathrm{m},(l),\dot{\bm u}^w} & \boldsymbol{\mathrm K}_{\mathrm{m},(l),\dot{\bm v}^f} + \displaystyle\frac{\alpha_f \gamma \Delta t_n}{\alpha_m}\boldsymbol{\mathrm K}_{\mathrm{m},(l),\dot{\bm u}^w} & \boldsymbol{\mathrm K}_{\mathrm{m},(l),\dot{p}^f} \\[4mm]
\bm 0 & \boldsymbol{\mathrm K}_{\mathrm{c},(l),\dot{\bm v}^f} & \boldsymbol{\mathrm K}_{\mathrm{c},(l),\dot{p}^f}
\end{bmatrix}
\begin{bmatrix}
\alpha_m \bm I & -\alpha_f \gamma \Delta t_n \bm I & \bm 0 \\[4mm]
\bm 0 & \bm I & \bm 0 \\[4mm]
\bm 0 & \bm 0 & \bm I
\end{bmatrix}.
\end{align*}
With the above decomposition, the original linear system for the Newton-Raphson method,
\begin{align*}
\boldsymbol{\mathrm K}_{(l)} \Delta \dot{\bm y}_{n+1,(l)} = -\boldsymbol{\mathrm R}_{(l)},
\end{align*}
can be solved to obtain the increments $\Delta \dot{\bm y}_{n+1,(l)} := \lbrace \Delta \dot{\bm u}_{n+1, (l)}^{w}, \Delta \dot{\bm v}_{n+1, (l)}^{f}, \Delta \dot{p}_{n+1, (l)}^{f}\rbrace^T$ at iteration number $l$ in the following two-stage segregated algorithm. In the first stage, intermediate increments are solved from
\begin{align}
\label{eq: segregated_stage-one}
\begin{bmatrix}
\bm I & \bm 0 & \bm 0 \\[4mm]
\displaystyle\frac{1}{\alpha_m}\boldsymbol{\mathrm K}_{\mathrm{m},(l),\dot{\bm u}^w} & \boldsymbol{\mathrm K}_{\mathrm{m},(l),\dot{\bm v}^f} + \displaystyle\frac{\alpha_f \gamma \Delta t_n}{\alpha_m}\boldsymbol{\mathrm K}_{\mathrm{m},(l),\dot{\bm u}^w} & \boldsymbol{\mathrm K}_{\mathrm{m},(l),\dot{p}^f} \\[4mm]
\bm 0 & \boldsymbol{\mathrm K}_{\mathrm{c},(l),\dot{\bm v}^f} & \boldsymbol{\mathrm K}_{\mathrm{c},(l),\dot{p}^f}
\end{bmatrix}
\begin{bmatrix}
\Delta \dot{\bm u}_{n+1, (l)}^{w*} \\[3mm]
\Delta \dot{\bm v}_{n+1, (l)}^{f*} \\[3mm]
\Delta \dot{p}_{n+1, (l)}^{f*}
\end{bmatrix} = -
\begin{bmatrix}
\boldsymbol{\mathrm R}_{\mathrm{k},(l)} \\[3mm]
\boldsymbol{\mathrm R}_{\mathrm{m}, (l)} \\[3mm]
\boldsymbol{\mathrm R}_{\mathrm{c}, (l)}
\end{bmatrix}.
\end{align}
In the second stage, the increments are obtained from the following system of equations,
\begin{align}
\label{eq: segregated_stage-two}
\begin{bmatrix}
\alpha_m \bm I & -\alpha_f \gamma \Delta t_n \bm I & \bm 0 \\[2mm]
\bm 0 & \bm I & \bm 0 \\[2mm]
\bm 0 & \bm 0 & \bm I
\end{bmatrix}
\begin{bmatrix}
\Delta \dot{\bm u}_{n+1, (l)}^{w} \\[2mm]
\Delta \dot{\bm v}_{n+1, (l)}^{f} \\[2mm]
\Delta \dot{p}_{n+1, (l)}^{f}
\end{bmatrix} =
\begin{bmatrix}
\Delta \dot{\bm u}_{n+1, (l)}^{w*} \\[2mm]
\Delta \dot{\bm v}_{n+1, (l)}^{f*} \\[2mm]
\Delta \dot{p}_{n+1, (l)}^{f*}
\end{bmatrix}.
\end{align}
From \eqref{eq: segregated_stage-one} and \eqref{eq: segregated_stage-two}, we make the following observations,
\begin{align*}
& \alpha_m \Delta \dot{\bm u}_{n+1, (l)}^{w} - \alpha_f \gamma \Delta t_n \Delta \dot{\bm v}_{n+1, (l)}^{f} = \Delta \dot{\bm u}_{n+1, (l)}^{w*} = -\boldsymbol{\mathrm R}_{\mathrm{k},(l)}, \quad
\Delta \dot{\bm v}_{n+1, (l)}^{f} = \Delta \dot{\bm v}_{n+1, (l)}^{f*}, \quad
\Delta \dot{p}_{n+1, (l)}^{f} = \Delta \dot{p}_{n+1, (l)}^{f*},
\end{align*}
with which we may reduce the linear systems in the segregated algorithm to
\begin{align}
\label{eq: segregated_stage-one_reduced}
\begin{bmatrix}
\boldsymbol{\mathrm K}_{\mathrm{m},(l),\dot{\bm v}^f} + \displaystyle\frac{\alpha_f \gamma \Delta t_n}{\alpha_m}\boldsymbol{\mathrm K}_{\mathrm{m},(l),\dot{\bm u}^w} & \boldsymbol{\mathrm K}_{\mathrm{m},(l),\dot{p}^f} \\[4mm]
\boldsymbol{\mathrm K}_{\mathrm{c},(l),\dot{\bm v}^f} & \boldsymbol{\mathrm K}_{\mathrm{c},(l),\dot{p}^f}
\end{bmatrix}
\begin{bmatrix}
\Delta \dot{\bm v}_{n+1, (l)}^{f} \\[4mm]
\Delta \dot{p}_{n+1, (l)}^{f}
\end{bmatrix} = -
\begin{bmatrix}
\boldsymbol{\mathrm R}_{\mathrm{m}, (l)} - \displaystyle\frac{1}{\alpha_m} \boldsymbol{\mathrm K}_{\mathrm{m},(l),\dot{\bm u}^w} \boldsymbol{\mathrm R}_{\mathrm{k},(l)} \\[4mm]
\boldsymbol{\mathrm R}_{\mathrm{c}, (l)}
\end{bmatrix},
\end{align}
\begin{align}
\label{eq: segregated_stage-two_reduced}
\Delta \dot{\bm u}_{n+1, (l)}^{w} = \frac{\alpha_f \gamma \Delta t_n}{\alpha_m} \Delta \dot{\bm v}_{n+1, (l)}^{f} - \frac{1}{\alpha_m}\boldsymbol{\mathrm R}_{\mathrm{k},(l)}.
\end{align}
The segregated algorithm therefore consists of solving \eqref{eq: segregated_stage-one_reduced} for $\lbrace \Delta \dot{\bm v}_{n+1, (l)}^{f}, \Delta \dot{p}_{n+1, (l)}^{f}\rbrace^T$, then subsequently obtaining $\Delta \dot{\bm u}_{n+1, (l)}^{w}$ from the algebraic update \eqref{eq: segregated_stage-two_reduced}. Furthermore, it has been shown in Proposition 5 of \cite{Liu2018} that 
\begin{align*}
\boldsymbol{\mathrm R}_{\mathrm{k},(l)} = \bm 0 \quad \mbox{ for } l \geq 2
\end{align*}
holds true for any given update $\Delta \dot{\bm v}_{n+1, (l)}^{f}$ in \eqref{eq: segregated_stage-two_reduced}, prompting us to set $\boldsymbol{\mathrm R}_{\mathrm{k},(l)} = \bm 0$ for all $l \geq 1$ in \eqref{eq: segregated_stage-one_reduced}. While this may lead to inconsistent updates of $\Delta \dot{\bm v}_{n+1, (l)}^{f}$ and $\Delta \dot{p}_{n+1, (l)}^{f}$ for $l=1$, we have observed no deterioration of the overall Newton-Raphson algorithm's convergence rate in our collective experience. Interested readers are referred to Appendix B of \cite{Liu2018} for more details on the numerical analysis. For notational simplicity, we denote the block matrices in \eqref{eq: segregated_stage-one_reduced} as
\begin{align}
\label{eq: predictor_multi_correct_notation_for_block_matrices}
\boldsymbol{\mathrm A}_{(l)} := \boldsymbol{\mathrm K}_{\mathrm{m},(l),\dot{\bm v}^f} + \displaystyle\frac{\alpha_f \gamma \Delta t_n}{\alpha_m}\boldsymbol{\mathrm K}_{\mathrm{m},(l),\dot{\bm u}^w}, \quad \boldsymbol{\mathrm B}_{(l)} := \boldsymbol{\mathrm K}_{\mathrm{m},(l),\dot{p}^f}, \quad \boldsymbol{\mathrm C}_{(l)} := \boldsymbol{\mathrm K}_{\mathrm{c},(l),\dot{\bm v}^f}, \quad \boldsymbol{\mathrm D}_{(l)} := \boldsymbol{\mathrm K}_{\mathrm{c},(l),\dot{p}^f}.
\end{align}
We can now summarize our above discussion in the following segregated predictor multi-corrector algorithm.

\begin{myenv}{Segregated predictor multi-corrector algorithm}
\noindent \textbf{Predictor stage:} Set
\begin{align*}
\bm y_{n+1, (0)} = \bm y_n, \quad \dot{\bm y}_{n+1, (0)} = \frac{\gamma - 1}{\gamma} \dot{\bm y_n}.
\end{align*}

\noindent \textbf{Multi-corrector stage:} Repeat the following steps for $l=1, 2, ..., l_{max}$
\begin{enumerate}
    \item Evaluate the solution vector and its time derivative at intermediate time steps,
    \begin{align*}
    & \bm y_{n+\alpha_f, (l)} = \bm y_n + \alpha_f \left( \bm y_{n+1, (l-1)} - \bm y_n \right), \quad \dot{\bm y}_{n+\alpha_m, (l)} = \dot{\bm y}_n + \alpha_m \left(\dot{\bm y}_{n+1, (l-1)} - \dot{\bm y}_n \right).
    \end{align*}
    \item Assemble the residual vector $\boldsymbol{\mathrm R}_{(l)}$ using $\dot{\bm y}_{n+\alpha_m, (l)}$ and $\bm y_{n+\alpha_f, (l)}$.
    \item Let $\| \boldsymbol{\mathrm R}_{(l)} \|_{\mathfrak l_2}$ denote the $\mathfrak l_2$-norm of the residual vector, and let $\mathrm{tol}_{\mathrm{R}}$ and $\mathrm{tol}_{\mathrm{A}}$ denote the prescribed relative and absolute tolerances, respectively. If either of the following stopping criteria
    \begin{align*}
    \frac{\| \boldsymbol{\mathrm R}_{(l)} \|_{\mathfrak l_2}}{\| \boldsymbol{\mathrm R}_{(0)} \|_{\mathfrak l_2}} \leq \mathrm{tol}_{\mathrm{R}}, \qquad
    \| \boldsymbol{\mathrm R}_{(l)} \|_{\mathfrak l_2} \leq \mathrm{tol}_{\mathrm{A}},
    \end{align*}
    is satisfied, then set 
    \begin{align*}
    \bm y_{n+1} = \bm y_{n+1, (l-1)}, \quad \dot{\bm y}_{n+1} = \dot{\bm y}_{n+1, (l-1)},
    \end{align*}
    and exit the multi-corrector stage. Otherwise, continue to step 4.
    \item Assemble the following sub-tangent matrices,
    \begin{align*}
    & \boldsymbol{\mathrm A}_{(l)} := \boldsymbol{\mathrm K}_{\mathrm{m},(l),\dot{\bm v}^f} + \displaystyle\frac{\alpha_f \gamma \Delta t_n}{\alpha_m}\boldsymbol{\mathrm K}_{\mathrm{m},(l),\dot{\bm u}^w}, &&
    \boldsymbol{\mathrm B}_{(l)} := \boldsymbol{\mathrm K}_{\mathrm{m},(l),\dot{p}^f}, \\
    & \boldsymbol{\mathrm C}_{(l)} := \boldsymbol{\mathrm K}_{\mathrm{c},(l),\dot{\bm v}^f}, &&
    \boldsymbol{\mathrm D}_{(l)} := \boldsymbol{\mathrm K}_{\mathrm{c},(l),\dot{p}^f}.
    \end{align*}
    \item Solve the following linear system for $\Delta \dot{\bm v}_{n+1, (l)}^{f}$ and $\Delta \dot{p}_{n+1, (l)}^{f}$,
    \begin{align}
    \label{eq:pred_multi_correct_linear_system}
    \begin{bmatrix}
    \boldsymbol{\mathrm A}_{(l)} & \boldsymbol{\mathrm B}_{(l)} \\[1mm]
    \boldsymbol{\mathrm C}_{(l)} & \boldsymbol{\mathrm D}_{(l)}
    \end{bmatrix}
    \begin{bmatrix}
    \Delta \dot{\bm v}_{n+1, (l)}^{f} \\[1mm]
    \Delta \dot{p}_{n+1, (l)}^{f}
    \end{bmatrix} = -
    \begin{bmatrix}
    \boldsymbol{\mathrm R}_{\mathrm{m}, (l)} \\[1mm]
    \boldsymbol{\mathrm R}_{\mathrm{c}, (l)}
    \end{bmatrix}.
    \end{align}
    \item Obtain $\Delta \dot{\bm u}_{n+1, (l)}^{w}$ from $\Delta \dot{\bm v}_{n+1, (l)}^{f}$ via the relation \eqref{eq: segregated_stage-two_reduced}, that is,
    \begin{align*}
    \Delta \dot{\bm u}_{n+1, (l)}^{w} = \frac{\alpha_f \gamma \Delta t_n}{\alpha_m} \Delta \dot{\bm v}_{n+1, (l)}^{f} - \frac{1}{\alpha_m}\boldsymbol{\mathrm R}_{\mathrm{k},(l)}.
    \end{align*}
    \item Update the solution vector and its time derivative as
    \begin{align*}
    & \bm y_{n+1, (l)} = \bm y_{n+1, (l-1)} + \gamma \Delta t_n \Delta \dot{\bm y}_{n+1, (l)}, \quad \dot{\bm y}_{n+1, (l)} = \dot{\bm y}_{n+1, (l-1)} + \Delta\dot{\bm y}_{n+1, (l)}.
    \end{align*}
\end{enumerate}
\end{myenv}

\noindent In this work, unless otherwise specified, we set the tolerances to $\mathrm{tol}_{\mathrm R} = \mathrm{tol}_{\mathrm A} = 10^{-6}$ and the maximum number of nonlinear iterations to $l_{max} = 20$.

\begin{remark}
\label{remark:zero_kinematic_residual}
We have chosen $\boldsymbol{\mathrm R}_{\mathrm{k},(l)} = \bm 0$ for all $l \geq 1$ in \eqref{eq: segregated_stage-one_reduced} to simplify the formation of the right-hand side of the linear system. We note that the wall displacement update \eqref{eq: segregated_stage-two_reduced} still requires a consistent definition of $\boldsymbol{\mathrm R}_{\mathrm{k},(l)}$, as stagnation or divergence may otherwise be observed. Numerical evidence will be documented for patient-specific clinical cases in Section \ref{subsec:zero_kinematic_residual_numerical_evidence}.
\end{remark}

\begin{remark}
In comparison to the consistent tangent matrix for the incompressible Navier-Stokes equations, only block matrix $\boldsymbol{\mathrm A}_{(l)}$ has been modified to include the wall stiffness term $\alpha_f \gamma \Delta t_n\boldsymbol{\mathrm K}_{\mathrm{m},(l),\dot{\bm u}^w} / \alpha_m$. As was shown in our prior analysis \cite{Liu2020}, $\boldsymbol{\mathrm A}_{(l)}$ additionally consists of contributions from the transient, convection, viscous, and subgrid scale modeling terms as well as multiple rank-one modifications from coupling with reduced models. Particular attention will thus be paid to approximate $\boldsymbol{\mathrm A}_{(l)}$ in our design of the iterative solution method, as discussed in the next section. 
\end{remark}

\begin{remark}
We further note that for time steps of practical interest, the use of the `frozen-coefficient' tangent matrix was previously deemed necessary for achieving stability in the first few time steps \cite{Jansen2000,Johan1991a}. Nonetheless, we implement the consistent tangent matrix here as in our previous studies \cite{Liu2018, Liu2020, Liu2020b} without stability issues, thereby achieving rapid quadratic convergence.
\end{remark}

\section{Iterative solution method}
\label{sec:iterative_solution_method}
In this section, we consider the linear system \eqref{eq:pred_multi_correct_linear_system} arising in the aforementioned segregated predictor multi-corrector algorithm, which often comprises the most time-consuming part of the overall algorithm. We focus on a linear system 
\begin{align}
\label{eq:general-linear-system}
\boldsymbol{\mathcal A} \bm x = \bm r
\end{align}
exhibiting the following $2\times 2$ block structure,
\begin{align*}
\boldsymbol{\mathcal A} :=
\begin{bmatrix}
\boldsymbol{\mathrm A} & \boldsymbol{\mathrm B} \\[0.3mm]
\boldsymbol{\mathrm C} & \boldsymbol{\mathrm D}
\end{bmatrix}, \quad
\bm x :=
\begin{bmatrix}
\bm x_{\bm v} \\[0.3em]
\bm x_{p}
\end{bmatrix},
\quad
\bm r :=
\begin{bmatrix}
\bm r_{\bm v} \\[0.3em]
\bm r_{p}
\end{bmatrix}.
\end{align*}
As is clear from our derivation of the consistent tangent matrix $\boldsymbol{\mathcal A}$ in the previous section, the segregated algorithm allows the implicit solver to retain the same block structure as that of the incompressible Navier-Stokes equations \cite{Liu2020}. From \eqref{eq: predictor_multi_correct_notation_for_block_matrices}, we observe that block matrices $\boldsymbol{\mathrm B}$, $\boldsymbol{\mathrm C}$, and $\boldsymbol{\mathrm D}$ are in fact identical to their counterparts in the incompressible Navier-Stokes equations, and the block matrix $\boldsymbol{\mathrm A}$ is only modified by an additional term representing the wall contribution, scaled by the time step size and parameters in the generalized-$\alpha$ method. The consistent tangent matrix $\boldsymbol{\mathcal A}$ can be factorized as $\boldsymbol{\mathcal A} = \boldsymbol{\mathcal L} \boldsymbol{\mathcal D} \boldsymbol{\mathcal U}$, with
\begin{align}
\label{eq:ISM_A_LDU_block_factorization}
\boldsymbol{\mathcal L} = 
\begin{bmatrix}
\boldsymbol{\mathrm I} & \boldsymbol{\mathrm O} \\[0.3em]
\boldsymbol{\mathrm C} \boldsymbol{\mathrm A}^{-1} & \boldsymbol{\mathrm I}
\end{bmatrix},
\qquad
\boldsymbol{\mathcal D} =
\begin{bmatrix}
\boldsymbol{\mathrm A} & \boldsymbol{\mathrm O} \\[0.3em]
\boldsymbol{\mathrm O} & \boldsymbol{\mathrm S}
\end{bmatrix},
\qquad
\boldsymbol{\mathcal U} =
\begin{bmatrix}
\boldsymbol{\mathrm I} & \boldsymbol{\mathrm A}^{-1} \boldsymbol{\mathrm B} \\[0.3em]
\boldsymbol{\mathrm O} & \boldsymbol{\mathrm I}
\end{bmatrix},
\end{align}
where $\boldsymbol{\mathrm S} := \boldsymbol{\mathrm D} - \boldsymbol{\mathrm C} \boldsymbol{\mathrm A}^{-1} \boldsymbol{\mathrm B}$ is the Schur complement of $\boldsymbol{\mathrm A}$. The above block factorization immediately implies a solution procedure for the linear system $\boldsymbol{\mathcal A} \bm x = \bm r$. Applying $\boldsymbol{\mathcal L}^{-1}$ to both sides of \eqref{eq:general-linear-system} transforms the linear system to $\boldsymbol{\mathcal D} \boldsymbol{\mathcal U} \bm x = \boldsymbol{\mathcal L}^{-1} \bm r$, which can be written explicitly as
\begin{align}
\label{eq: ISM_DU_equations}
\begin{bmatrix}
\boldsymbol{\mathrm A} & \boldsymbol{\mathrm B} \\[0.3em]
\boldsymbol{\mathrm O} & \boldsymbol{\mathrm S}
\end{bmatrix}
\begin{bmatrix}
\bm x_{\bm v} \\[0.3em]
\bm x_{p}
\end{bmatrix}
=
\begin{bmatrix}
\bm r_{\bm v} \\[0.3em]
\bm r_{p} - \boldsymbol{\mathrm C} \boldsymbol{\mathrm A}^{-1} \bm r_{\bm v}
\end{bmatrix}.
\end{align}
The so-called Schur complement reduction (SCR) procedure \cite{Benzi2005,May2008} solves \eqref{eq: ISM_DU_equations} via back substitution and therefore involves solving smaller systems associated with $\boldsymbol{\mathrm A}$ and $\boldsymbol{\mathrm S}$. Given the dense structure of $\boldsymbol{\mathrm S}$ stemming from the presence of $\boldsymbol{\mathrm A}^{-1}$, solving a linear system associated with $\boldsymbol{\mathrm S}$ to high precision is, however, prohibitively expensive. The SCR procedure can therefore be applied as a preconditioning technique for an iterative solution method, such that the smaller systems need not be solved to high precision. The action of the preconditioner $\boldsymbol{\mathcal P}$ for the linear system $\boldsymbol{\mathcal A} \bm x = \bm r$ is defined in Algorithm \ref{algorithm:exact_block_factorization}, where the smaller systems associated with $\boldsymbol{\mathrm A}$ and $\boldsymbol{\mathrm S}$ are solved by the GMRES algorithm preconditioned by $\boldsymbol{\mathrm P}_{\mathrm A}$ and $\boldsymbol{\mathrm P}_{\mathrm S}$, respectively. The stopping criteria for the two iterative solvers include relative tolerances $\delta_{\mathrm A}$ and $\delta_{\mathrm S}$ and maximum iteration numbers $\mathrm n^{\textup{max}}_{\mathrm A}$ and $\mathrm n^{\textup{max}}_{\mathrm S}$, respectively.

\begin{algorithm}[H]
\caption{The action of $\boldsymbol{\mathcal P}^{-1}$ on a vector $\bm s := [ \bm s_{\bm v}; \bm s_p]^T$ with the output being $\bm y := [ \bm y_{\bm v}; \bm y_p]^T$.}
\label{algorithm:exact_block_factorization}
\begin{algorithmic}[1]
\State \texttt{Solve for an intermediate velocity $\hat{\bm y}_{\bm v}$ from the equation}
\begin{align}
\label{eq:seg_sol_int_disp}
\boldsymbol{\mathrm A} \hat{\bm y}_{\bm v} = \bm s_{\bm v}
\end{align}
\texttt{by GMRES preconditioned by $\boldsymbol{\mathrm P}_{\mathrm A}$ with $\delta_{\mathrm A}$ and $\mathrm n^{\textup{max}}_{\mathrm A}$ prescribed.}
\State \texttt{Update the continuity residual by $\bm s_{p} \gets  \bm s_{p} - \boldsymbol{\mathrm C} \hat{\bm y}_{\bm v}$.}
\State \texttt{Solve for $\bm y_p$ from the equation}
\begin{align}
\label{eq:seg_sol_pres}
\boldsymbol{\mathrm S} \bm y_p = \bm s_{p}
\end{align}
\texttt{by GMRES preconditioned by $\boldsymbol{\mathrm P}_{\mathrm S}$ with $\delta_{\mathrm S}$ and $\mathrm n^{\textup{max}}_{\mathrm S}$ prescribed.}
\algorithmiccomment{The action of $\boldsymbol{\mathrm S}$ on a vector in the GMRES iteration will be defined in Algorithm \ref{algorithm:matrix_free_mat_vec_for_S}.}
\State \texttt{Update the momentum residual by $\bm s_{\bm v} \gets \bm s_{\bm v} - \boldsymbol{\mathrm B} \bm y_{p}$.}
\State \texttt{Solve for $\bm y_{\bm v}$ from the equation}
\begin{align}
\label{eq:seg_sol_disp}
\boldsymbol{\mathrm A} \bm y_{\bm v} = \bm s_{\bm v}
\end{align}
\texttt{by GMRES preconditioned by $\boldsymbol{\mathrm P}_{\mathrm A}$ with $\delta_{\mathrm A}$ and $\mathrm n^{\textup{max}}_{\mathrm A}$ prescribed.}
\end{algorithmic}
\end{algorithm}
\noindent Since the preconditioner $\boldsymbol{\mathcal P}$ is defined through an algorithm, its algebraic definition may vary over iterations. The flexible GMRES (FGMRES) algorithm \cite{Saad1993} is thus applied as the outer iterative method for the tangent matrix $\boldsymbol{\mathcal A}$ with a corresponding relative tolerance $\delta$ and maximum iteration number $\mathrm n^{\textup{max}}$. In this work, we set the relative tolerances $\delta_{\mathrm A} = 10^{-5}$, $\delta_{\mathrm S} = 10^{-2}$ and maximum iteration numbers $\mathrm n^{\textup{max}}_{\mathrm A} = \mathrm n^{\textup{max}}_{\mathrm S} = 100$, $\mathrm n^{\textup{max}} = 200$ unless otherwise specified. While the FGMRES algorithm minimizes the residual over the generated subspace, its convergence is not guaranteed in general, as the approximation subspace is not a standard Krylov subspace. Nonetheless, given this flexibility in preconditioner variation over iterations, the robustness and efficiency of the overall iterative method can be well balanced with a proper design of $\boldsymbol{\mathcal P}$. We therefore turn our attention to the design of $\boldsymbol{\mathrm P}_{\mathrm A}$ and $\boldsymbol{\mathrm P}_{\mathrm S}$ for the associated GMRES algorithms.

The block matrix $\boldsymbol{\mathrm A}$ consists of a discrete convection-diffusion-reaction operator, subgrid scale modeling terms, rank-one modifications from reduced models coupled to the outflow boundaries to represent the downstream vasculature \cite{Liu2020} (see also Section \ref{subsec:coupling-with-reduced-models}), and the stiffness matrix of the vascular wall (see \eqref{eq: predictor_multi_correct_notation_for_block_matrices}). The class of multigrid methods, which has been proposed as a scalable and robust preconditioner for elliptic problems \cite{Ieary2000}, presents an attractive choice for $\boldsymbol{\mathrm P}_{\mathrm A}$, particularly when the mesh size demands extremely large-scale parallel computations. Compared to the geometric multigrid method \cite{Wesseling2001}, the algebraic multigrid method (AMG) can more conveniently be integrated with numerical codes developed for unstructured grids. 

The Schur complement $\boldsymbol{\mathrm S}$ is implicitly defined through the four block matrices. While constructing and storing $\boldsymbol{\mathrm S}$ is rather cost-prohibitive, the GMRES algorithm only necessitates the construction of a Krylov subspace via repeated matrix-vector multiplication operations. The action of $\boldsymbol{\mathrm S}$ on a given vector can therefore be computed with the four block matrices on the fly in a so-called ``matrix-free" fashion as outlined in Algorithm \ref{algorithm:matrix_free_mat_vec_for_S}, which is used to construct the Krylov subspace in Step $3$ of Algorithm \ref{algorithm:exact_block_factorization}. Inspired by the SIMPLE algorithm, the preconditioner $\boldsymbol{\mathrm P}_{\mathrm S}$ is formed by BoomerAMG \cite{Falgout2002} based on a sparse approximation of $\boldsymbol{\mathrm S}$ given by $\hat{\boldsymbol{\mathrm S}} := \boldsymbol{\mathrm D} - \boldsymbol{\mathrm C} \left(\textup{diag}\left(\boldsymbol{\mathrm A}\right)\right)^{-1} \boldsymbol{\mathrm B}$.

\begin{algorithm}[H]
\caption{The matrix-free algorithm for multiplying $\boldsymbol{\mathrm S}$ with a vector $\bm x_p$.}
\label{algorithm:matrix_free_mat_vec_for_S}
\begin{algorithmic}[1]
\State \texttt{Compute the matrix-vector multiplication $\hat{\bm x}_p \gets \boldsymbol{\mathrm D} \bm x_p$.}
\State \texttt{Compute the matrix-vector multiplication $\bar{\bm x}_p \gets \boldsymbol{\mathrm B} \bm x_p$.}
\State \texttt{Solve for $\tilde{\bm x}_p$ from the linear system}
\begin{align}
\label{eq:S_inner_A_eqn}
\boldsymbol{\mathrm A} \tilde{\bm x}_p = \bar{\bm x}_p
\end{align}
\texttt{by GMRES preconditioned by $\boldsymbol{\mathrm P}_{\mathrm A}$ with $\delta_{\mathrm I}$ and $\mathrm n^{\textup{max}}_{\mathrm A}$ prescribed.} 

\algorithmiccomment{The action of $\boldsymbol{\mathrm A}^{-1}$ on $\bar{\bm x}_p$ is approximated by solving \eqref{eq:S_inner_A_eqn} with the given stopping criteria.}
\State \texttt{Compute the matrix-vector multiplication $\bar{\bm x}_p \gets \boldsymbol{\mathrm C} \tilde{\bm x}_p$.}
\State \Return $\hat{\bm x}_p - \bar{\bm x}_p$.
\end{algorithmic}
\end{algorithm}

We note that replacing equation \eqref{eq:seg_sol_pres} in Algorithm \ref{algorithm:exact_block_factorization} by $\hat{\boldsymbol{\mathrm S}} \bm y_p = \bm s_{p}$ renders the FGMRES algorithm similar to the SIMPLE algorithm, which does not require solving equation \eqref{eq:S_inner_A_eqn}. The preconditioner stated in Algorithm \ref{algorithm:exact_block_factorization} may therefore be regarded as a generalization of the SIMPLE algorithm, in which the matrix-free algorithm stated in Algorithm \ref{algorithm:matrix_free_mat_vec_for_S} is invoked to attain an improved approximation of the Schur complement.

\begin{remark}
The matrix-free technique is often invoked in Krylov subspace methods when assembling or storing a matrix is inconvenient or expensive. Often, a sparse approximation to the unassembled matrix, like $\hat{\boldsymbol{\mathrm S}}$, is also designed to provide a preconditioner to accelerate the Krylov iteration. This approach can also be found in higher-order methods \cite{Brown2010,Davydov2020} and Jacobian-free nonlinear solvers \cite{Knoll2004}.
\end{remark}

\section{Verification by the Womersley solution}
\label{sec:verification}
In this section, we present two verification studies using the Womersley solutions describing pulsatile flow in an axisymmetric cylindrical pipe, first with rigid walls and subsequently with thin, linear elastic walls. Furthermore, in the case of the rigid pipe, we use analytical solutions for pressure, velocity, and wall shear stress (WSS) to perform spatial convergence studies. All parameters are reported in centimeter-gram-second units in this work.

\subsection{Womersley flow in a rigid pipe}
\label{section:womersley_rigid}
The Womersley solution for pulsatile flow in a rigid pipe describes axisymmetric, fully developed flow subject to a pressure gradient with both steady and oscillatory contributions. The pressure can be expressed with the following Fourier series,
\begin{align}
\label{eq:womersley_rigid_pressure}
p = p_{\textup{ref}} + \left( k_0 + \sum_{n=1}^{N} k_n e^{\iota n\omega t} \right) z,
\end{align}
where $z$ is the longitudinal coordinate along the length of the pipe, $p_{\textup{ref}}$ is the reference pressure at the $z=0$ surface, $k_0$ is the steady zeroth mode of the pressure gradient, $k_n$ is the $n$-th Fourier coefficient in the oscillatory component of the pressure gradient, $\iota$ is the solution to $\iota^2=-1$, $T_p$ is the period of oscillation, and $\omega:=2\pi / T_p$ is the fundamental frequency. Whereas $k_0$ produces steady forward flow, the oscillatory component of the pressure gradient drives a phase-shifted oscillatory flow with zero net flow over $T_p$. Per the assumptions of axisymmetric and fully developed flow, the velocity is identically zero in the radial and circumferential directions and takes the following analytical form in the axial direction,
\begin{align}
\label{eq:womersley_rigid_velo}
v_z = \frac{k_0}{4\mu^f}\left(r^2 - R^2\right) + \sum\limits_{n=1}^{N} \frac{\iota k_n}{\rho^f n\omega} \left( 1 - \frac{J_0(\iota^{\frac{3}{2}} \alpha_n \frac{r}{R})}{J_0(\iota^{\frac{3}{2}}\alpha_n)} \right) e^{\iota n \omega t},
\end{align}
wherein $r:= \sqrt{x^2+y^2}$, $R$ is the pipe radius, $J_0$ is the zeroth-order Bessel function of the first kind, and $\alpha_n := R \sqrt{\rho^f n\omega / \mu^f}$ is the Womersley number for the $n$-th Fourier mode. The only nonzero component of WSS takes the corresponding form,
\begin{align}
\label{eq:womersley_rigid_wss}
\tau_{zr} = \sigma^f_{\mathrm{dev},zr}|_{r=R} = \frac{k_0 R}{2} - \sum\limits_{n=1}^{N} \frac{k_n R}{\iota^{\frac{3}{2}}\alpha_n} \frac{J_1(\iota^{\frac{3}{2}}\alpha_n)}{J_0(\iota^{\frac{3}{2}}\alpha_n)} e^{\iota n \omega t}.
\end{align}
The complex forms of $p$, $v_z$, and $\tau_{zr}$ in \eqref{eq:womersley_rigid_pressure}-\eqref{eq:womersley_rigid_wss} indicate the existence of two sets of real independent solutions. Here, we take the set of real components as the benchmark solution and represent a single oscillatory mode (i.e., $N=1$).

\begin{figure}
	\begin{center}
	\begin{tabular}{c}
\includegraphics[angle=0, trim=140 95 30 100, clip=true, scale=0.6]{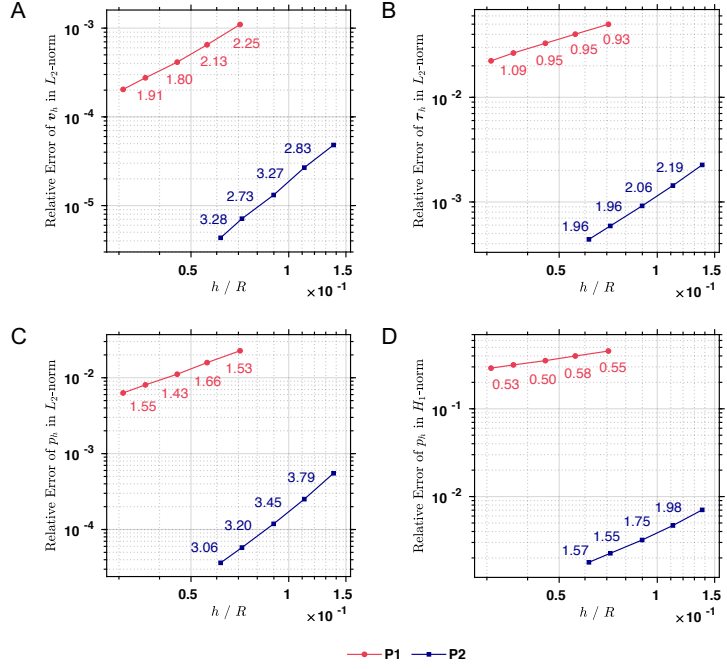}
\end{tabular}
\end{center}
\caption{Relative errors of (A) $\bm v_h$ in $L_2$ norm, (B) $\bm \tau_h$ in $L_2$ norm, (C) $p_h$ in $L_2$ norm, and (D) $p_h$ in $H_1$ norm for linear (P1) and quadratic (P2) tetrahedral elements with different mesh sizes $h$ normalized by the pipe radius $R$. Convergence rates computed from successive errors and mesh sizes are annotated.} 
\label{fig:rigid_womersley_convergence}
\end{figure}

To reflect typical physiological flows, we set the pipe radius $R$ to $0.3$; fluid density $\rho^f$ and viscosity $\mu^f$ to $1.0$ and $0.04$, respectively; period $T_p$ to $1.1$; reference pressure $p_{\textup{ref}}$ to $0$; and Fourier coefficients $k_0$ and $k_1$ to $-21.0469$ and $-33.0102+42.9332\iota$, respectively. Correspondingly, the fundamental frequency $\omega$ and Womersley number $\alpha_1$ were approximately $5.71$ and $3.59$, respectively. Furthermore, given the fully developed flow, we set a short pipe length of $0.3$. The no-slip boundary condition was prescribed on the wall, and traction boundary conditions were prescribed on both the inlet and outlet. Simulations were performed with uniform time steps using both linear (P1) and quadratic (P2) tetrahedral meshes of comparable numbers of nodes generated by MeshSim (Simmetrix, Inc., Clifton Park, NY, USA), and relative errors of velocity, pressure, and WSS were computed. To circumvent confounding errors from temporal discretization, temporal refinement was performed for each simulation until the first three significant digits of all computed errors were preserved across two temporal refinement levels.

Figure \ref{fig:rigid_womersley_convergence} plots the relative errors of velocity $\bm v_h$ and WSS $\bm \tau_h$ in the $L_2$ norm, and of pressure $p_h$ in the $L_2$ and $H_1$ norms. For three of the four computed errors (Figure \ref{fig:rigid_womersley_convergence} A, B, and D), we consistently observe theoretical rates, with P2 elements exhibiting spatial accuracy of one order higher than that of P1 elements. For P2 elements, however, the relative error of $p_h$ in the $L_2$ norm (Figure \ref{fig:rigid_womersley_convergence} C), converges faster than the theoretical rate of $2.5$. This is likely due to the fact that the analytical solution for pressure here is only linear in space and thus falls within a subspace smaller than the approximation space.

\subsection{Womersley flow in a thin-walled elastic pipe}
\label{section:womersley_def}
As in the rigid case, the Womersley solution for pulsatile flow in an elastic pipe describes axisymmetric flow subject to a pressure gradient with both steady and oscillatory contributions. Given the motion of the elastic pipe, however, the radial velocity of the fluid is no longer identically zero, and the pressure propagates down the pipe with a gradient dependent on both time $t$ and the longitudinal coordinate $z$. This wave propagation is in sharp contrast to the rigid case, in which the fluid oscillates in bulk. 

Let $c_n$ be the $n$-th complex-valued wave speed. Then under the long-wave approximation, namely that the wavelength $\lambda_n := c_{n} T_p = 2\pi c_{n} / (n \omega)$ is much larger than the pipe radius $R$, and the assumption that the wave speed $c_{n}$ is much larger than the fluid velocity, all nonlinear convective terms can be considered negligible, thereby reducing the Navier-Stokes equations to a set of linear equations. As in the rigid case in Section \ref{section:womersley_rigid}, the solution can then be represented as the summation of $N$ superimposed Fourier series. In this case, pressure can be expressed with Fourier coefficients $B_n$ as follows,
\begin{align}
\label{eq:womersley_def_pressure}
p = p_{\textup{ref}} + B_0 z + \sum_{n=1}^{N} B_n e^{\iota n\omega (t - z/c_n)},
\end{align}

In addition to a thin-walled assumption for the elastic pipe (i.e., $h^s \ll R$), the radial wall displacement can be assumed small such that the continuity of velocity at the fluid-solid interface can be imposed at the neutral position of the wall, $r=R$. The fluid velocity components in the longitudinal and radial directions, $v_z$ and $v_r$, can then be expressed with the same Fourier coefficients $B_n$,
\begin{align}
\label{eq:womersley_def_velo}
& v_z = \frac{B_0}{4\mu^f}\left(r^2 - R^2\right) + \sum\limits_{n=1}^{N} \frac{B_n}{\rho^f c_n} \left( 1 - G_n \frac{J_0(\iota^{\frac{3}{2}} \alpha_n \frac{r}{R})}{J_0(\iota^{\frac{3}{2}}\alpha_n)} \right) e^{\iota n \omega (t - z/c_n)}, \\
& v_r = \sum\limits_{n=1}^{N} \frac{\iota n \omega B_n  R}{2 \rho^f c_n^2} \left( \frac{r}{R} - G_n \frac{2 J_1(\iota^{\frac{3}{2}} \alpha_n \frac{r}{R})}{\iota^{\frac{3}{2}}\alpha_n J_0(\iota^{\frac{3}{2}}\alpha_n)} \right) e^{\iota n \omega (t - z/c_n)}, 
\end{align}
and the wall displacement components in the longitudinal and radial directions, $u_z$ and $u_r$, are
\begin{align}
\label{eq:womersley_def_disp}
u_z = \sum\limits_{n=1}^{N} \frac{\iota B_n}{\rho^f c_n n \omega}(G_n - 1) e^{\iota n \omega (t - z/c_n)}, \qquad u_r = \sum\limits_{n=1}^{N} \frac{B_n R}{2 \rho^f c_n^2}(1 - G_n g_n) e^{\iota n \omega (t - z/c_n)}.
\end{align}
The volumetric flow rate can be found by integrating $v_z$ over a cross-section of the pipe,
\begin{align}
\label{eq:womersley_def_flow}
Q = \int_0^R{2\pi r v_z dr} = \frac{-\pi B_0 R^4}{8 \mu^f} + \sum\limits_{n=1}^{N} \frac{B_n \pi R^2}{\rho^f c_n} (1 - G_n g_n) e^{\iota n \omega (t - z/c_n)}.
\end{align}
In the above analytical forms, $G_n$ is the elasticity factor defined as
\begin{align*}
G_n := \frac{2 + \gamma_n (2\nu - 1)}{\gamma_n (2\nu - g_n)}, \qquad \gamma_n := \frac{E h^s}{\rho^f R (1 - \nu^2) c_n^2},
\end{align*}
and the wave speed $c_n$ can be determined from the following equation,
\begin{align}
\label{eq:womersley_def_frequency_eqn}
(g_n - 1)(\nu^2 - 1) \gamma_n^2 + \left( \frac{\rho^s h^s}{\rho^f R}(g_n - 1) + (2\nu - 0.5) g_n - 2 \right) \gamma_n + \frac{2 \rho^s h^s}{\rho^f R} + g_n = 0,
\end{align}
wherein
\begin{align*}
g_n := \frac{2 J_1(\iota^{\frac{3}{2}} \alpha_n)}{\iota^{\frac{3}{2}}\alpha_n J_0(\iota^{\frac{3}{2}}\alpha_n)}.
\end{align*}
Equation \eqref{eq:womersley_def_frequency_eqn}, commonly known as the frequency equation, is constructed by demanding a nontrivial solution to the coupled system of the fluid and elastic pipe \cite[Sec.~5.7]{Zamir2000}. Upon solving for $\gamma_n$ from \eqref{eq:womersley_def_frequency_eqn}, $c_n$ can be represented as
\begin{align*}
c_n = \sqrt{\frac{2}{(1-\nu^2)\gamma_n}} c_{\mathrm{inv}},
\end{align*}
where $c_{\mathrm{inv}}$ is the wave speed in inviscid flows, as given by the Moens-Korteweg formula,
\begin{align}
\label{eq:moens-korteweg-formula}
c_{\mathrm{inv}} = \sqrt{\frac{Eh^s}{2\rho^fR}}.
\end{align}
The consequence of this complex-valued wave speed $c_n$ can be understood from the following decomposition,
\begin{align*}
\frac{1}{c_n} = \frac{1}{c_{n}^{\mathrm{R}}} + \iota \frac{1}{c_{n}^{\mathrm{I}}}, \quad c_{n}^{\mathrm{R}} := \left( \mathrm{Re}\left[c_{n}^{-1} \right] \right)^{-1} , \quad c_{n}^{\mathrm{I}} := \left( \mathrm{Im} \left[ c_{n}^{-1} \right] \right)^{-1},
\end{align*}
wherein $c_{n}^{\mathrm{R}}$ and $c_{n}^{\mathrm{I}}$ are commonly referred to as the dispersion and attenuation coefficients, respectively representing differences in the wave frequency and amplitude from the inviscid case. As is clear from above, $c_n$ depends not only on properties of the fluid and the pipe, but also on the frequency of oscillations.

\begin{figure}
	\begin{center}
	\begin{tabular}{c}
\includegraphics[angle=0, trim=195 190 250 120, clip=true, scale=1.0]{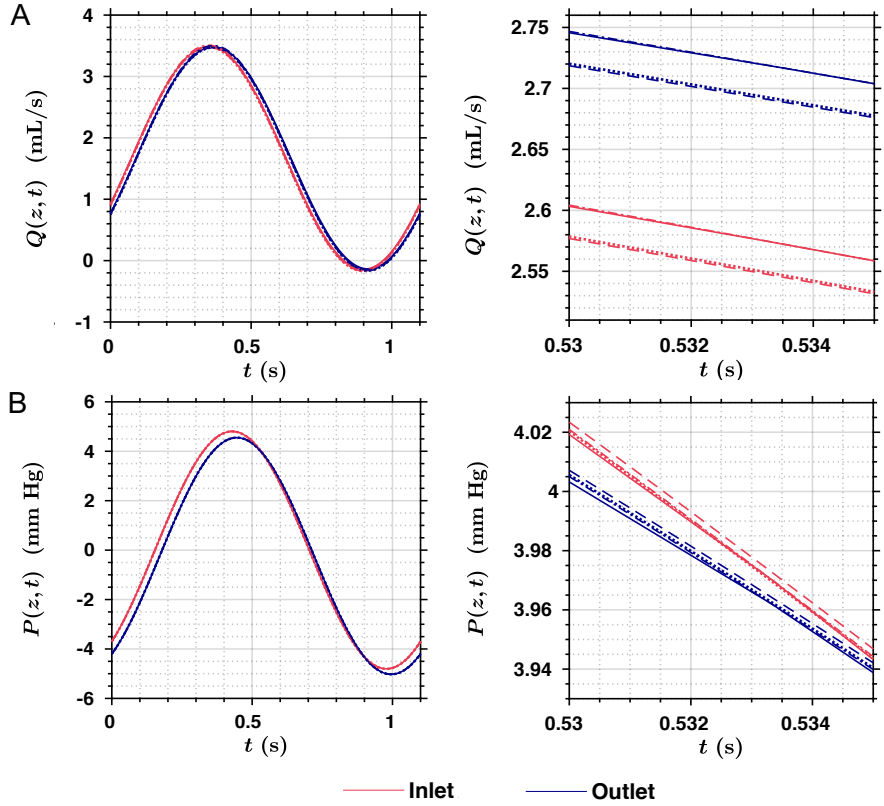}
\end{tabular}
\end{center}
\caption{Analytical (solid) and numerical solutions from CMM (dashed) and our reduced unified continuum formulation using either P1 (dotted) or P2 (dash-dotted) elements for the inlet and outlet (A) volumetric flow rates and (B) pressures over a period. Detailed views are shown in the right column.} 
\label{fig:def_womersley_cap_pres_flow}
\end{figure}

\begin{figure}
	\begin{center}
	\begin{tabular}{c}
\includegraphics[angle=0, trim=195 255 205 165, clip=true, scale=0.83]{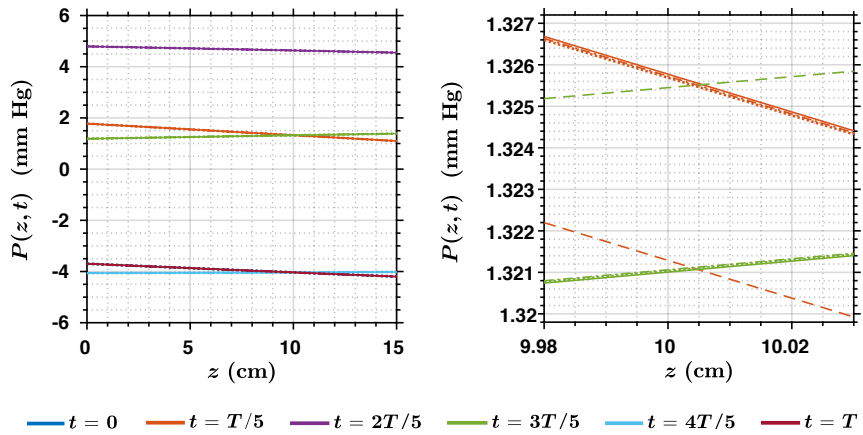}
\end{tabular}
\end{center}
\caption{Analytical (solid) and numerical solutions from CMM (dashed) and our reduced unified continuum formulation using either P1 (dotted) or P2 (dash-dotted) elements for the pressures along the longitudinal axis at different time instances. The solutions at $t=0$ and $t=T$ are overlaid as a result of temporal periodicity. A detailed view is shown on the right.} 
\label{fig:def_womersley_pres}
\end{figure}

\begin{figure}
\begin{center}
\begin{tabular}{c}
\includegraphics[angle=0, trim=270 90 260 88, clip=true, scale=1.2]{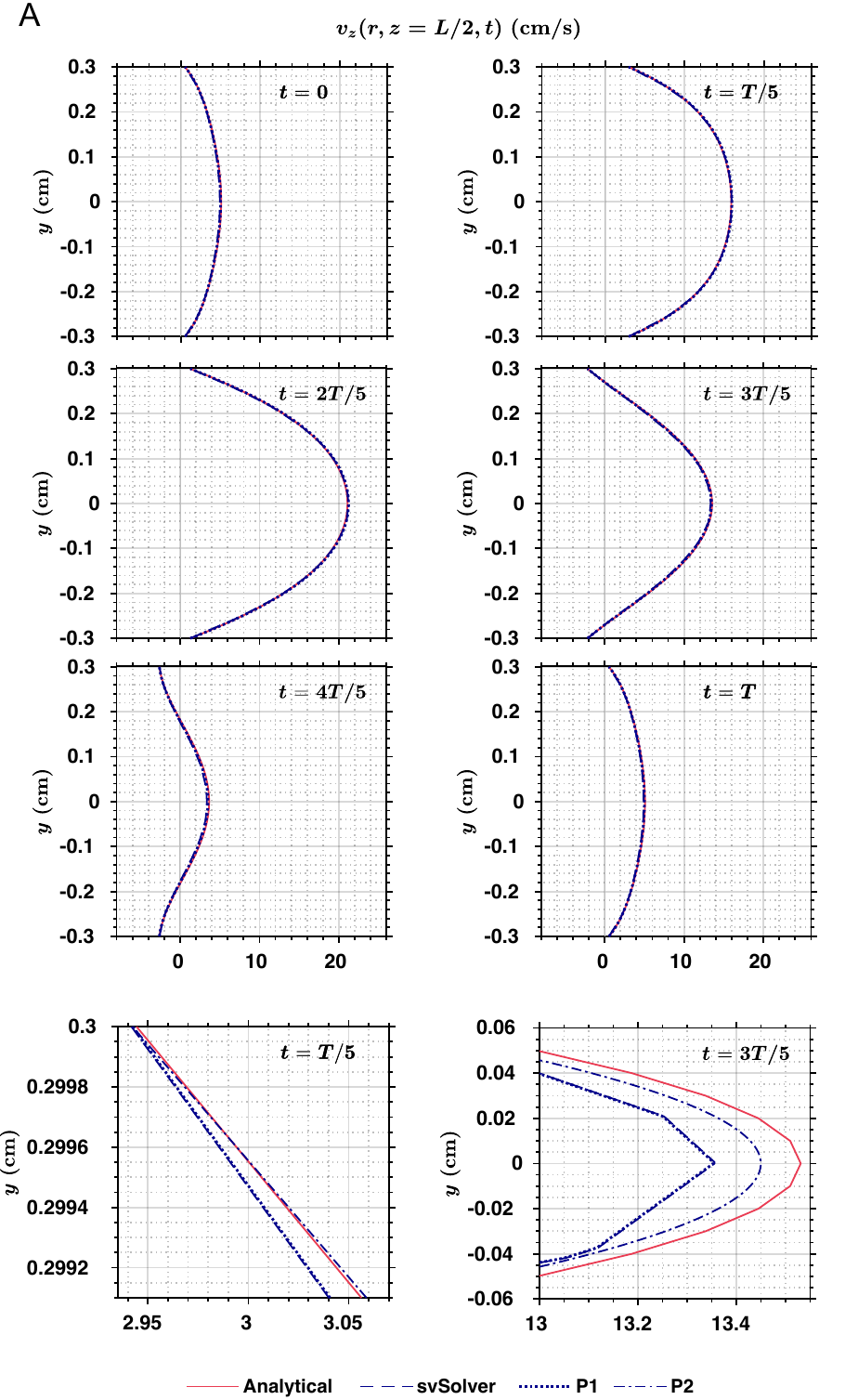}
\end{tabular}
\end{center}
\label{fig:def_womersley_axial_velo_profiles}
\end{figure}

\begin{figure}
\begin{center}
\begin{tabular}{c}
\includegraphics[angle=0, trim=270 90 260 88, clip=true, scale=1.2]{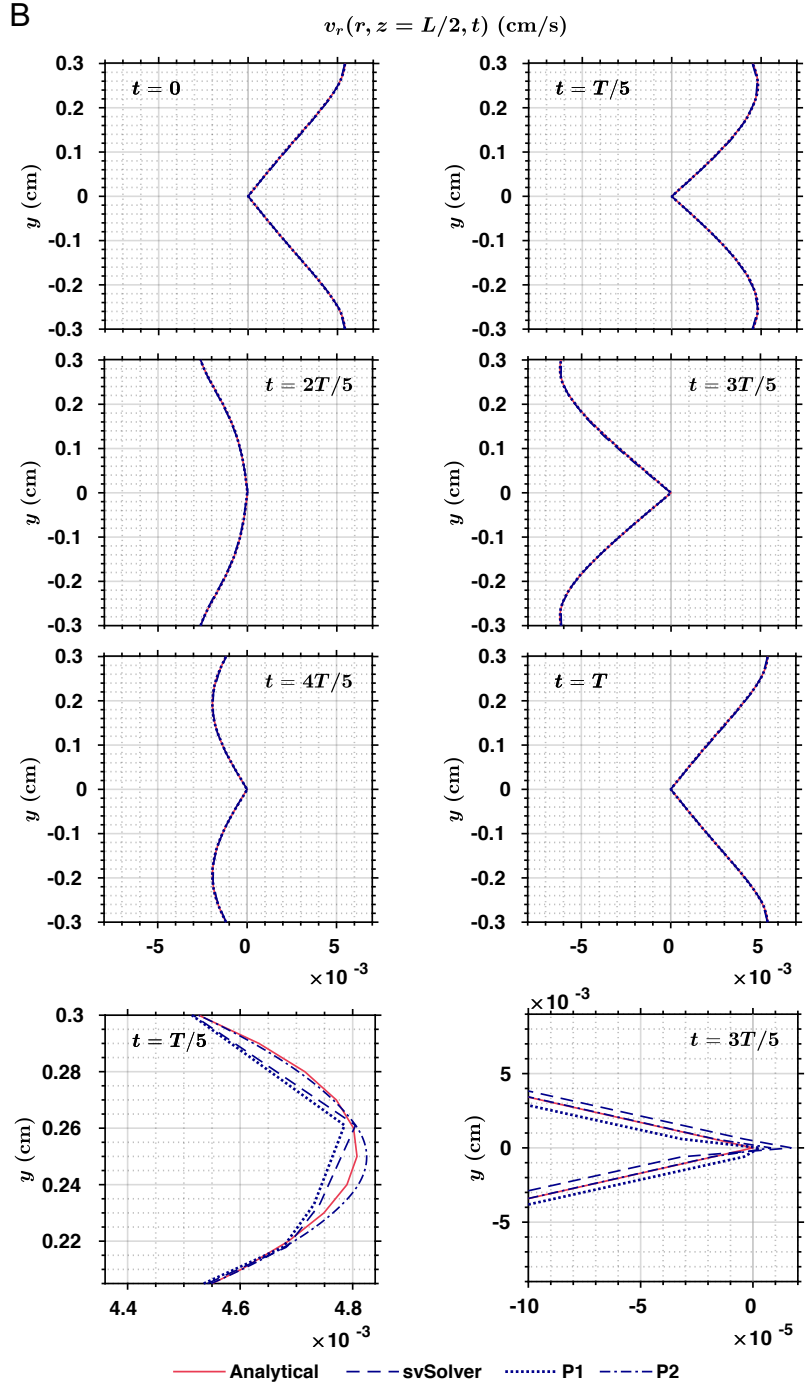}
\end{tabular}
\end{center}
\caption{Analytical and numerical solutions from CMM (dashed) and our reduced unified continuum formulation using either P1 or P2 elements for the (A) longitudinal and (B) radial velocity profiles along the $y$-axis on the $z=L/2$ surface at different time instances. Detailed views at $t=T/5$ and $t=3T/5$ are shown in the bottom row.} 
\label{fig:def_womersley_radial_velo_profiles}
\end{figure}

\begin{figure}
	\begin{center}
	\begin{tabular}{c}
\includegraphics[angle=0, trim=210 90 210 85, clip=true, scale=1.2]{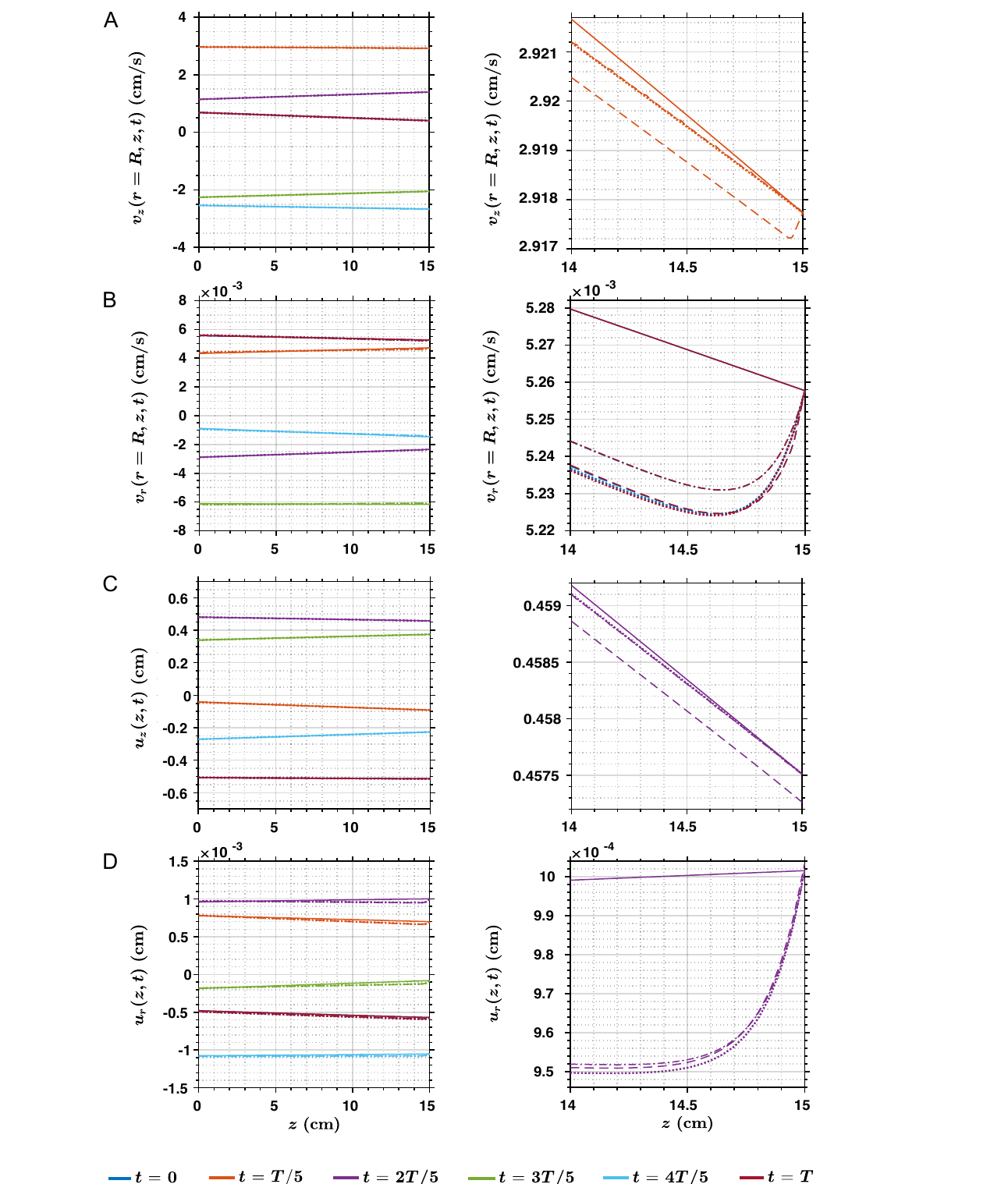}
\end{tabular}
\end{center}
\caption{Analytical (solid) and numerical solutions from CMM (dashed) and our reduced unified continuum formulation using either P1 (dotted) or P2 (dash-dotted) elements for the (A) longitudinal and (B) radial fluid velocities at the wall, and the (C) longitudinal and (D) radial wall displacements along the longitudinal axis at different time instances. The solutions at $t=0$ and $t=T$ are overlaid as a result of temporal periodicity. Detailed views are shown in the right column.} 
\label{fig:def_womersley_wall_disp_velo}
\end{figure}

We again considered only the real components as the benchmark solution and represented a single oscillatory mode ($N=1$). As in Section \ref{section:womersley_rigid}, we set the pipe radius $R$ to 0.3, fluid density $\rho^f$  to $1.0$, viscosity $\mu^f$ to $0.04$, period $T_p$ to $1.1$, and reference pressure $p_{\textup{ref}}$ to $0$. We further set the pipe length $L$ to 15, and considered uniform wall properties, including a wall density $\rho^s$ of 1.0, Poisson's ratio $\nu$ of 0.5, thickness $h^s$ of 0.06, and Young's modulus $E$ of $9.5678 \times 10^6$, which yielded a wave speed $c_1$ of $886.31 + 29.786\iota$. In order to achieve the same volumetric flow rate as in Section \ref{section:womersley_rigid}, the Fourier coefficients $B_0$ and $B_1$ were set to $-21.0469$ and $-4926.29-4092.54\iota$, respectively. Given these parameters, we may examine the validity of the invoked assumptions. At the fundamental frequency, the real component of the wave speed $c_{1}^{\mathrm{R}} = 887.31$ is much larger than the maximum longitudinal velocity $\mathrm{max}\lbrace v_z \rbrace = 21.0701$. Correspondingly, the real component of the leading wavelength $\lambda_1^{\mathrm{R}} := c_{1}^{\mathrm{R}} T_p = 2\pi c_{1}^{\mathrm{R}} / \omega$ is $976.05$, three orders of magnitude larger than $R=0.3$, thereby satisfying the long wave approximation assumption. We further verify for the elastic pipe that both the thickness $h^s = 0.06$ and the maximum radial wall displacement $\mathrm{max}\lbrace u_r\rbrace = 0.0010$ are much smaller than $R$.

To account for the truncation of the semi-infinite domain used in Womersley's derivation to a finite domain, Cartesian velocity components were prescribed on the boundary nodes of the wall at $z=0$ and $z=L$ in the following form,
\begin{align*}
\bm v|_{r=R} := \{v_x, v_y, v_z\}|_{r=R} = \{ v_r \mathrm{cos}\theta, v_r \mathrm{sin}\theta, v_z \}|_{r=R},
\end{align*}
wherein $\theta$ is the four-quadrant inverse tangent of the point $(x,y)$.\footnote{ This function is commonly denoted as $\theta := \mathrm{atan2}\left( y, x \right)$ in programming languages.} Traction boundary conditions were prescribed on both the inlet and outlet surfaces, where the traction $\bm h^f$ was constructed from the pressure in \eqref{eq:womersley_rigid_pressure} and the following Cartesian velocity gradients,
\begin{alignat*}{3}
\frac{\partial v_x}{\partial x} &= \mathrm{cos}^2\theta \frac{\partial v_r}{\partial r} + \frac{\mathrm{sin}^2\theta}{r} v_r, \qquad
 &\frac{\partial v_x}{\partial y} &= \mathrm{sin}\theta \mathrm{cos}\theta \left( \frac{\partial v_r}{\partial r} - \frac{v_r}{r}  \right), \qquad
 &\frac{\partial v_x}{\partial z} &= \mathrm{cos}\theta \frac{\partial v_r}{\partial z}, \\
\frac{\partial v_y}{\partial x} &= \mathrm{sin}\theta \mathrm{cos}\theta \left( \frac{\partial v_r}{\partial r} - \frac{v_r}{r} \right), \qquad
&\frac{\partial v_y}{\partial y} &= \mathrm{sin}^2\theta \frac{\partial v_r}{\partial r} + \frac{\mathrm{cos}^2\theta}{r} v_r, \qquad
&\frac{\partial v_y}{\partial z} &= \mathrm{sin}\theta \frac{\partial v_r}{\partial z}, \\
\frac{\partial v_z}{\partial x} &= \mathrm{cos}\theta \frac{\partial v_z}{\partial r}, \qquad
&\frac{\partial v_z}{\partial y} &= \mathrm{sin}\theta \frac{\partial v_z}{\partial r}, \qquad
& \frac{\partial v_z}{\partial z} &= \frac{\partial v_z}{\partial z},
\end{alignat*}
wherein
\begin{align*}
& \frac{\partial v_z}{\partial r} = \frac{B_0 r}{2\mu^f} + \sum_{n=1}^N \frac{\iota^{\frac{3}{2}} \alpha_n B_n G_n J_1(\iota^{\frac{3}{2}} \alpha_n \frac{r}{R})}{\rho^f c_n J_0(\iota^{\frac{3}{2}} \alpha_n)} e^{\iota n \omega (t - z/c_n)}, \displaybreak[2] \\
& \frac{\partial v_z}{\partial z} = \sum_{n=1}^N \frac{-\iota n \omega B_n}{\rho^f c_n^2} \left( 1 - G_n \frac{J_0(\iota^{\frac{3}{2}} \alpha_n \frac{r}{R})}{J_0(\iota^{\frac{3}{2}} \alpha_n)}\right) e^{\iota n \omega (t - \frac{z}{c_n})}, \displaybreak[2] \\
& \frac{\partial v_r}{\partial r} = \sum_{n=1}^N\frac{\iota n \omega_n B_n}{2 \rho^f c_n^2}\left( 1 - G_n \frac{2 \left(J_1(\iota^{\frac{3}{2}} \alpha_n \frac{r}{R}) - J_2(\iota^{\frac{3}{2}} \alpha_n \frac{r}{R}) \right)}{\iota^{\frac{3}{2}} \alpha_n \frac{r}{R} J_0(\iota^{\frac{3}{2}} \alpha_n)} \right) e^{\iota n \omega (t - z/c_n)}, \displaybreak[2] \\
& \frac{\partial v_r}{\partial z} = \sum_{n=1}^N\frac{n^2 \omega^2 B_n R}{2 \rho^f c_n^3} \left( \frac{r}{R} - G_n \frac{2 J_1(\iota^{\frac{3}{2}} \alpha_n \frac{r}{R})}{\iota^{\frac{3}{2}} \alpha_n J_0(\iota^{\frac{3}{2}} \alpha_n)} \right) e^{\iota n \omega (t - z/c_n)}. 
\end{align*}
We note that our choice of boundary conditions differs from the approach adopted in verification studies for CMM, in which only the normal components of the tractions are prescribed on the inlet and outlet surfaces via impedance boundary conditions \cite{Figueroa2006a,Filonova2019}.

Simulations were performed over three periods with uniform time steps using linear and quadratic tetrahedral meshes, both of 284,400 elements and respectively 53,879 and 404,473 nodes. The linear tetrahedral mesh was additionally used to make comparisons against svSolver \cite{svsolver}, the CMM implementation in SimVascular. For each simulation, only the final period was analyzed. Given the assumptions and scaling analyses invoked in the derivations, the analytical solutions \eqref{eq:womersley_def_velo}-\eqref{eq:womersley_def_disp} are only approximate solutions to the FSI problem presented in Section \ref{sec:formulation} and thereby preclude any spatial convergence analyses. We show comparisons of analytical and numerical solutions for the volumetric flow rates and pressures (Figures \ref{fig:def_womersley_cap_pres_flow}, \ref{fig:def_womersley_pres}), longitudinal and radial fluid velocity profiles (Figure \ref{fig:def_womersley_radial_velo_profiles}), and longitudinal and radial wall velocity and displacement (Figure \ref{fig:def_womersley_wall_disp_velo}). We note that all numerical results are nearly indistinguishable from the analytical solutions. Differences, however, can be observed in the detailed views, where the P2 results are in closer agreement to the analytical solutions compared to the P1 results. In Figures \ref{fig:def_womersley_cap_pres_flow}B and \ref{fig:def_womersley_pres}, we observe that CMM yields larger discrepancies in pressure than our proposed method, likely due to the different treatment of pressure in the temporal discretization (see Remark \ref{remark:gen-alpha-pressure}) \cite{Liu2020a}. Across all numerical cases, the fluid velocity is in good agreement with the analytical solutions and presents axisymmetric profiles along the radial direction. This is in sharp contrast to existing CMM verification results in the literature exhibiting a notable lack of axisymmetry in the radial velocity profiles \cite{Filonova2019} that could be attributed to the outlet impedance boundary condition \cite{Figueroa2006a,Filonova2019}, which neglects viscous traction components. Contrary to the fluid quantities, larger discrepancies can be observed in the wall displacement and velocity. These discrepancies, which were not mitigated upon mesh refinement, can be attributed to the assumptions inherent in the theory.

\section{Physiological modeling techniques}
\label{sec:practical_modeling_techniques}
In this section, we briefly present a suite of practical techniques for appropriate modeling of physiological phenomena in clinical applications. Specifically, these techniques pertain to vascular wall thickness heterogeneity, in vivo tissue prestressing, and boundary conditions reflecting distal vasculature. 

\subsection{Spatially varying vascular wall thickness}
\begin{figure}
	\begin{center}
	\begin{tabular}{c}
\includegraphics[angle=0, trim=135 95 150 95, clip=true, scale=0.63]{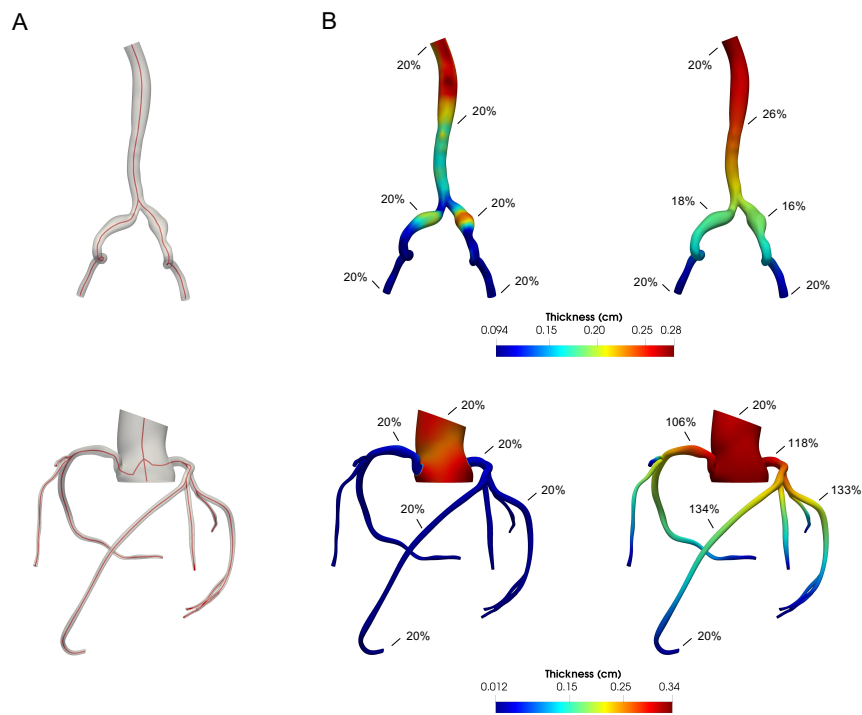}
\end{tabular}
\end{center}
\caption{(A) Centerlines extracted for healthy models of the aortailiac (top) and coronary arteries (bottom). (B) Spatially varying wall thickness distributions obtained from the centerline-based (left) and Laplacian (right) approaches. The wall thickness is precisely 20\% of the local radius \textit{everywhere} in the centerline-based approach, but \textit{only at the inlets and outlets} in the Laplacian approach. Local thickness-to-radius ratios are annotated at multiple sites to highlight the Laplacian approach's deviation from the prescribed thickness-to-radius ratio and its resulting non-physiological wall thickness distribution.} 
\label{fig:var_wall_thickness}
\end{figure}

The most commonly employed imaging modalities, such as computed tomography or magnetic resonance angiography, do not adequately resolve vascular wall thicknesses for most applications of clinical interest. While intravascular ultrasound is a notable exception, it is only performed in a small subset of clinical cases, primarily in the coronary vasculature. As a result, spatially varying distributions of vascular wall thicknesses must frequently be prescribed with limited knowledge, commonly with an assumed local thickness-to-radius ratio \cite{Humphrey2002}. In a previously proposed Laplacian approach \cite{Bazilevs2009a}, a Laplacian problem is solved with prescribed Dirichlet boundary conditions at the wall boundary nodes on all inlets and outlets \cite{Bazilevs2009a}. A similar approach has also since been adopted for prescribing cardiac fiber orientations in heart models \cite{Wong2014}. While the Laplacian approach effectively generates smooth distributions of wall thicknesses, it fails to capture sharp local changes in geometry as often occur in disease, yielding physiological thicknesses near the inlets and outlets but significant deviations from the desired thickness-to-radius ratio elsewhere. For example, non-physiological wall thicknesses up to 134\% of the local radii are prescribed in the coronary arteries near the ostia on the aortic root (Figure \ref{fig:var_wall_thickness}B).

For more refined control over the local thickness, we instead adopt a centerline-based approach similar to the one used in \cite{Xiao2013}, in which centerlines for all inlet-outlet pairs are extracted using the Vascular Modeling Toolkit \cite{vmtk-website,Antiga2008}. Upon specifying a global distribution over the entire wall, we can overwrite thicknesses with distinct local distributions for arbitrary sub-domains of the wall. We summarize our approach in Algorithm \ref{algorithm:var_wall_prop}, in which the conglomeration of all vessel centerlines is referred to as the \textit{global centerline}. We note that for geometries with sharp changes in radius, simply computing the local radius as the shortest distance to the global centerline could yield values based not on the corresponding vessel centerline of interest, but rather on an alternate centerline adjoined to the vessel centerline of interest. It is therefore sometimes helpful to extract individual vessel-specific centerlines from the global centerline prior to overwriting thicknesses in vessel sub-domains. In this work, we set a thickness-to-radius ratio of 20\% \cite{Humphrey2002}, which is precisely satisfied everywhere (Figure \ref{fig:var_wall_thickness}B).

\begin{algorithm}[H]
\caption{Centerline-based assignment of spatially varying vascular wall thickness.}
\label{algorithm:var_wall_prop}
\begin{algorithmic}[1]
\State \texttt{Extract the global centerline from the wall surface \cite{Antiga2002}}
\For{ \texttt{each node on the wall} }
    \State \texttt{Compute the radius $r$ as the shortest distance to the global centerline}
    \State \texttt{Compute the thickness} $h^s \gets x\% \ r$
\EndFor
\For{ \texttt{each sub-domain with a distinct local distribution} }
    \State \texttt{Extract the local centerline from the global centerline}
    \For{ \texttt{each node on the sub-domain} }
        \State \texttt{Compute the radius $r$ as the shortest distance to the local centerline}
        \State \texttt{Compute the thickness $h^s$ as desired}
    \EndFor
\EndFor
\end{algorithmic}
\end{algorithm}

\subsection{Tissue prestressing}
The semi-discrete FSI formulation \eqref{eq:semidiscrete-fsi-couple}-\eqref{eq:semidiscrete-fsi-continuity} assumes the in vivo vascular wall configuration at imaging to be stress-free, yet vascular walls withstand physiological loading. An internal stress state, termed the prestress, must exist to balance the in vivo blood pressure and viscous traction. In contrast to approaches that seek to determine a stress-free configuration \cite{Tezduyar2008,Nama2020}, here we generate the prestress $\bm \sigma_0$ via a fixed-point algorithm similar to the one proposed for an ALE formulation \cite{Hsu2011, Baeumler2020}. Given a prestress field $\bm \sigma_0$, the wall momentum balance \eqref{eq:semi-discrete-u-reformulated} in the FSI formulation can correspondingly be modified as
\begin{align}
\label{eq:semi-discrete-u-with-fluid-traction}
\mathbf B^w_{\mathrm{m}} \left( \bm w_h^f ;  \dot{\bm y}_h, \bm y_h\right) := & \int_{\Gamma_I} \bm w_h^f \cdot \rho^s h^s \left( \frac{d \bm v_h^f}{d t} - \bm b^s \right) d\Gamma  + \int_{\Gamma_I} h^s \bm \epsilon(\bm w_h^f) : \Big( \bm \sigma^s(\bm u_h^w) + \bm \sigma_0 \Big) d\Gamma \nonumber \\
&- \int_{\partial \Gamma_I \cap \Gamma^h_s } h^s \bm w_h^f \cdot \bm h^s d\Gamma.
\end{align}
To determine $\bm \sigma_0$, we consider the following variational problem for the vascular wall. Given the body force per unit mass $\bm b^s$, boundary traction $\bm h^s$, and fluid boundary traction $\bm h^f$, find $\bm u_h^w \in \mathcal S^w_{\bm u}$ and $\bm v^w_h \in \mathcal S^w_{\bm v}$, such that for $\forall \bm w^f_h \in \mathcal{V}^f_{\bm v}$,
\begin{align}
\label{eq:semi-discrete-u-prestress-kinematics}
\bm 0 =& \frac{d \bm u^w_h}{dt} - \bm v^w_h, \quad \mbox{ and } \quad
0 = \mathbf B^w_{\mathrm{m}} \left( \bm w_h^f ;  \dot{\bm y}_h, \bm y_h\right) + \int_{\Gamma_I} \bm w_h^f \cdot \bm h^f d\Gamma,
\end{align}
where $\mathcal S^w_{\bm u}$ and $\mathcal{V}^f_{\bm v}$ are as previously defined in Section \ref{subsec:semi-discrete-formulation}, and $\mathcal S^w_{\bm v}$ is a suitable trial solution space for the wall velocity. Using the prestress generation algorithm summarized below, $\bm \sigma_0$ is then determined such that equations \eqref{eq:semi-discrete-u-prestress-kinematics} are satisfied under the imaged wall configuration. We denote the prestress at the $m$-th iteration as $\bm \sigma_{0,(m)}$ and the maximum number of iterations as $m_{\mathrm{max}}$.

\begin{myenv}{Prestress generation algorithm}
\noindent \textbf{Initialization:} Set $\bm \sigma_{0, (0)} = \bm 0$ and $\bm u^w_{0} = \bm 0$.

\noindent \textbf{Fixed-point iteration:} Repeat the following steps for $m=0, 1, ..., m_{\mathrm{max}}$.
\begin{enumerate}
	\item Set $\bm \sigma_0 = \bm \sigma_{0, (m)}$ and $\bm u^w_m = \bm 0$.
    \item From $t_m$ to $t_{m+1}$, solve the variational problem \eqref{eq:semi-discrete-u-prestress-kinematics} for $\bm u^w_{m+1}$ and $\bm v^w_{m+1}$ using the backward Euler method for temporal discretization.
    \item Update the prestress tensor as $\bm \sigma_{0, (m+1)} = \bm \sigma^s(\bm u^w_{m+1}) + \bm \sigma_{0, (m)}$.
    \item Let $\mathrm{tol}_{\mathrm{P}}$ denote a prescribed tolerance. If the stopping criterion $\| \bm u^w_{m+1} \|_{\mathfrak l_2} \leq \mathrm{tol}_{\mathrm{P}}$ is satisfied, then set $\bm \sigma_0 = \bm \sigma_{0,(m+1)}$ and exit the fixed-point iteration.
\end{enumerate}
\end{myenv}

\begin{remark}
\label{remark:prestress_diastolic_traction}
To minimize cardiac motion artifacts, cardiac images are commonly acquired at diastole via electrocardiogram gating. The fluid boundary traction $\bm h^f$ at diastole can then be obtained from a separate rigid-wall CFD simulation prescribed with a steady diastolic inflow rate and outlet resistances tuned to achieve the corresponding diastolic pressures and flow splits.
\end{remark}

\subsection{Coupling with reduced models}
\label{subsec:coupling-with-reduced-models}
As alluded to in Section \ref{sec:iterative_solution_method}, zero-dimensional models representing the downstream vasculature are frequently coupled to outlets of the three-dimensional domain \cite{Moghadam2013, VignonClementel2006, VignonClementel2010}. While we restrict our attention to Neumann coupling with zero-dimensional models in which the boundary traction is a function of the flow rate at the corresponding outlet surface only, we note that more generally, any arbitrary combination of Neumann (Dirichlet-to-Neumann) and Dirichlet (Neumann-to-Dirichlet) inlets and outlets and their corresponding system of (nonlinear) ordinary differential equations can be considered \cite{Kung2020}. We consider the Neumann boundary $\Gamma^f_h$ to consist of $n_{\mathrm{out}}$ non-overlapping planar outlet surfaces,
\begin{align*}
\Gamma^f_h = \bigcup_{i=1}^{\mathrm n_{\mathrm{out}}} \Gamma^f_{\mathrm{out},i}, \qquad \overline{\Gamma^f_{\mathrm{out,i}}} \cap \overline{\Gamma^f_{\mathrm{out,j}}} = \emptyset, \mbox{ for } 1 \leq i,j \leq \mathrm n_{\mathrm{out}} \mbox{ and } i \neq j.
\end{align*}
Let $\mathcal F^k$ be a functional operator and $Q^k(t)$ be the volumetric flow rate through the outlet surface $\Gamma^f_{\mathrm{out,k}}$,
\begin{align*}
Q^k(t) := \int_{\Gamma^f_{\mathrm{out},k}} \bm v^f(t) \cdot \bm n d\Gamma.
\end{align*}
The boundary traction on $\Gamma^f_{\mathrm{out},k}$ is then given by $\bm h^f = -P^k(t)\bm n$, where $P^k(t) = \mathcal F^k(Q^k(t))$, and the term \eqref{eq:vms_traction} can be written explicitly as
\begin{align*}
\mathbf B_{\mathrm{m}}^{\mathrm{h}} \left( \bm w_h^f ;  \dot{\bm y}_h^f, \bm y_h^f \right) := - \int_{\Gamma_h^f} \bm w_h^f \cdot \bm h^f d\Gamma = \sum_{k=1}^{n_{\mathrm{out}}} \mathcal F^k(Q^k(t)) \int_{\Gamma^f_{\mathrm{out},k}} \bm w_h^f \cdot \bm n d\Gamma.
\end{align*}
The corresponding contribution to the block matrix $\boldsymbol{\mathrm A}_{(l)}$ in \eqref{eq: predictor_multi_correct_notation_for_block_matrices} is then a weighted sum of rank-one matrices. Readers are referred to \cite[Section~2.4]{Liu2020} for more details of the consistent tangent matrix. The three-element Windkessel model and coronary model, two commonly used zero-dimensional models for Neumann coupling in cardiovascular simulations, are reviewed in \ref{sec:appendix}.

\section{Clinical applications}
\label{sec:clinical_applications}
In this section, we apply our combined FSI technology and practical modeling techniques to two patient-specific models, one of the pulmonary arteries of a healthy 9-year-old male and the other of the coronary arteries of a healthy 24-year-old male. Linear tetrahedral meshes were generated with three boundary layers each, at a thickness gradation ratio of $0.5$. Patient-specific inflow waveforms were prescribed with parabolic velocity profiles. Outlet boundary conditions were tuned to achieve target inlet systolic and diastolic pressures as well as assumed flow splits (Table \ref{table:model_characteristics}). Specifically, in the pulmonary model, all outlets were coupled to RCR models, and flow was assumed to be evenly distributed to the left and right lungs; in the coronary model, the aortic outlet was coupled to an RCR model while all remaining outlets were coupled to coronary models. Furthermore, $4$\% of the flow was distributed to the coronary arteries, with a $60$\%-$40$\% split for the left and right coronary arteries. Consistent with Section \ref{section:womersley_def}, we adopt the centimeter-gram-second units, and we set the fluid density $\rho^f$ to $1.0$, fluid viscosity $\mu^f$ to $0.04$, and wall density $\rho^s$ to $1.0$. Unless otherwise specified, we set the wall Poisson's ratio $\nu$ to $0.5$, and the Young's modulus $E$ to $1.3 \times 10^6$ uniformly for the pulmonary arteries \cite{Yang2019}, $7.0 \times 10^6$ for the aortic root, and $1.15 \times 10^7$ for the coronary arteries \cite{Ramachandra2016}. The time step size was chosen to be $T_p / 2000$. As discussed in Remark \ref{remark:prestress_diastolic_traction}, we generated initial conditions for each FSI simulation by first running a rigid-wall CFD simulation to generate solution fields at the diastolic pressure. The prestress generation algorithm was subsequently used to obtain the prestress $\bm \sigma_0$ balancing the diastolic fluid boundary traction under zero wall displacement relative to the imaged configuration.

\begin{table*}[htbp]
\footnotesize
\begin{center}
\tabcolsep=0.25cm
\renewcommand{\arraystretch}{1.2}
\begin{tabular}{c | c c c c c c c c}
\hline
\hline
\multirow{2}{*}{\textbf{Model}} & \multirow{2}{*}{\textbf{Sex}} & \vtop{\hbox{\strut \textbf{Age}}\hbox{\strut \textbf{(year)}}} & \vtop{\hbox{\strut \bm{$R_{\mathrm{in}}$}}\hbox{\strut \textbf{(cm)}}} & \vtop{\hbox{\strut \bm{$P_{\mathrm{in}}$}}\hbox{\strut \textbf{(mm Hg)}}} & \vtop{\hbox{\strut \textbf{CO}}\hbox{\strut \textbf{(L/min)}}} & \multirow{2}{*}{\textbf{Flow Split}} & \vtop{\hbox{\strut \bm{$T_p$}}\hbox{\strut \textbf{(s)}}} & \multirow{2}{*}{\textbf{Outlets}} \\
\hline
\vtop{\hbox{\strut Pulmonary}\hbox{\strut Arteries}} & \multirow{2}{*}{M} & \multirow{2}{*}{$9$} & \multirow{2}{*}{$1.24$} & \multirow{2}{*}{$24 \ / \ 8$} & \multirow{2}{*}{$4.59$} & \vtop{\hbox{\strut $50$\% LPA}\hbox{\strut $50$\% RPA}} & \multirow{2}{*}{$0.811$} & \multirow{2}{*}{$46$ RCR} \\
\hline
\multirow{3}{*}{\vtop{\hbox{\strut Coronary}\hbox{\strut Arteries}}} & \multirow{3}{*}{M} & \multirow{3}{*}{$24$} & \multirow{3}{*}{$1.40$} & \multirow{3}{*}{$123 \ / \ 81$} & \multirow{3}{*}{$3.78$} & \vtop{\hbox{\strut $96$\% AO}\hbox{\strut $2.4$\% LCOR}\hbox{\strut $1.6$\% RCOR}} & \multirow{3}{*}{$1.43$} & \multirow{3}{*}{\vtop{\hbox{\strut $1$ RCR}\hbox{\strut $25$ coronary}}} \\
\hline
\hline
\end{tabular}
\end{center}
\caption{Model characteristics. $R_{\mathrm{in}}$: inlet radius; $P_{\mathrm{in}}$: target inlet pressure; CO: prescribed cardiac output; $T_p$: cardiac period; LPA: left pulmonary artery; RPA: right pulmonary artery; AO: aorta; LCOR: left coronary artery; RCOR: right coronary artery.}
\label{table:model_characteristics}
\end{table*}

All results reported in this section were obtained using the TaiYi supercomputer, a Lenovo system equipped with Intel Xeon Gold 6148 processors interconnected by a 100 GB/s Intel Omni-Path network. Each processor consists of 40 CPUs and 192 GB RAM and operates at a clock rate of 2.4GHz \cite{Taiyi-machine-details}.

\subsection{Linear solver robustness}
\label{subsection:solver_robustness}
We examined the linear solver performance under varying wall properties. For this test, we set the relative tolerance $\delta = 10^{-8}$ for the stopping criterion, and the maximum number of iterations for the outer, intermediate, and inner solvers $\mathrm n^{\textup{max}} = \mathrm n^{\textup{max}}_{\mathrm A} = \mathrm n^{\textup{max}}_{\mathrm S} = 200$. Relative tolerances $\delta_{\mathrm A}$ and $\delta_{\mathrm S}$ were jointly varied from $10^{-6}$ to $10^{-2}$, and $\delta_{\mathrm I} = \sqrt{\delta_{\mathrm A}}$. As described in Section \ref{sec:iterative_solution_method}, the preconditioners $\boldsymbol{\mathrm P}_{\mathrm A}$ and $\boldsymbol{\mathrm P}_{\mathrm S}$ were formed by BoomerAMG based on $\boldsymbol{\mathrm A}$ and $\hat{\boldsymbol{\mathrm S}}$, respectively. 

\begin{figure}
	\begin{center}
	\begin{tabular}{cc}
\includegraphics[angle=0, trim=90 90 130 90, clip=true, scale = 0.3]{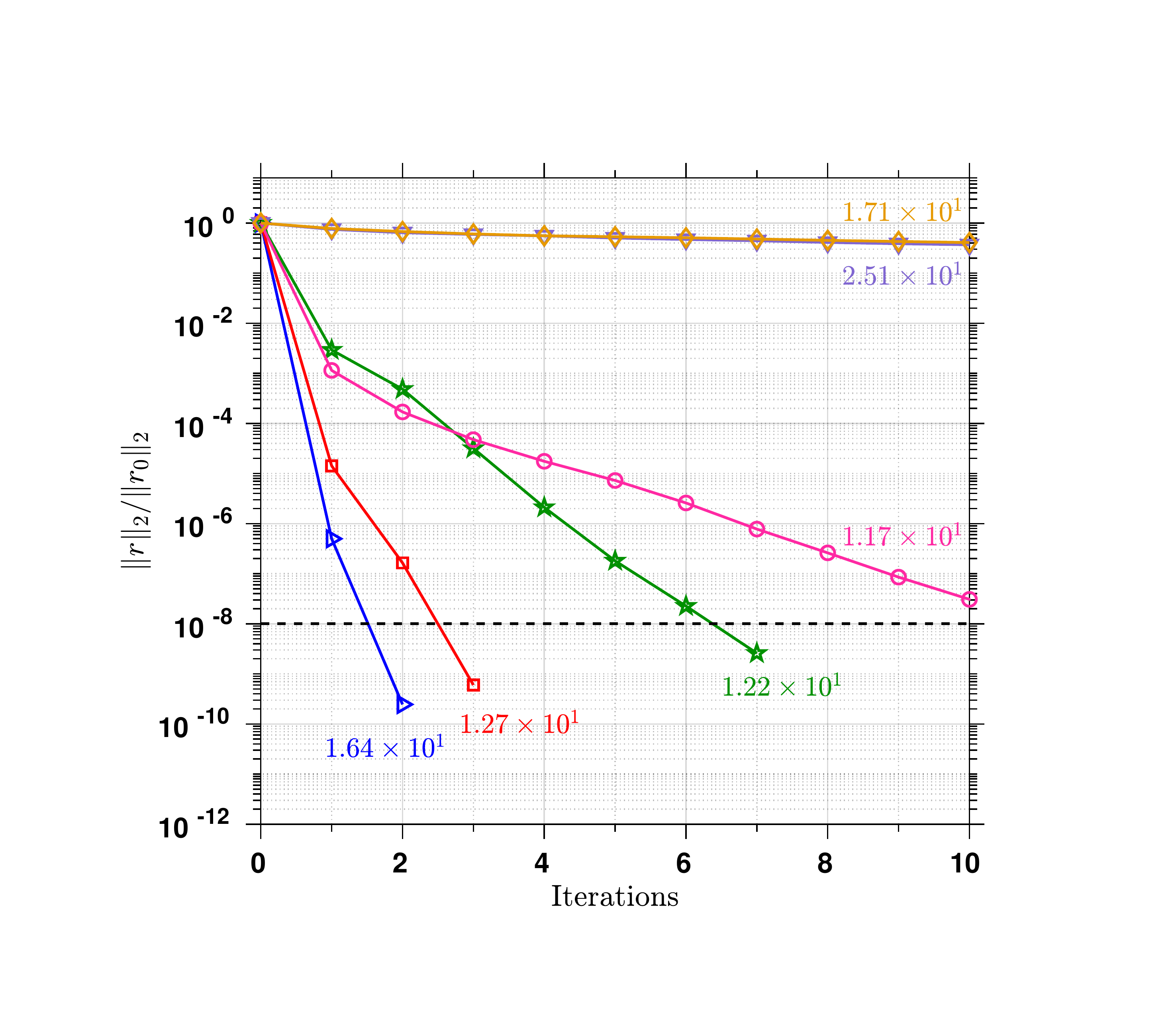} &
\includegraphics[angle=0, trim=90 90 130 90, clip=true, scale = 0.3]{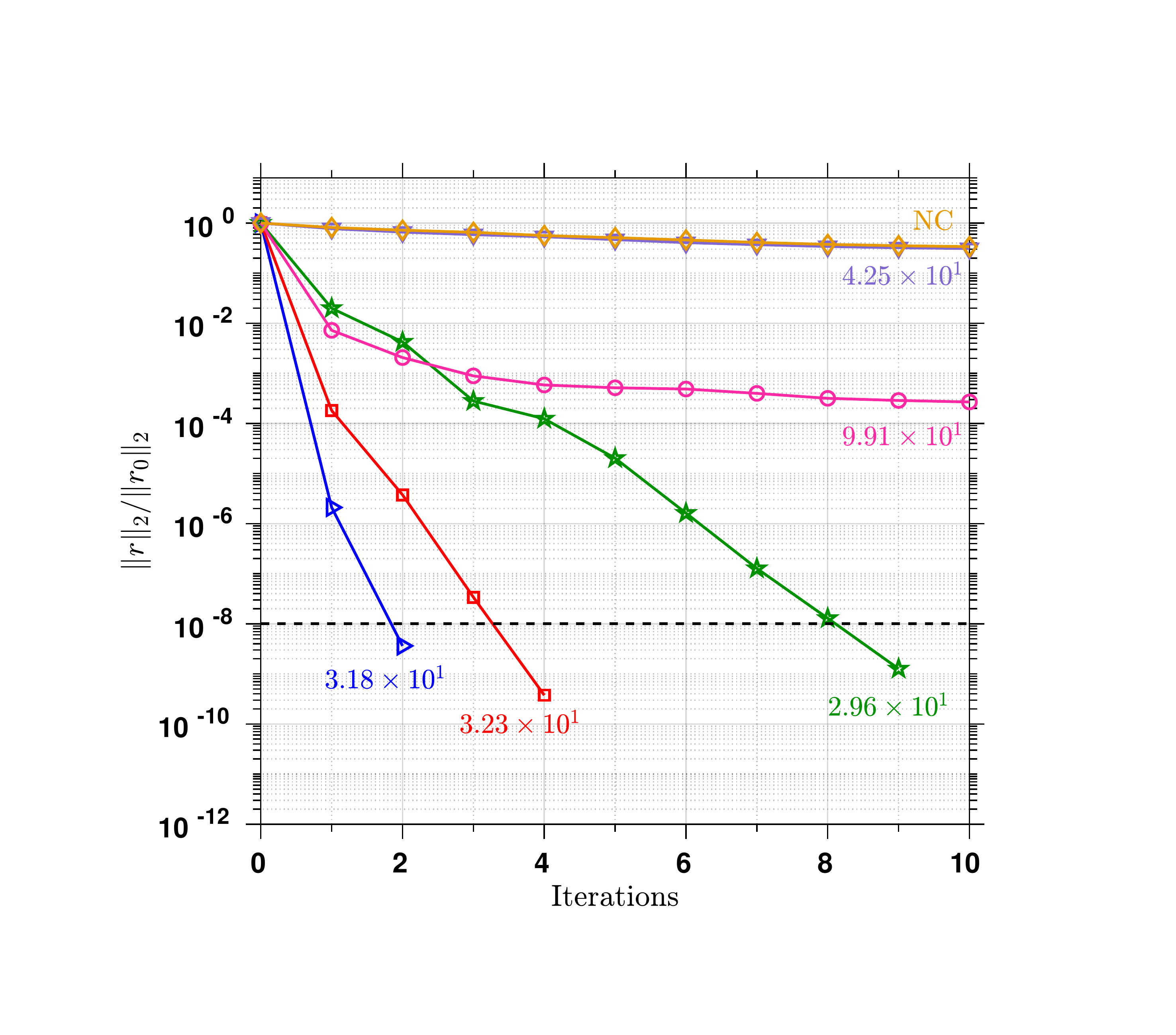} \\
(A) $E=1.3\times 10^5$ & (B) $E=1.3\times 10^6$ \\
\includegraphics[angle=0, trim=90 90 130 90, clip=true, scale = 0.3]{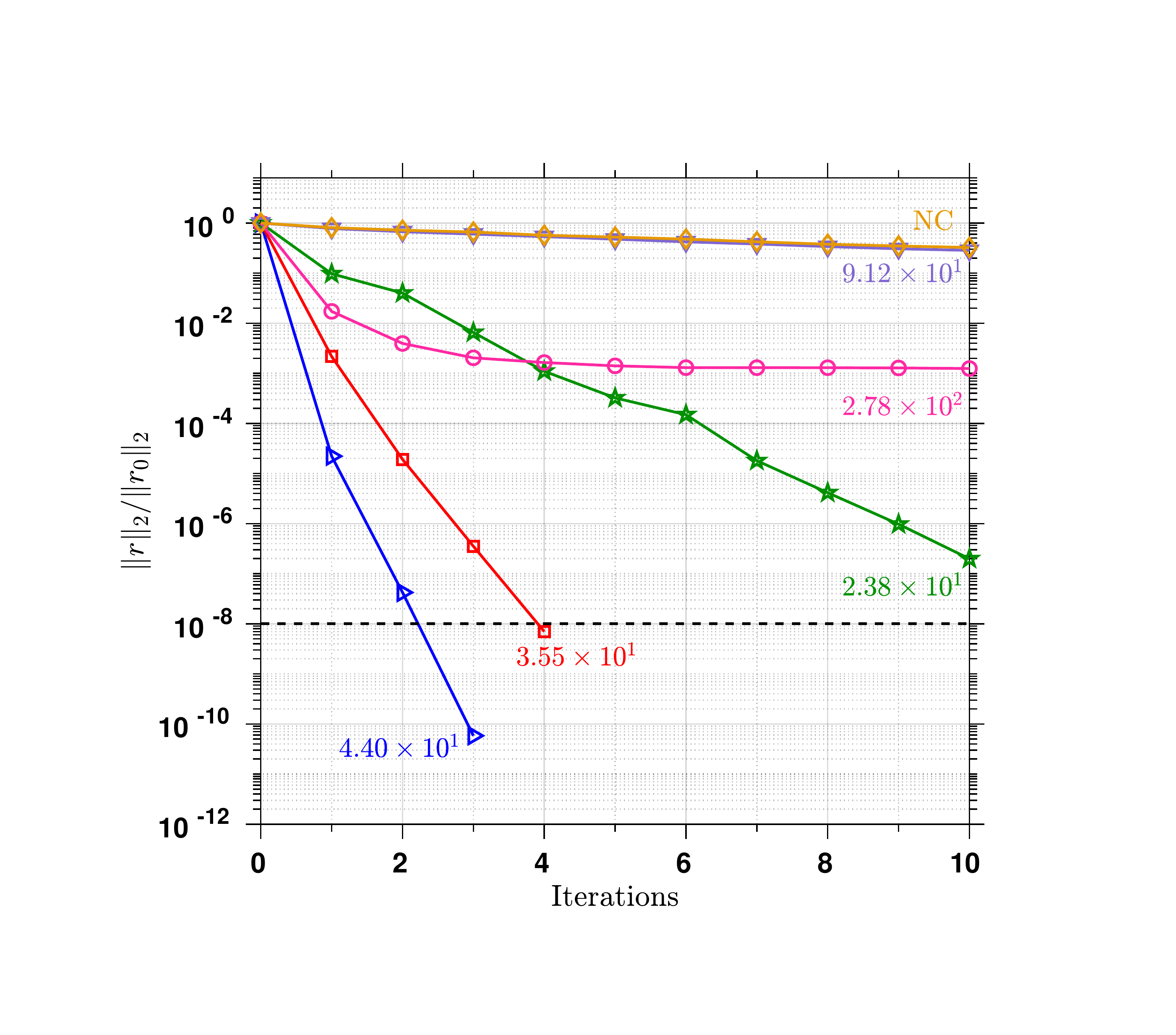} &
\includegraphics[angle=0, trim=90 90 130 90, clip=true, scale = 0.3]{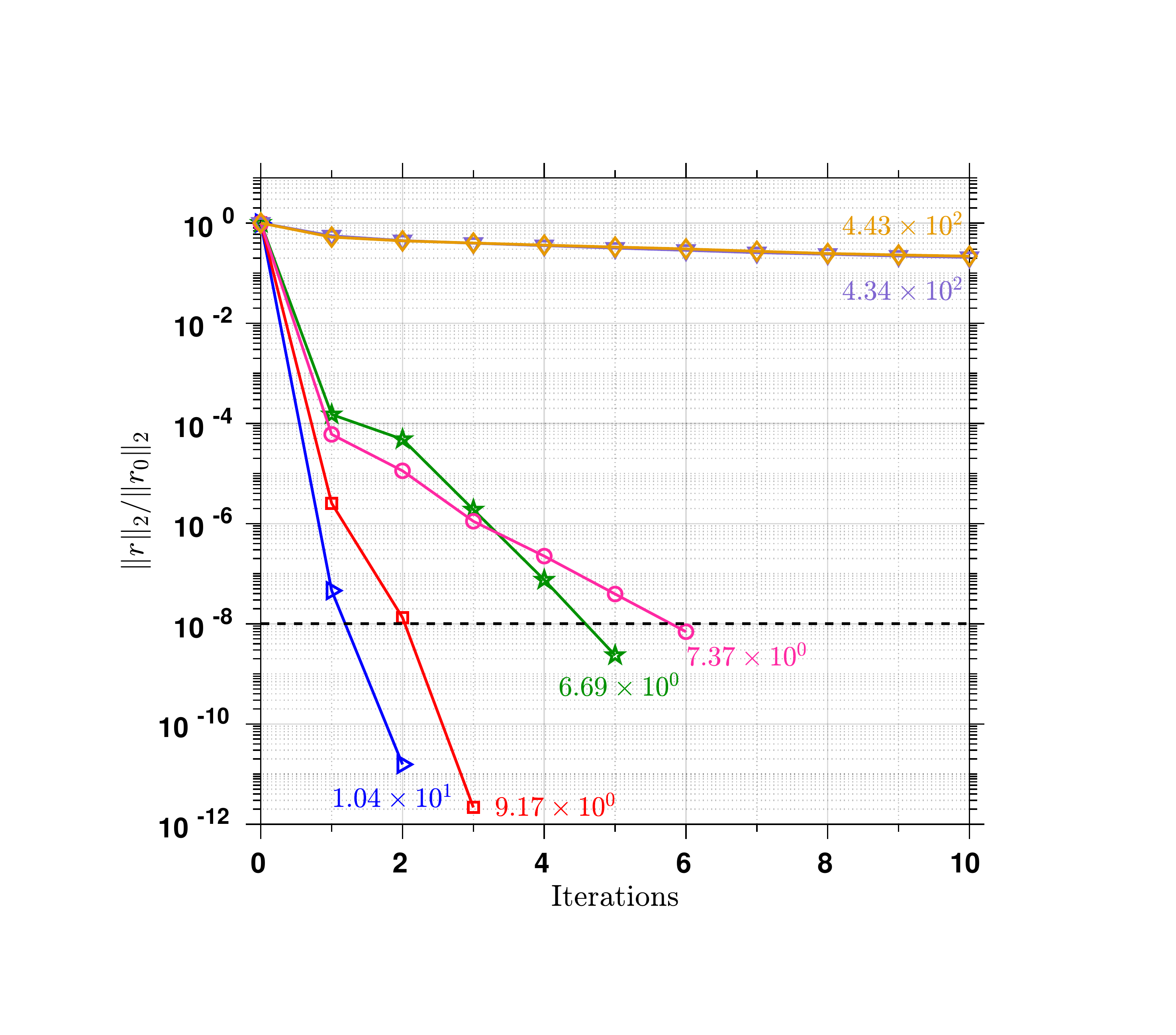} \\
(C) $E=1.3\times 10^7$ & (D) Rigid Wall \\
\multicolumn{2}{c}{ \includegraphics[angle=0, trim=420 120 840 170, clip=true, scale = 0.45]{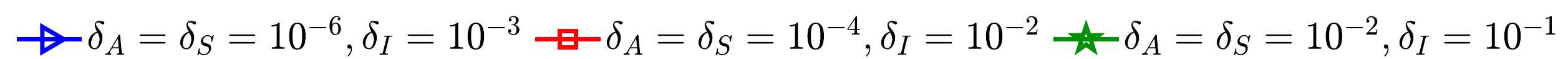}} \\
\multicolumn{2}{c}{ \includegraphics[angle=0, trim=1140 120 320 170, clip=true, scale = 0.45]{./figures/conv_history_legend.pdf}}
\end{tabular}
\caption{Convergence history for the pulmonary arterial model with varying values of the prescribed Young's modulus. A rigid-wall CFD simulation is also included for comparison. The latter three items in the legend correspond to the three alternative linear solver options investigated. The horizontal dashed black line demarcates the prescribed stopping criterion for the relative error $\delta = 10^{-8}$. CPU times (s) for the linear solver averaged over ten time steps are annotated. NC: no convergence within the prescribed maximum number of iterations.} 
\label{fig:conv_history_pul}
\end{center}
\end{figure}

We additionally compared three other linear solver options. In the first alternative, we applied the block preconditioner without invoking the inner solver, that is, replacing \eqref{eq:seg_sol_pres} in Algorithm \ref{algorithm:exact_block_factorization} with $\hat{\boldsymbol{\mathrm S}} \bm y_p = \bm s_{p}$. In the second alternative, we applied the additive Schwarz method to $\boldsymbol{\mathcal A}$, using the incomplete LU factorization for the subdomain solver. In the third alternative, we applied the Jacobi preconditioner to $\boldsymbol{\mathcal A}$. For the latter two alternatives, we increased $\mathrm n^{\textup{max}} = \mathrm n^{\textup{max}}_{\mathrm A} = \mathrm n^{\textup{max}}_{\mathrm S}$ to $1 \times 10^4$, as significantly more iterations are generally required for convergence.

The pulmonary arterial mesh consists of $2.11 \times 10^6$ linear tetrahedral elements and $3.97 \times 10^5$ nodes, corresponding to $1.59 \times 10^6$ degrees of freedom in the associated linear system. Solver performance was investigated with varying values of the prescribed Young's modulus over three orders of magnitude, namely $E=1.3\times 10^5$, $1.3 \times 10^6$, and $1.3 \times 10^7$. The rigid-wall CFD simulation was also included as the extreme case of an infinitely large Young's modulus. Simulations were performed on a single node with $16$ CPUs. Figure \ref{fig:conv_history_pul} depicts all convergence histories and further annotates the CPU time for the linear solver averaged over ten time steps. We observe that with increasing Young's moduli, all preconditioners require increasingly more iterations and time to converge. This can be understood from the wall contribution $\boldsymbol{\mathrm K}_{\mathrm{m},(l),\dot{\bm u}^w}$ in \eqref{eq: predictor_multi_correct_notation_for_block_matrices}$_1$, in which an increased wall stiffness engenders stronger heterogeneity for the block matrix $\boldsymbol{\mathrm A}$. In contrast, no wall contribution is present in the rigid-wall case, and all block preconditioners require fewer iterations and less time to converge. In addition, the additive Schwarz and Jacobi preconditioners closely resemble each other in convergence behavior and are evidently less robust than the block preconditioners, signifying the importance of leveraging the block structure of $\boldsymbol{\mathcal A}$.

\begin{figure}
	\begin{center}
	\begin{tabular}{cc}
\includegraphics[angle=0, trim=90 90 130 90, clip=true, scale = 0.33]{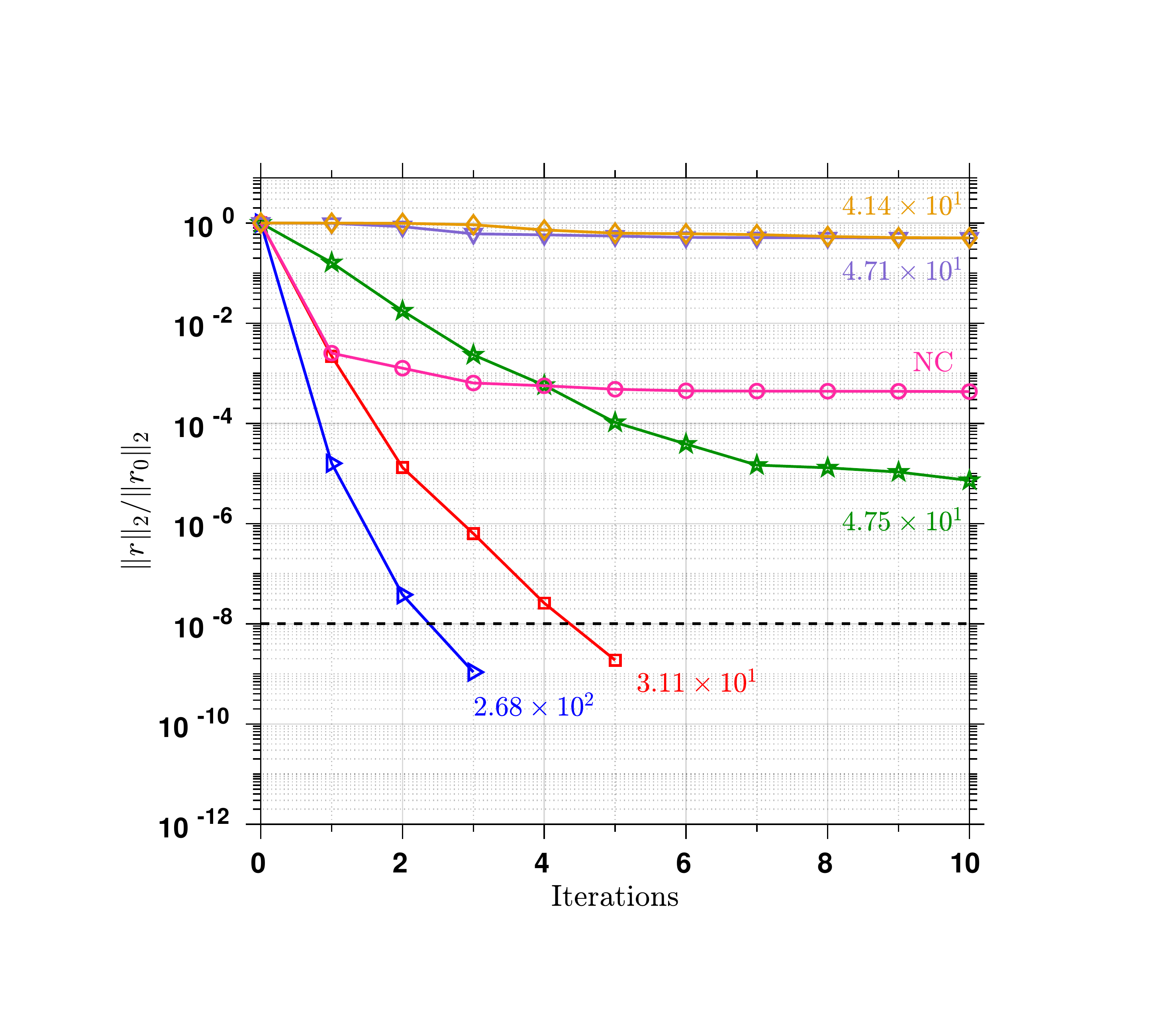} &
\includegraphics[angle=0, trim=90 90 130 90, clip=true, scale = 0.33]{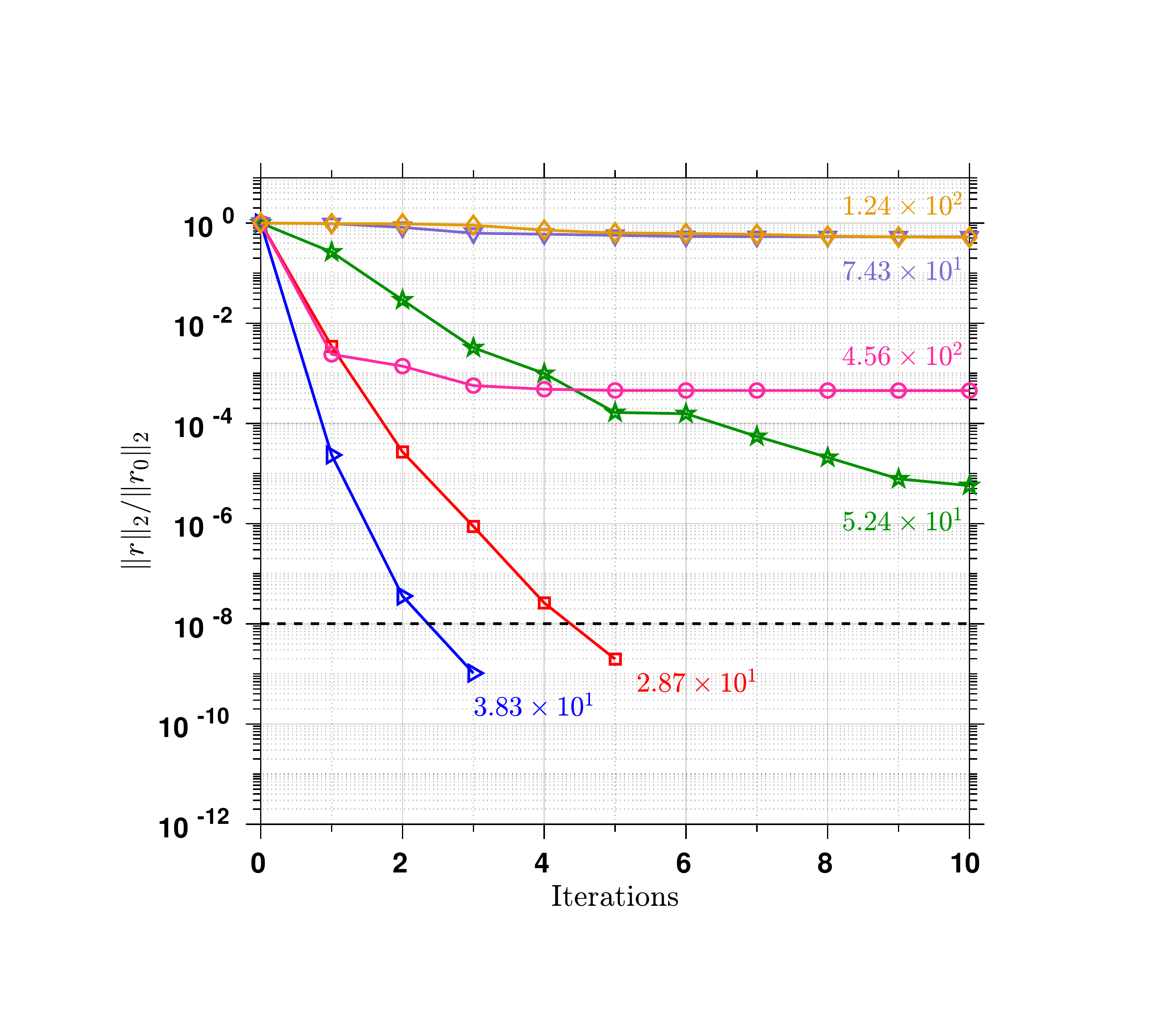} \\
(A) $\nu = 0.3$ & (B) $\nu = 0.5$  \\
\multicolumn{2}{c}{ \includegraphics[angle=0, trim=420 120 840 170, clip=true, scale = 0.45]{./figures/conv_history_legend.pdf}} \\
\multicolumn{2}{c}{ \includegraphics[angle=0, trim=1140 120 320 170, clip=true, scale = 0.45]{./figures/conv_history_legend.pdf}}\end{tabular}
\caption{Convergence history for the coronary arterial model with two different values of the Poisson's ratio. The latter three items in the legend correspond to the three alternative linear solver options investigated. The horizontal dashed black line demarcates the prescribed stopping criterion for the relative error $\delta = 10^{-8}$. CPU times (s) for the linear solver averaged over ten time steps are annotated. NC: no convergence within the prescribed maximum number of iterations.} 
\label{fig:conv_history_cor}
\end{center}
\end{figure}

The coronary arterial mesh consists of $1.66 \times 10^6$ elements and $3.15 \times 10^5$ nodes, corresponding to $1.26 \times 10^6$ degrees of freedom in the associated linear system. Solver performance was investigated for two values of the Poisson's ratio $\nu = 0.5$ and $0.3$. Simulations were performed on a single processor with $48$ CPUs. Figure \ref{fig:conv_history_cor} depicts all convergence histories and again annotates the CPU time for the linear solver averaged over ten time steps. Only minor differences are observed between the two cases, suggesting a smaller impact of $\nu$ on the linear system as compared to $E$.

\subsection{Fixed-size scalability}
\begin{figure}
	\begin{center}
\includegraphics[angle=0, trim=120 80 130 100, clip=true, scale = 0.45]{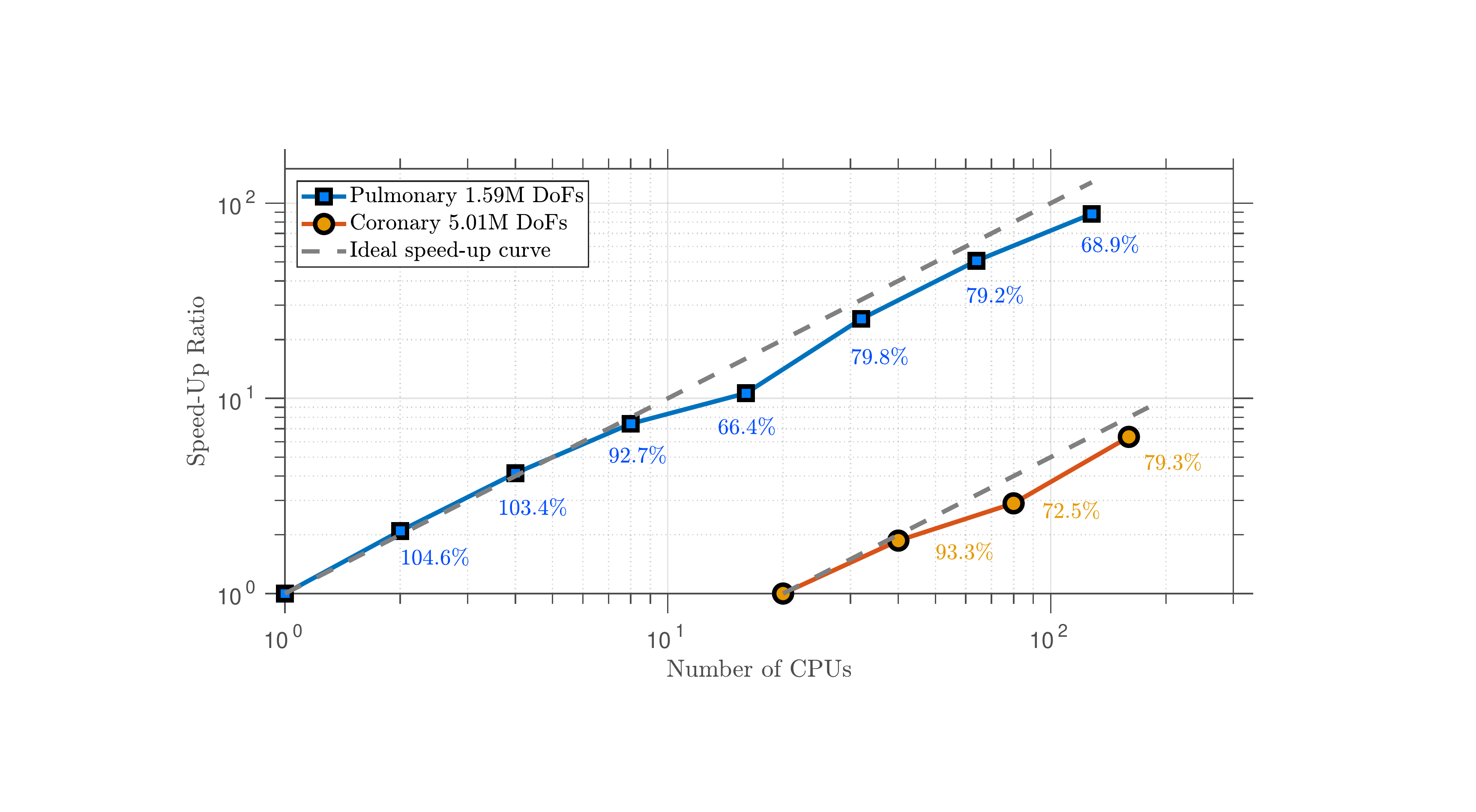}
\caption{Fixed-size scalability of our solution strategy. Annotated efficiency rates are computed from the total runtime.} 
\label{fig:strong_scaling}
\end{center}
\end{figure}

We examined the parallel performance of our proposed solution strategy, setting the relative tolerances $\delta = 10^{-8}$, $\delta_{\mathrm A}=\delta_{\mathrm S}=10^{-4}$, and $\delta_{\mathrm I}=10^{-2}$. While the same pulmonary arterial mesh was used as in Section \ref{subsection:solver_robustness}, we used a finer coronary mesh of $6.44 \times 10^6$ linear elements and $1.25 \times 10^6$ nodes, corresponding to $5.01 \times 10^6$ degrees of freedom in the associated linear system. Speed-up ratios were calculated based on a serial simulation for the pulmonary mesh and a parallel simulation with $20$ CPUs for the coronary mesh. Each job was run for $20$ time steps. For the pulmonary arterial mesh, super-optimal parallel efficiency was observed for $2$ and $4$ CPUs (Figure \ref{fig:strong_scaling}), likely a consequence of more efficient utilization of the cache in these scenarios.

\subsection{Performance of the segregated predictor multi-corrector algorithm}
\label{subsec:zero_kinematic_residual_numerical_evidence}
As discussed above in Remark \ref{remark:zero_kinematic_residual}, we have conveniently chosen $\boldsymbol{\mathrm R}_{\mathrm{k},(l)} = \bm 0$ for all $l \geq 1$ to allow for the simplified right-hand side in the linear system \eqref{eq:pred_multi_correct_linear_system} in the segregated predictor multi-corrector algorithm. In Table \ref{table:model_residuals}, we document the nonlinear residual $\boldsymbol{\mathrm R}_{(l)}$ and kinematic residual $\boldsymbol{\mathrm R}_{\mathrm{k},(l)}$ at all Newton-Raphson iterations within two time steps of the cardiac cycle for each model, one at peak systole and the other at mid-diastole. We note that $\boldsymbol{\mathrm R}_{\mathrm{k},(l)} < 10^{-12}$ beginning with $l=1$ and is driven close to machine precision for $l \geq 2$, closely agreeing with our prior analysis of the segregated algorithm \cite{Liu2019a}. We also note the expected quadratic convergence of the relative nonlinear residual in the first two iterations of the Newton-Raphson procedure. The convergence rate from the second to the third iteration is slightly reduced, likely a consequence of the linear solver accuracy.

\begin{table}[htbp]
\footnotesize
\begin{center}
\tabcolsep=0.3cm
\renewcommand{\arraystretch}{1.3}
\begin{tabular}{c | c c c c}
\hline
\hline
Model & $n$ & $l$ & $\boldsymbol{\mathrm R}_{(l)}/\boldsymbol{\mathrm R}_{(0)}$ & $\boldsymbol{\mathrm R}_{\mathrm{k},(l)}$ \\
\hline
\multirow{6}{*}{Pulmonary} & \multirow{3}{*}{$491$} & $1$ & $3.40 \times 10^{-1}$  & $4.73 \times 10^{-15}$  \\
& & $2$ & $1.81 \times 10^{-5}$  & $4.51 \times 10^{-16}$ \\
& &  $3$ & $1.09 \times 10^{-7}$ & $6.78 \times 10^{-21}$\\
\cline{2-5}
& \multirow{3}{*}{$1246$} & $1$ & $2.23 \times 10^{-1}$ & $3.08 \times 10^{-15}$  \\
& & $2$ & $3.71 \times 10^{-5}$ & $2.35\times 10^{-16}$  \\
& & $3$ & $1.08 \times 10^{-7}$ & $3.50\times 10^{-18}$   \\
\hline
\multirow{6}{*}{Coronary} & \multirow{3}{*}{$446$} & $1$ & $2.64 \times 10^{0}$  & $2.49 \times 10^{-13}$ \\
& & $2$ & $4.22 \times 10^{-4}$  & $4.93 \times 10^{-14}$ \\
& &  $3$ & $2.78 \times 10^{-7}$ & $4.44 \times 10^{-16}$ \\
\cline{2-5}
& \multirow{3}{*}{$1223$} & $1$ & $2.28 \times 10^{0}$ & $6.45 \times 10^{-14}$  \\
& & $2$ & $8.51 \times 10^{-5}$ & $1.18\times 10^{-14}$  \\
& & $3$ & $4.12 \times 10^{-8}$ & $1.94\times 10^{-18}$  \\
\hline 
\hline
\end{tabular}
\end{center}
\caption{The nonlinear residual $\boldsymbol{\mathrm R}_{(l)}$ and kinematic residual $\boldsymbol{\mathrm R}_{\mathrm{k},(l)}$ at all nonlinear iterations within time steps corresponding to peak systole and mid-diastole.}
\label{table:model_residuals}
\end{table}

\subsection{Simulation with higher-order elements}
\begin{figure}
	\begin{center}
	\begin{tabular}{c}
\includegraphics[angle=0, trim=0 90 380 95, clip=true, scale=1.0]{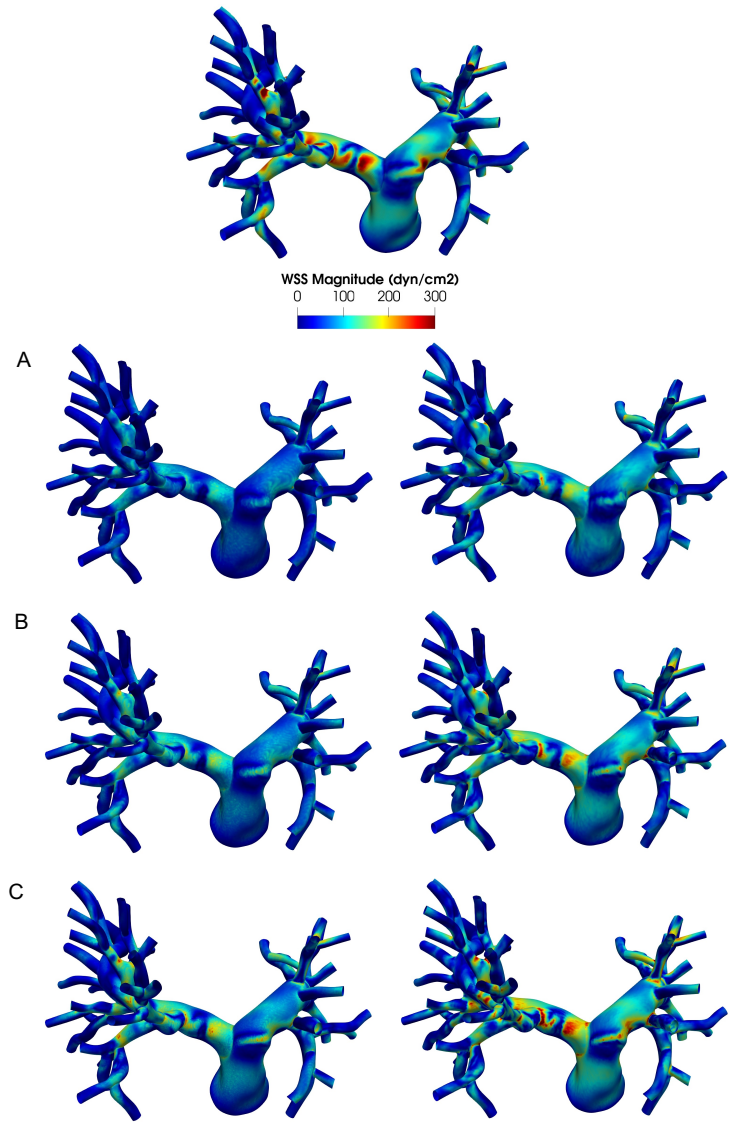}
\end{tabular}
\end{center}
\caption{Spatial convergence of peak systolic WSS using isotropic linear (left) and quadratic (right) tetrahedral meshes of the pulmonary arterial model at three refinement levels: (A) $4.0 \times 10^5$ nodes, (B) $8.0 \times 10^5$ nodes, (C) $1.6 \times 10^6$ nodes.} 
\label{fig:systolic-wss-spatial-convergence}
\end{figure}

Given limitations of the meshing software MeshSim, we were unable to generate a suite of spatially homogeneous quadratic tetrahedral meshes with boundary layers that would be of tractable computational cost. We therefore investigated the spatial convergence of peak systolic WSS with isotropic pulmonary arterial meshes of linear and quadratic tetrahedral elements at three refinement levels of comparable numbers of nodes ($4.0 \times 10^5$, $8.0 \times 10^5$, and $1.6 \times 10^6$ nodes). For each model, the coarsest isotropic mesh was chosen to match the number of nodes in the linear tetrahedral mesh with three boundary layers used in Section \ref{subsection:solver_robustness}. WSS results from the boundary layer mesh were verified to be mesh independent and taken as reference values. Consistent with observations from our spatial convergence study in Section \ref{section:womersley_rigid}, quadratic elements resolve WSS more accurately (Figure \ref{fig:systolic-wss-spatial-convergence}). We do, however, note the presence of undesirable oscillations yielding sharp local gradients and local WSS over/underestimations. 

\subsection{Patient-specific simulations}
\begin{figure}
	\begin{center}
	\begin{tabular}{c}
\includegraphics[angle=0, trim=190 90 150 90, clip=true, scale=1.1]{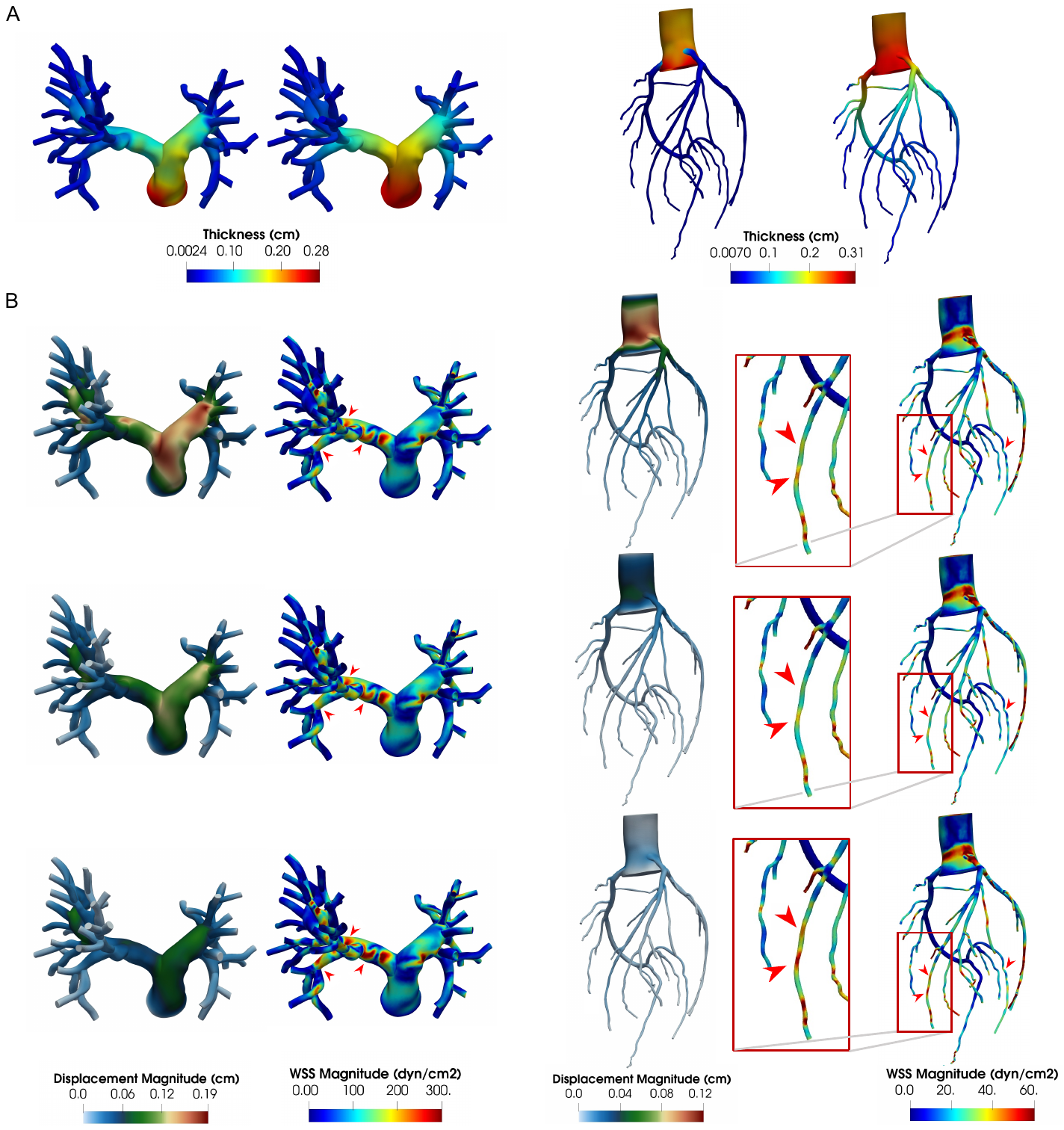}
\end{tabular}
\end{center}
\caption{Effects of different wall thickness distributions and tissue prestressing on patient-specific pulmonary (left) and coronary (right) arterial models. (A) Centerline-based (left) and Laplacian wall thickness distributions. (B) Peak systolic wall displacement (left) and wall shear stress (WSS; right) magnitudes for unprestressed simulations with centerline-based thickness (top), prestressed simulations with centerline-based thickness (middle), and prestressed simulations with Laplacian-based thickness (bottom). Red arrows and detailed views are included to highlight WSS differences across the three cases.} 
\label{fig:patient-specific_systolic_comparisons}
\end{figure}

In order to assess our proposed techniques for variable wall thickness assignment and tissue prestressing, three simulations were performed for each model: (i) an unprestressed simulation with centerline-based thickness, (ii) a prestressed simulation with centerline-based thickness, (iii) a prestressed simulation with Laplacian-based thickness. In the centerline-based approach, the local thickness was prescribed to be $20$\% of the local radius everywhere; in the Laplacian approach, the thickness was prescribed to be $20$\% of the corresponding cap radius at all wall boundary nodes (Figure \ref{fig:patient-specific_systolic_comparisons}A). Simulations were performed over three cardiac cycles with uniform time steps and verified for convergence to a limit cycle. For each simulation, only the final cardiac cycle was analyzed. 

The wall displacement and WSS distributions at peak systole were compared across the three cases (Figure \ref{fig:patient-specific_systolic_comparisons}B). Relative to Simulation (ii), failure to consider tissue prestressing in Simulation (i) \textit{overestimates} the maximum wall displacement magnitude by $37.9\%$ ($0.189$ vs. $0.137$ cm) in the pulmonary arterial model and by $162\%$ ($0.119$ vs. $0.0453$ cm) in the coronary arterial model. It also overestimates the mean displacement magnitude by $29.4$\% ($0.0616$ vs. $0.0476$ cm) over the main pulmonary arterial (MPA) bifurcation, by $159$\% ($0.01829$ vs. $0.00707$ cm) over the aortic root, and by $183$\% ($0.0267$ vs. $0.00942$ cm) over the left coronary artery (LCOR). While prestressing yields significantly different displacements in both models, our results suggest that prestressing is particularly critical in the systemic circulation where the diastolic pressure is an order of magnitude larger than that in the pulmonary circulation. In contrast, the Laplacian-based thickness in Simulation (iii) \textit{underestimates} the mean displacement magnitude by $29.8$\% ($0.0334$ vs. $0.0453$ cm) over the MPA bifurcation, by $45.3$\% ($0.00387$ vs. $0.00707$ cm) over the aortic root, and by $45.7$\% ($0.00511$ vs. $0.00942$ cm) over the LCOR. We note that while we prescribed Young's moduli that were previously determined to yield cross-sectional relative area changes observed from PC-MRI, tissue prestressing was not considered in these prior studies \cite{Yang2019,Ramachandra2016}. More compliant wall properties would thus need to be considered to achieve the same relative area changes under prestressing. 

While local discrepancies in peak systolic WSS are also observed across the three cases (as highlighted by the red arrows in Figure \ref{fig:patient-specific_systolic_comparisons}(B), these drastic discrepancies in displacement do not produce discrepancies in volumetric flow rates or spatially averaged WSS quantities. In fact, the mean WSS magnitude only differs by up to $2.03$\% over the MPA bifurcation, $0.174\%$ over the aorta, and $0.149$\% over the LCOR. Despite the rationale behind our proposed modeling techniques, the merits of Simulation (ii) remain to be assessed with in vivo and/or in vitro validation data.

\section{Conclusions}
\label{sec:conclusion}
In this work, we derived a reduced unified continuum formulation for vascular FSI and presented strong verification of our numerical methodology against Womersley's deformable wall theory using both linear and quadratic tetrahedral elements. Compared to the unified continuum ALE formulation, our reduced theory invokes three assumptions for the vascular wall to achieve monolithic FSI coupling in the Eulerian frame for small-strain problems. The residual-based VMS formulation is adopted for spatial discretization, and the generalized-$\alpha$ method is adopted for temporal discretization such that velocity and pressure are uniformly second-order accurate in time, a significant improvement over the predominant dichotomous approach. Block preconditioning of a monolithically coupled FSI system is also performed for the first time. Using two patient-specific models, we demonstrated the fixed-size scalability and enhanced robustness of our nested block preconditioner as compared to alternative preconditioners for vascular FSI applications. To appropriately model physiological phenomena, we further outlined a centerline-based approach for wall thickness assignment and a fixed-point algorithm for prestressing the vascular wall at the imaged configuration. Validation of our combined FSI methodology against in vivo and/or in vitro data will be examined in future studies.

\section*{Acknowledgements}
This work was supported by the National Institutes of Health [grant numbers 1R01HL121754, 1R01HL123689, R01EB01830204], Southern University of Science and Technology [startup grant number Y01326127], the National Natural Science Foundation of China [grant number 12172160], and the Guangdong-Hong Kong-Macao Joint Laboratory for Data-Driven Fluid Mechanics and Engineering Applications [grant number 2020B1212030001]. Ingrid S. Lan was supported by the National Science Foundation (NSF) Graduate Research Fellowship and the Stanford Graduate Fellowship in Science and Engineering. Computational resources were provided by the Stanford Research Computing Center, the Extreme Science and Engineering Discovery Environment supported by NSF [grant number ACI-1053575], and the Center for Computational Science and Engineering at Southern University of Science and Technology.

\bibliographystyle{elsarticle-num}
\bibliography{reduced-formulation}

\begin{thebibliography}{100}
\expandafter\ifx\csname url\endcsname\relax
  \def\url#1{\texttt{#1}}\fi
\expandafter\ifx\csname urlprefix\endcsname\relax\def\urlprefix{URL }\fi
\expandafter\ifx\csname href\endcsname\relax
  \def\href#1#2{#2} \def\path#1{#1}\fi

\bibitem{Hirt1974}
C.~Hirt, A.~Amsden, J.~Cook, An arbitrary {L}agrangian-{E}ulerian computing
  method for all flow speeds, Journal of Computational Physics 14 (1974)
  227--253.

\bibitem{Hughes1981}
T.~Hughes, W.~Liu, T.~Zimmermann, Lagrangian-{E}ulerian finite element
  formulation for incompressible viscous flows, Computer Methods in Applied
  Mechanics and Engineering 29 (1981) 329--349.

\bibitem{Donea1982}
J.~Donea, S.~Giuliani, J.~Halleux, An arbitrary {L}agrangian-{E}ulerian
  finite-element method for transient dynamic fluid structure interactions,
  Computer Methods in Applied Mechanics and Engineering 33 (1982) 689--723.

\bibitem{Peskin1972}
C.~Peskin, Flow patterns around heart valves: {A} numerical method, Journal of
  Computational Physics 10 (1972) 252--271.

\bibitem{Mittal2005}
R.~Mittal, G.~Iaccarino, Immersed boundary methods, Annual Review of Fluid
  Mechanics 37 (2005) 239--261.

\bibitem{Baaijens2001}
F.~Baaijens, A fictitious domain/mortar element method for fluid-structure
  interaction, International Journal for Numerical Methods in Fluids 35 (2001)
  743--761.

\bibitem{Borazjani2013}
I.~Borazjani, Fluid-structure interaction, immersed boundary-finite element
  method simulations of bio-prosthetic heart valves, Computer Methods in
  Applied Mechanics and Engineering 257 (2013) 103--116.

\bibitem{Griffith2009}
B.~Griffith, X.~Luo, D.~McQueen, C.~Peskin, Simulating the fluid dynamics of
  natural and prosthetic heart valves using the immersed boundary method,
  International Journal of Applied Mechanics 1 (2009) 137--177.

\bibitem{Kamensky2015}
D.~Kamensky, M.~Hsu, D.~Schillinger, J.~Evans, A.~Aggarwal, Y.~Bazilevs,
  M.~Sacks, T.~Hughes, An immersogeometric variational framework for
  fluid–structure interaction: {A}pplication to bioprosthetic heart valves,
  Computer Methods in Applied Mechanics and Engineering 284 (2015) 1005--1053.

\bibitem{Hart2003}
J.~de~Hart, G.~Peters, P.~Schreurs, F.~Baaijens, A three-dimensional
  computational analysis of fluid-structure interaction in the aortic valve,
  Journal of Biomechanics 36 (2003) 103--12.

\bibitem{Loon2006}
R.~van Loon, P.~Anderson, F.~van~de Vosse, A fluid-structure interaction method
  with solid-rigid contact for heart valve dynamics, Journal of Computational
  Physics 217 (2006) 806--823.

\bibitem{Wu2014}
Y.~Wu, X.~Cai, A fully implicit domain decomposition based {ALE} framework for
  three-dimensional fluid-structure interaction with application in blood flow
  computation, Journal of Computational Physics 258 (2014) 524--537.

\bibitem{Hsu2014}
M.~Hsu, D.~Kamensky, Y.~Bazilevs, M.~Sacks, T.~Hughes, Fluid-structure
  interaction analysis of bioprosthetic heart valves: significance of arterial
  wall deformation, Computational Mechanics 54 (2014) 1055--1071.

\bibitem{Liu2018}
J.~Liu, A.~Marsden, A unified continuum and variational multiscale formulation
  for fluids, solids, and fluid-structure interaction, Computer Methods in
  Applied Mechanics and Engineering 337 (2018) 549--597.

\bibitem{Liu2020b}
J.~Liu, W.~Yang, I.~Lan, A.~Marsden, Fluid-structure interaction modeling of
  blood flow in the pulmonary arteries using the unified continuum and
  variational multiscale formulation, Mechanics Research Communications 107
  (2020) 103556.

\bibitem{Fernandez2011}
M.~Fern\'{a}ndez, Coupling schemes for incompressible fluid-structure
  interaction: implicit, semi-implicit and explicit, SeMA Journal 55 (2011)
  59--108.

\bibitem{Nobile2013}
F.~Nobile, M.~Pozzoli, C.~Vergara, Time accurate partitioned algorithms for the
  solution of fluid-structure interaction problems in haemodynamics, Computers
  \& Fluids 86 (2013) 470--482.

\bibitem{Piperno2000}
S.~Piperno, C.~Farhat, {D}esign of {E}fficient {P}artitioned {P}rocedures for
  the {T}ransient {S}olution of {A}eroelastic {P}roblems, European Journal of
  Computational Mechanics 9 (2000) 655--680.

\bibitem{Badia2008}
S.~Badia, A.~Quaini, A.~Quarteroni, Modular vs. non-modular preconditioners for
  fluid-structure systems with large added-mass effect, Computer Methods in
  Applied Mechanics and Engineering 197 (2008) 4216--4232.

\bibitem{Causin2005}
P.~Causin, J.~Gerbeau, F.~Nobil, Added-mass effect in the design of partitioned
  algorithms for fluid-structure problems, Computer Methods in Applied
  Mechanics and Engineering 194 (2005) 4506--4527.

\bibitem{Guidoboni2009}
G.~Guidoboni, R.~Glowinski, N.~Cavallini, S.~Canic, Stable loosely-coupled-type
  algorithm for fluid-structure interaction in blood flow, Journal of
  Computational Physics 228 (2009) 6916--6937.

\bibitem{Badia2008a}
S.~Badia, F.~Nobile, C.~Vergara, Fluid-structure partitioned procedures based
  on {R}obin transmission conditions, Journal of Computational Physics 227
  (2008) 7027--7051.

\bibitem{Kadapa2021}
C.~Kadapa, Insights into the performance of loosely-coupled {FSI} schemes based
  on {R}obin boundary conditions, arXiv preprint arXiv:2105.14831 (2021).

\bibitem{Figueroa2006}
C.~Figueroa, I.~Vignon-Clementel, K.~Jansen, T.~Hughes, C.~Taylor, A coupled
  momentum method for modeling blood flow in three-dimensional deformable
  arteries, Computer Methods in Applied Mechanics and Engineering 195 (2006)
  5685--5706.

\bibitem{Nama2020}
N.~Nama, M.~Aguirre, J.~Humphrey, C.~Figueroa, A nonlinear rotation-free shell
  formulation with prestressing for vascular biomechanics, Scientific Reports
  10 (2020) 17528.

\bibitem{Womersley1955}
J.~Womersley, Oscillatory motion of a viscous liquid in a thin-walled elastic
  tube—{I}: {T}he linear approximation for long waves, The London, Edinburgh,
  and Dublin Philosophical Magazine and Journal of Science 46 (1955) 199--221.

\bibitem{Zamir2000}
M.~Zamir, The physics of pulsatile flow, Springer, New York, 2000.

\bibitem{Updegrove2017}
A.~Updegrove, N.~Wilson, J.~Merkow, H.~Lan, A.~Marsden, S.~Shadden,
  Sim{V}ascular: {A}n {O}pen {S}ource {P}ipeline for {C}ardiovascular
  {S}imulation, Annals of Biomedical Engineering 45 (2017) 525--541.

\bibitem{Lan2018}
H.~Lan, A.~Updegrove, N.~M. Wilson, G.~D. Maher, S.~C. Shadden, A.~L. Marsden,
  A {R}e-{E}ngineered {S}oftware {I}nterface and {W}orkflow for the
  {O}pen-{S}ource {S}im{V}ascular {C}ardiovascular {M}odeling {P}ackage,
  Journal of Biomechanical Engineering 140~(2) (2018) 0245011--02450111.

\bibitem{Arthurs2020}
C.~Arthurs, R.~Khlebnikov, A.~Melville, A.~Gomez, D.~Dillon-Murphy, F.~Cuomo,
  M.~Vieira, J.~Schollenberger, S.~Lynch, C.~Tossas-Betancourt, et~al.,
  Crimson: An open-source software framework for cardiovascular integrated
  modelling and simulation, bioRxiv (2020).

\bibitem{Williams2010}
A.~Williams, B.~Koo, T.~Gundert, P.~Fitzgerald, J.~L. Jr., Local hemodynamic
  changes caused by main branch stent implantation and subsequent virtual side
  branch balloon angioplasty in a representative coronary bifurcation, Journal
  of Applied Physiology 109 (2010) 532--540.

\bibitem{Gundert2011}
T.~Gundert, S.~Shadden, A.~Williams, B.~Koo, J.~Feinstein, J.~L. Jr., A rapid
  and computationally inexpensive method to virtually implant current and
  next-generation stents into subject-specific computational fluid dynamics
  models, Annals of Biomedical Engineering 39 (2011) 1423--1437.

\bibitem{Taylor2013}
C.~Taylor, T.~Fonte, J.~Min, Computational {F}luid {D}ynamics {A}pplied to
  {C}ardiac {C}omputed {T}omography for {N}oninvasive {Q}uantification of
  {F}ractional {F}low {R}eserve: {S}cientific {B}asis, Journal of the American
  College of Cardiology 61 (2013) 2233--2241.

\bibitem{Coogan2010}
J.~Coogan, F.~Chan, C.~Taylor, J.~Feinstein, Computational fluid dynamic
  simulations of aortic coarctation comparing the effects of surgical‐ and
  stent‐based treatments on aortic compliance and ventricular workload,
  Catheterization and Cardiovascular Interventions 77 (2010) 680--691.

\bibitem{Yang2010}
W.~Yang, J.~Feinstein, A.~Marsden, Constrained optimization of an idealized
  {Y}-shaped baffle for the {F}ontan surgery at rest and exercise, Computer
  Methods in Applied Mechanics and Engineering 199 (2010) 2135--2149.

\bibitem{Yang2016}
W.~Yang, J.~Feinstein, I.~Vignon-Clementel, Adaptive outflow boundary
  conditions improve post-operative predictions after repair of peripheral
  pulmonary artery stenosis, Biomechanics and Modeling in Mechanobiology 15
  (2016) 1345--1353.

\bibitem{Yang2017}
W.~Yang, F.~Hanley, F.~Chan, A.~Marsden, I.~Vignon‐Clementel, J.~Feinstein,
  Computational simulation of postoperative pulmonary flow distribution in
  {A}lagille patients with peripheral pulmonary artery stenosis, Congenital
  Heart Disease 13 (2017) 241--250.

\bibitem{Kung2011}
E.~Kung, A.~Les, C.~Figueroa, F.~Medina, K.~Arcaute, R.~Wicker, M.~McConnell,
  C.~Taylor, In vitro validation of finite element analysis of blood flow in
  deformable models, Annals of Biomedical Engineering 39 (2011) 1947--1960.

\bibitem{Kung2011a}
E.~Kung, A.~Les, F.~Medina, R.~Wicker, M.~McConnell, C.~Taylor, In vitro
  validation of finite-element model of {AAA} hemodynamics incorporating
  realistic outlet boundary conditions, Journal of Biomechanical Engineering
  (2011).

\bibitem{Figueroa2006a}
C.~Figueroa, A {C}oupled-{M}omentum {M}ethod to {M}odel {B}low {F}low and
  {V}essel {D}eformation in {H}uman {A}rteries: {A}pplications in {D}isease
  {R}esearch and {S}imulation-{B}ased {M}edical {P}lanning, Ph.D. thesis,
  Stanford University (2006).

\bibitem{Filonova2019}
V.~Filonova, C.~Arthurs, I.~Vignon-Clementel, C.~Figueroa, Verification of the
  coupled-momentum method with {W}omersley's {D}eformable {W}all analytical
  solution, International Journal for Numerical Methods in Biomedical
  Engineering 36 (2019) e3266.

\bibitem{Bazilevs2007}
Y.~Bazilevs, V.~Calo, J.~Cottrell, T.~Hughes, A.~Reali, G.~Scovazzi,
  Variational multiscale residual-based turbulence modeling for large eddy
  simulation of incompressible flows, Computer Methods in Applied Mechanics and
  Engineering 197 (2007) 173--201.

\bibitem{Chung1993}
J.~Chung, G.~Hulbert, A {T}ime {I}ntegration {A}lgorithm for {S}tructural
  {D}ynamics {W}ith {I}mproved {N}umerical {D}issipation: {T}he
  {G}eneralized-$\alpha$ {M}ethod, Journal of Applied Mechanics 60 (1993)
  371--375.

\bibitem{Jansen2000}
K.~Jansen, C.~Whiting, G.~Hulbert, A generalized-$\alpha$ method for
  integrating the filtered {N}avier-{S}tokes equations with a stabilized finite
  element method, Computer Methods in Applied Mechanics and Engineering 190
  (2000) 305--319.

\bibitem{Bazilevs2008}
Y.~Bazilevs, V.~Calo, T.~Hughes, Y.~Zhang, Isogeometric fluid-structure
  interaction: theory, algorithms, and computations, Computational Mechanics 43
  (2008) 3--–37.

\bibitem{Bazilevs2012}
Y.~Bazilevs, K.~Takizawa, T.~Tezduyar, Computational {F}luid-{S}tructure
  {I}nteraction: {M}ethods and {A}pplications, John Wiley \& Sons, Ltd, 2012.

\bibitem{Joshi2018}
V.~Joshi, R.~Jaiman, A hybrid variational {A}llen‐{C}ahn/{ALE} scheme for the
  coupled analysis of two‐phase fluid‐structure interaction, International
  Journal for Numerical Methods in Engineering 117 (2018) 405--429.

\bibitem{Kang2012}
S.~Kang, H.~Choi, J.~Yoo, Investigation of fluid-structure interactions using a
  velocity‐linked {P}2/{P}1 finite element method and the
  generalized‐$\alpha$ method, International Journal for Numerical Methods in
  Engineering 90 (2012) 1529--1548.

\bibitem{Liu2020a}
J.~Liu, I.~Lan, O.~Tikenogullari, A.~Marsden, A note on the accuracy of the
  generalized-$\alpha$ scheme for the incompressible {N}avier-{S}tokes
  equations, International Journal for Numerical Methods in Engineering 122
  (2021) 638--651.

\bibitem{Newmark1959}
N.~Newmark, A method of computation for structural dynamics, Proceedings of the
  American Society of Civil Engineers: Journal of the Engineering Mechanics
  Division (1959) 67--94.

\bibitem{Liu2020}
J.~Liu, W.~Yang, M.~Dong, A.~Marsden, The nested block preconditioning
  technique for the incompressible {N}avier-{S}tokes equations with emphasis on
  hemodynamic simulations, Computer Methods in Applied Mechanics and
  Engineering 367 (2020) 113122.

\bibitem{Hughes1987}
T.~Hughes, The Finite Element Method: Linear Static and Dynamic Finite Element
  Analysis, Prentice-Hall, 1987.

\bibitem{Liu2019}
J.~Liu, A.~Marsden, Z.~Tao, An energy-stable mixed formulation for isogeometric
  analysis of incompressible hyperelastodynamics, International Journal for
  Numerical Methods in Engineering 120 (2019) 937--963.

\bibitem{Colciago2014}
C.~Colciago, S.~Deparis, A.~Quarteroni, Comparisons between reduced order
  models and full 3{D} models for fluid-structure interaction problems in
  haemodynamics, Journal of Computational and Applied Mathematics 265 (2014)
  120--138.

\bibitem{Hulbert2017}
G.~Hulbert, Encyclopedia of Computational Mechanics, John Wiley \& Sons, Ltd,
  2017, Ch. Computational Structural Dynamics.

\bibitem{Alastruey2012}
J.~Alastruey, K.~Parker, S.~Sherwin, 11th {I}nternational {C}onference on
  {P}ressure {S}urges, Virtual PiE Led t/a BHR Group, 2012, Ch. Arterial pulse
  wave haemodynamics, pp. 401--443.

\bibitem{Bazilevs2009a}
Y.~Bazilevs, M.~Hsu, D.~Benson, S.~Sankaran, A.~Marsden, Computational
  fluid-structure interaction: methods and application to a total cavopulmonary
  connection, Computational Mechanics 45 (2009) 77--89.

\bibitem{Bischoff2004}
M.~Bischoff, K.-U. Bletzinger, W.~Wall, E.~Ramm, Encyclopedia of Computational
  Mechanics, John Wiley \& Sons, Ltd., 2004, Ch. Models and Finite Elements for
  Thin-Walled Structures, pp. 59--137.

\bibitem{Bergan1985}
P.~Bergan, C.~Felippa, A triangular membrane element with rotational degrees of
  freedom, Computer Methods in Applied Mechanics and Engineering 50 (1985)
  25--69.

\bibitem{Jun2018}
H.~Jun, K.~Yoon, P.~Lee, K.~Bathe, The {MITC}3+ shell element enriched in
  membrane displacements byinterpolation covers, Computer Methods in Applied
  Mechanics and Engineering 337 (2018) 458--480.

\bibitem{Debes1995}
J.~Debes, Y.~Fung, Biaxial mechanics of excised canine pulmonary arteries,
  American Journal of Physiology 269 (1995) H433--H442.

\bibitem{Zhou1997}
J.~Zhou, Y.~Fung, The degree of nonlinearity and anisotropy of blood vessel
  elasticity, Proceedings of the National Academy of Sciences of the United
  States of America 94 (1997) 14255--14260.

\bibitem{Franca1992}
L.~Franca, S.~Frey, Stabilized finite element methods: {II}. {T}he
  incompressible {N}avier-{S}tokes equations, Computer Methods in Applied
  Mechanics and Engineering 99 (1992) 209--233.

\bibitem{Bazilevs2009b}
Y.~Bazilevs, J.~Gohean, T.~Hughes, R.~Moser, Y.~Zhang, Patient-specific
  isogeometric fluid-structure interaction analysis of thoracic aortic blood
  flow due to implantation of the {J}arvik 2000 left ventricular assist device,
  Computer Methods in Applied Mechanics and Engineering 198 (2009) 3534--3550.

\bibitem{Moghadam2011}
M.~Moghadam, Y.~Bazilevs, T.~Hsia, I.~Vignon-Clementel, A.~Marsden, {Modeling
  Of Congenital Hearts Alliance (MOCHA)}, A comparison of outlet boundary
  treatments for prevention of backflow divergence with relevance to blood flow
  simulations, Computational Mechanics 48 (2011) 277--291.

\bibitem{Taylor1998}
C.~Taylor, T.~Hughes, C.~Zarins, Finite element modeling of blood flow in
  arteries, Computer Methods in Applied Mechanics and Engineering 158 (1998)
  155--196.

\bibitem{Gresho1998}
P.~Gresho, R.~Sani, Incompressible flow and the finite element method. {V}olume
  1: {A}dvection-diffusion and isothermal laminar flow, John Wiley \& Sons,
  Inc., New York, NY (United States), 1998.

\bibitem{Bazilevs2007a}
Y.~Bazilevs, C.~Michler, V.~Calo, T.~Hughes, Weak {D}irichlet boundary
  conditions for wall-bounded turbulent flows, Computer Methods in Applied
  Mechanics and Engineering 196 (2007) 4853--4862.

\bibitem{Colomes2015}
O.~Colom{\'e}s, S.~Badia, R.~Codina, J.~Principe, Assessment of variational
  multiscale models for the large eddy simulation of turbulent incompressible
  flows, Computer Methods in Applied Mechanics and Engineering 285 (2015)
  32--63.

\bibitem{Brooks1982}
A.~Brooks, T.~Hughes, Streamline upwind/{P}etrov-{G}alerkin formulations for
  convection dominated flows with particular emphasis on the incompressible
  {N}avier-{S}tokes equations, Computer Methods in Applied Mechanics and
  Engineering 32 (1982) 199--259.

\bibitem{Hughes2000}
T.~Hughes, L.~Mazzei, K.~Jansen, Large eddy simulation and the variational
  multiscale method, Computing and Visualization in Science 3 (2000) 47--59.

\bibitem{Pauli2017}
L.~Pauli, M.~Behr, On stabilized space-time {FEM} for anisotropic meshes:
  {I}ncompressible {N}avier-{S}tokes equations and applications to blood flow
  in medical devices, International Journal for Numerical Methods in Fluids 85
  (2017) 189--209.

\bibitem{Danwitz2019}
M.~von Danwitz, V.~Karyofylli, N.~Hosters, M.~Behr, Simplex space-time meshes
  in compressible flow simulations, International Journal for Numerical Methods
  in Fluids 91 (2019) 29--48.

\bibitem{Figueroa2017}
C.~Figueroa, C.~Taylor, A.~Marsden, Encyclopedia of {C}omputational
  {M}echanics, John Wiley \& Sons, Ltd, 2017, Ch. Blood Flow.

\bibitem{Taylor2009}
C.~Taylor, C.~Figueroa, Patient-specific modeling of cardiovascular mechanics,
  Annual Review of Biomedical Engineering 11 (2009).

\bibitem{Simo1992}
J.~Simo, N.~Tarnow, K.~Wong, Exact energy-momentum conserving algorithms and
  symmetric schemes for nonlinear dynamics, Computer Methods in Applied
  Mechanics and Engineering 100 (1992) 63--116.

\bibitem{Hilber1978}
H.~Hilber, T.~Hughes, Collocation, dissipation and `overshoot' for time
  integration schemes in structural dynamics, Earthquake Engineering \&
  Structural Dynamics 6 (1978) 99--117.

\bibitem{Kadapa2017}
C.~Kadapa, W.~Dettmer, D.~Peri\'{c}, On the advantages of using the first-order
  generalised-alpha scheme for structural dynamic problems, Computers \&
  Structures 193 (2017) 226--238.

\bibitem{Moghadam2013}
M.~Moghadam, I.~Vignon-Clementel, R.~Figliola, A.~Marsden, {Modeling Of
  Congenital Hearts Alliance (MOCHA) Investigators}, A modular numerical method
  for implicit 0{D}/3{D} coupling in cardiovascular finite element simulations,
  Journal of Computational Physics 244 (2013) 63--79.

\bibitem{Scovazzi2016}
G.~Scovazzi, B.~Carnes, X.~Zeng, S.~Rossi, A simple, stable, and accurate
  linear tetrahedral finite element for transient, nearly, and fully
  incompressible solid dynamics: a dynamic variational multiscale approach,
  International Journal for Numerical Methods in Engineering 106 (2016)
  799--839.

\bibitem{Liu2019a}
J.~Liu, A.~Marsden, A robust and efficient iterative method for
  hyper-elastodynamics with nested block preconditioning, Journal of
  Computational Physics 383 (2019) 72--93.

\bibitem{Johan1991a}
Z.~Johan, T.~Hughes, A globally convergent matrix-free algorithm for implicit
  time-marching schemes arising in finite element analysis in fluids, Computer
  Methods in Applied Mechanics and Engineering 87 (1991) 281--304.

\bibitem{Benzi2005}
M.~Benzi, G.~Golub, J.~Liesen, Numerical solution of saddle point problems,
  Acta numerica 14 (2005) 1--137.

\bibitem{May2008}
D.~May, L.~Moresi, Preconditioned iterative methods for {S}tokes flow problems
  arising in computational geodynamics, Physics of the Earth and Planetary
  Interiors 171 (2008) 33--47.

\bibitem{Saad1993}
Y.~Saad, A flexible inner-outer preconditioned {GMRES} algorithm, SIAM Journal
  on Scientific Computing 14 (1993) 461--469.

\bibitem{Ieary2000}
A.~Ieary, R.~Falgout, V.~Henson, J.~Jones, T.~Manteuffel, S.~McCormick,
  G.~Miranda, J.~Ruge, Robustness and scalability of algebraic multigrid, SIAM
  Journal on Scientific Computing 21~(5) (2000) 1886--1908.

\bibitem{Wesseling2001}
P.~Wesseling, C.~Oosterlee, Geometric multigrid with applications to
  computational fluid dynamics, Journal of computational and applied
  mathematics 128~(1-2) (2001) 311--334.

\bibitem{Falgout2002}
R.~Falgout, U.~Yang, hypre: {A} library of high performance preconditioners,
  in: International Conference on Computational Science, Springer, 2002, pp.
  632--641.

\bibitem{Brown2010}
J.~Brown, Efficient nonlinear solvers for nodal high-order finite elements in
  3{D}, Journal of Scientific Computing 45 (2010) 48--63.

\bibitem{Davydov2020}
D.~Davydov, J.~Pelteret, D.~Arndt, M.~Kronbichler, P.~Steinmann, A matrix-free
  approach for finite-strain hyperelastic problems using geometric multigrid,
  International Journal for Numerical Methods in Engineering 121 (2020)
  2874--2895.

\bibitem{Knoll2004}
D.~Knoll, D.~Keyes, Jacobian-free {N}ewton-{K}rylov methods: a survey of
  approaches and applications, Journal of Computational Physics 193 (2004)
  357--397.

\bibitem{svsolver}
sv{S}olver {G}it{H}ub repository,
  \url{github.com/SimVascular/svSolver/blob/master/Code/FlowSolvers/ThreeDSolver/svSolver}.

\bibitem{Humphrey2002}
J.~Humphrey, Cardiovascular {S}olid {M}echanics: {C}ells, {T}issues, and
  {O}rgans, Springer Science + Business Media, 2002.

\bibitem{Wong2014}
J.~Wong, E.~Kuhl, Generating fibre orientation maps in human heart models using
  {P}oisson interpolation, Computer Methods in Biomechanics and Biomedical
  Engineering 17~(11) (2014) 1217--1226.

\bibitem{Xiao2013}
N.~Xiao, J.~Humphrey, C.~Figueroa, Multi-{S}cale {C}omputational {M}odel of
  {T}hree-{D}imensional {H}emodynamics within a {D}eformable {F}ull-{B}ody
  {A}rterial {N}etwork, Journal of Computational Physics 244 (2013) 22--40.

\bibitem{vmtk-website}
The {V}ascular {M}odeling {T}ool{K}it, \url{www.vmtk.org}.

\bibitem{Antiga2008}
L.~Antiga, M.~Piccinelli, L.~Botti, B.~Ene-Iordache, A.~Remuzzi, D.~Steinman,
  An image-based modeling framework for patient-specific computational
  hemodynamics, Medical \& Biological Engineering \& Computing 46 (2008) 1097.

\bibitem{Antiga2002}
L.~Antiga, Patient-{S}pecific {M}odeling of {G}eometry and {B}lood {F}low in
  {L}arge {A}rteries, Ph.D. thesis, Polytechnic University of Milan (2002).

\bibitem{Tezduyar2008}
T.~Tezduyar, S.~Sathe, M.~Schwaab, B.~Conklin, Arterial fluid mechanics
  modeling with the stabilized space–time fluid–structure interaction
  technique, International Journal for Numerical Methods in Fluids 57~(5)
  (2008) 601--629.

\bibitem{Hsu2011}
M.-C. Hsu, Y.~Bazilevs, Blood vessel tissue prestress modeling for vascular
  fluid–structure interaction simulation, Finite Elements in Analysis and
  Design 47~(6) (2011) 593--599.

\bibitem{Baeumler2020}
K.~B\"{a}umler, V.~Vedula, A.~Sailer, J.~Seo, P.~Chiu, G.~Mistelbauer, F.~Chan,
  M.~Fischbein, A.~Marsden, D.~Fleischmann, Fluid-structure interaction
  simulations of patient-specific aortic dissection, Biomechanics and Modeling
  in Mechanobiology https://doi.org/10.1007/s10237-020-01294-8 (2020).

\bibitem{VignonClementel2006}
I.~Vignon-Clementel, C.~Figueroa, K.~Jansen, C.~Taylor, Outflow boundary
  conditions for three-dimensional finite element modeling of blood flow and
  pressure in arteries, Computer Methods in Applied Mechanics and Engineering
  195 (2006) 3776--3796.

\bibitem{VignonClementel2010}
I.~Vignon-Clementel, C.~Figueroa, K.~Jansen, C.~Taylor, Outflow boundary
  conditions for 3{D} simulations of non-periodic blood flow and pressure
  fields in deformable arteries., Computer Methods in Biomechanics and
  Biomedical Engineering 13 (2010) 625--640.

\bibitem{Kung2020}
E.~Kung, C.~Corsini, A.~Marsden, I.~Vignon-Clementel, G.~Pennati, R.~Figliola,
  T.-Y. Hsia, {Modeling Of Congenital Hearts Alliance (MOCHA) Investigators},
  Multiscale {M}odeling of {S}uperior {C}avopulmonary {C}irculation:
  {H}emi-{F}ontan and {B}idirectional {G}lenn {A}re {E}quivalent, Seminars in
  Thoracic and Cardiovascular Surgery 32~(4) (2020) 883--892.

\bibitem{Yang2019}
W.~Yang, M.~Dong, M.~Rabinovitch, F.~P. Chan, A.~L. Marsden, J.~A. Feinstein,
  Evolution of hemodynamic forces in the pulmonary tree with progressively
  worsening pulmonary arterial hypertension in pediatric patients, Biomechanics
  and Modeling in Mechanobiology 18~(3) (2019) 779--796.

\bibitem{Ramachandra2016}
A.~Ramachandra, A.~Kahn, A.~Marsden, Patient-{S}pecific {S}imulations {R}eveal
  {S}ignificant {D}ifferences in {M}echanical {S}timuli in {V}enous and
  {A}rterial {C}oronary {G}rafts, Journal of Cardiovascular Translational
  Research 9~(279--290) (2016) 4.

\bibitem{Taiyi-machine-details}
{T}ai{Y}i {S}ystem, \url{https://www.top500.org/system/179572/}, accessed:
  2021-09-25.

\bibitem{Kim2010}
H.~Kim, I.~Vignon-Clementel, J.~Coogan, C.~Figueroa, K.~Jansen, C.~Taylor,
  Patient-specific modeling of blood flow and pressure in human coronary
  arteries, Annals of Biomedical Engineering 38~(10) (2010) 3195--3209.

\bibitem{Sankaran2012}
S.~Sankaran, M.~Esmaily~Moghadam, A.~M. Kahn, E.~E. Tseng, J.~M. Guccione,
  A.~L. Marsden, Patient-specific multiscale modeling of blood flow for
  coronary artery bypass graft surgery, Ann Biomed Eng 40~(10) (2012)
  2228--2242.

\bibitem{Gutierrez2019}
N.~Gutierrez, M.~Mathew, B.~McCrindle, J.~Tran, A.~Kahn, J.~Burns, A.~Marsden,
  Hemodynamic {V}ariables in {A}neurysms are {A}ssociated with {T}hrombotic
  {R}isk in {C}hildren with {K}awasaki {D}isease, International Journal of
  Cardiology 281 (2019) 15--21.

\end{thebibliography}

\appendix

\section{Zero-dimensional models for Neumann coupling}
\label{sec:appendix}

\subsection{Three-element Windkessel model}
The three-element Windkessel model (Figure \ref{fig:reduced_models}A), commonly known as the RCR model and described with the differential-algebraic equations below, models the compliance of the downstream vasculature with a capacitance $C^k$, the resistance of the downstream arteries with a proximal resistance $R_{\mathrm{p}}^k$, and the resistance of the downstream capillaries and veins with a distal resistance $R_{\mathrm{d}}^k$,
\begin{align*}
& \frac{dP_{\mathrm{c}}^k}{dt} = \frac{Q^k}{C^k} - \frac{P_{\mathrm{c}}^k - P_{\mathrm{d}}^k}{R_{\mathrm{d}}^k C^k}, \qquad
P^k = P_{\mathrm{c}}^k + R_{\mathrm{p}}^k Q^k,
\end{align*}
wherein $P_{\mathrm{c}}^k$ is the pressure proximal to the capacitance $C^k$, and $P_{\mathrm{d}}^k$ is the distal reference pressure. 

\subsection{Coronary model}
The coronary model (Figure \ref{fig:reduced_models}B) similarly models the compliance of the downstream vasculature with a capacitance $C^k$ and the resistance of the downstream arteries, capillaries, and veins with corresponding resistances $R_{\mathrm{a}}^k$, $R_{\mathrm{c}}^k$, and $R_{\mathrm{v}}^k$, respectively,
\begin{align*}
\frac{dP_{\mathrm{c}}^k}{dt} = \frac{Q^k}{C^k} - \frac{P_{\mathrm{c}}^k - P_{\mathrm{c},\mathrm{im}}^k}{R_{\mathrm{c}}^k C^k}, \quad
\frac{dP_{\mathrm{c},\mathrm{im}}^k}{dt} = \frac{P_{\mathrm{c}}^k - P_{\mathrm{c},\mathrm{im}}^k}{R_{\mathrm{c}}^k C_{\mathrm{im}}^k} - \frac{P_{\mathrm{c},\mathrm{im}}^k - P_{\mathrm{d}}^k}{R_{\mathrm{v}}^k C_{\mathrm{im}}^k} + \frac{dP_{\mathrm{im}}^k}{dt}, \quad 
P^k = P_{\mathrm{c}}^k + R_{\mathrm{a}}^k Q^k.
\end{align*}
The addition of an intramyocardial capacitance $C_{\mathrm{im}}^k$ (with a corresponding proximal pressure $P_{\mathrm{c},\mathrm{im}}^k$) and intramyocardial pressure $P_{\mathrm{im}}^k$, however, are distinctive features that capture the out-of-phase behavior of coronary flow relative to aortic flow. Namely, the increased resistance due to ventricular contraction yields restricted coronary flow during systole. Depending on the location of the coronary artery of interest, $P_{\mathrm{im}}^k$ is prescribed as either the left or right ventricular pressure over time. Readers may refer to \cite{Kim2010,Sankaran2012, Gutierrez2019} for more details.

\setcounter{figure}{0} 

\begin{figure}
	\begin{center}
	\begin{tabular}{c}
\includegraphics[angle=0, trim=150 365 320 170, clip=true, scale=1.1]{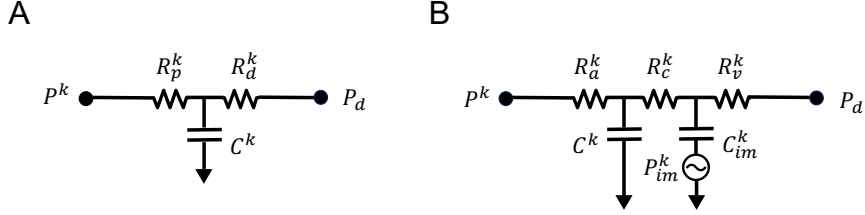}
\end{tabular}
\end{center}
\caption{Schematics of the (A) three-element Windkessel and (B) coronary outlet models commonly used in cardiovascular simulations.} 
\label{fig:reduced_models}
\end{figure}

\end{document}